\pdfoutput=1
\documentclass[12pt,a4paper]{article}
\usepackage[latin1]{inputenc}
\usepackage[doublespacing]{setspace}
\usepackage{amsmath}
\usepackage{graphicx}
\usepackage{xcolor,colortbl}
\usepackage{fullpage}
\usepackage{amsmath,amssymb,bm,bbm}
\usepackage{mathtools}
\colorlet{shadecolor}{lightgray!20}
\usepackage{natbib}
\usepackage{amsfonts}
\usepackage{amssymb}
\usepackage{authblk}
\usepackage{geometry}
\usepackage{pdflscape}
\usepackage{blkarray}
\definecolor{Gray}{gray}{0.9}
\definecolor{textGray}{gray}{0.95}
\definecolor{titleGray}{gray}{.85}
 
\usepackage[framemethod=tikz]{mdframed}
\newcounter{boxCount}
\renewcommand{\theboxCount}{\arabic{boxCount}}
\makeatletter
\def\textBox{
  \@ifnextchar[\textBox@opt{\textBox@noopt}}
\def\textBox@opt[#1]#2{
  \protected@edef\@currentlabelname{#1}
  \protected@edef\@currentlabel{#1}
  \refstepcounter{boxCount}
  \begin{mdframed}[
    innerlinewidth=0.5pt,
    innerleftmargin=10pt,
    innerrightmargin=10pt,
    innertopmargin = 10pt,
    innerbottommargin=10pt,
    skipabove=\dimexpr\topsep+\ht\strutbox\relax,
    roundcorner=5pt,
    frametitle={#2},
    frametitlerule=true,
    frametitlerulewidth=1pt,
    backgroundcolor=textGray,
    frametitlebackgroundcolor=titleGray
    ]
}
\def\textBox@noopt#1{
  \textBox@opt[#1]{#1}
}
\def\endtextBox{
  \end{mdframed}
}
\makeatother

\usepackage{caption}
\newcounter{boxTable}
\renewcommand{\theboxTable}{B\arabic{boxTable}}
\newenvironment{boxTable}[1][]{\refstepcounter{boxTable} \singlespacing \hspace{-.8cm} \textbf{Table \theboxTable.} }

\usepackage{totcount}
\newtotcounter{citnum}  
\def\oldbibitem{} \let\oldbibitem=\bibitem
\def\bibitem{\stepcounter{citnum}\oldbibitem}

\begin{document}
\sloppy
 
\begin{titlepage}
\begin{singlespacing}
\begin{center}
\huge{\textbf{Uncovering ecological state dynamics with hidden Markov models}}\\[1cm]
\large{Brett T.\ McClintock}\\
\normalsize{NOAA National Marine Fisheries Service, U.S.A.\\ brett.mcclintock@noaa.gov}\\[0.2cm]
\large{Roland Langrock}\\
\normalsize{Department of Business Administration and Economics, Bielefeld University\\ roland.langrock@uni-bielefeld.de}\\[0.2cm]
\large{Olivier Gimenez}\\
\normalsize{CNRS Centre d'Ecologie Fonctionnelle et Evolutive, France\\ olivier.gimenez@cefe.cnrs.fr}\\[0.2cm]
\large{Emmanuelle Cam}\\
\normalsize{Laboratoire des Sciences de l'Environnement Marin, Institut Universitaire Europ\'een de la Mer, Univ. Brest, CNRS, IRD, Ifremer, France\\ Emmanuelle.Cam@univ-brest.fr}\\[0.2cm]
\large{David L.\ Borchers}\\
\normalsize{School of Mathematics and Statistics, University of St Andrews\\ dlb@st-andrews.ac.uk}\\[0.2cm]
\large{Richard Glennie}\\
\normalsize{School of Mathematics and Statistics, University of St Andrews\\ rg374@st-andrews.ac.uk}\\[0.2cm]
\large{Toby A.\ Patterson}\\
\normalsize{CSIRO Oceans and Atmosphere, Australia\\ toby.patterson@csiro.au}\\[0.45cm]
\end{center}
\end{singlespacing}

\end{titlepage}

\onehalfspacing

\begin{abstract}
\noindent Ecological systems can often be characterised by changes among a finite set of underlying states pertaining to individuals, populations, communities, or entire ecosystems through time. Owing to the inherent difficulty of empirical field studies, ecological state dynamics operating at any level of this hierarchy can often be unobservable or ``hidden''. Ecologists must therefore often contend with incomplete or indirect observations that are somehow related to these underlying processes. By formally disentangling state and observation processes based on simple yet powerful mathematical properties that can be used to describe many ecological phenomena, hidden Markov models (HMMs) can facilitate inferences about complex system state dynamics that might otherwise be intractable. However, while HMMs are routinely applied in other disciplines, they have only recently begun to gain traction within the broader ecological community. We provide a gentle introduction to HMMs, establish some common terminology, and review the immense scope of HMMs for applied ecological research. We also provide a supplemental tutorial on some of the more technical aspects of HMM implementation and interpretation. By illustrating how practitioners can use a simple conceptual template to customise HMMs for their specific systems of interest, revealing methodological links between existing applications, and highlighting some practical considerations and limitations of these approaches, our goal is to help establish HMMs as a fundamental inferential tool for ecologists.
\end{abstract}

\section{Introduction}  

\begin{figure}
    \centering
    \includegraphics[width=\textwidth]{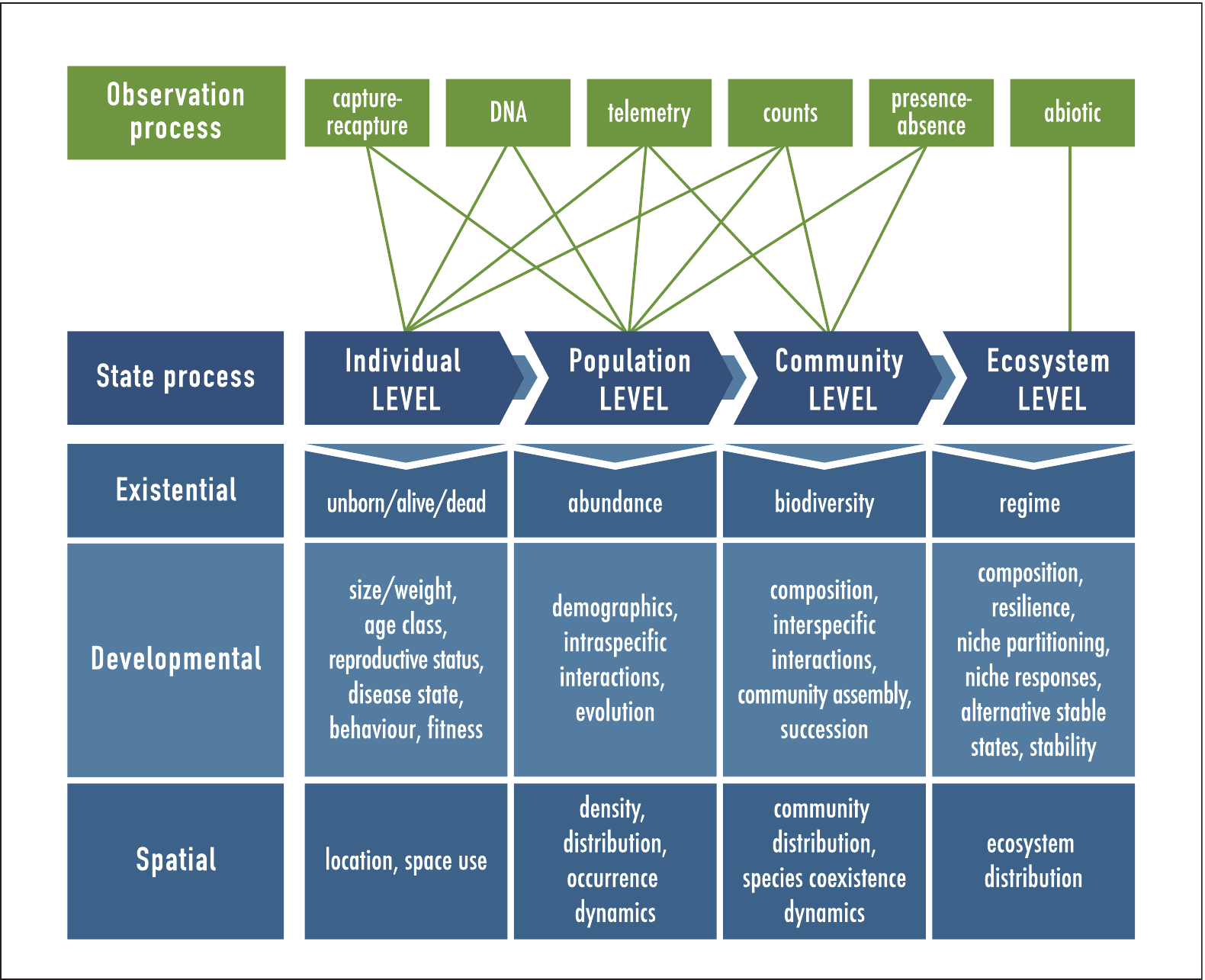}
    \caption{System state processes that can be difficult to observe directly, but can be uncovered from common ecological observation processes using hidden Markov models. The state process (blue) can pertain to any level within the ecological hierarchy (``Individual'', ``Population'', ``Commmunity'', or ``Ecosystem'') and for convenience is categorised as primarily ``Existential'', ``Developmental'', or ``Spatial'' in nature. The observation process (green) can provide information about state processes at different levels of the hierarchy (green lines) and includes capture-recapture, DNA sampling, animal-borne telemetry, count surveys, presence-absence surveys, and/or abiotic measurements. Observation and state processes from lower levels can be integrated for inferences at higher levels. For example, community-level biodiversity data could be combined with environmental data to describe ecosystem-level processes.
    }
    \label{fig:EcoStates}
\end{figure}

Ecological systems can often be characterised by changes among underlying system states through time. These state dynamics can pertain to individuals (e.g.\ birth, death), populations (e.g.\ increasing, decreasing), metapopulations (e.g.\ colonisation, extinction), communities (e.g.\ succession), or entire ecosystems (e.g.\ regime shifts). Gaining an understanding of state dynamics at each level of this hierarchy is a central goal of ecology and fundamental to studies of climate change, biodiversity, species distribution and density, habitat and patch selection, population dynamics, behaviour, evolution, and many other phenomena \citep[][]{BegonEtAl2006}. However, inferring ecological state dynamics is challenging for several reasons, including: 1) these complex systems often display non-linear, non-monotonic, non-stationary, and non-Gaussian behaviour \citep[][]{SchefferEtAl2001,TuckerAnand2005,Wood2010,PedersenEtAl2011,FasioloEtAl2016}; 2) changes in underlying states and dynamics can be rapid and drastic, but also gradual and more subtle \citep[][]{BeisnerEtAl2003,SchefferEtAl2003,FolkeEtAl2004}; and 3) the actual state of an ecological entity, be it an individual plant or animal, or a population or community, can often be difficult or impossible to observe directly \citep[][]{MartinEtAl2005,KerySchmidt2008,RoyleDorazio2008,ChenEtAl2013,KellnerSwihart2014}. Ecologists must therefore often contend with pieces of evidence believed to be informative of the state of an unobservable system at a particular point in time (see Fig.\ \ref{fig:EcoStates}). 

Whether for management, conservation, or empirical testing of ecological theory, there is a need for inferential methods that seek to uncover the relationships between factors driving such systems, and thereby predict them in quantitative terms. Hidden Markov models (HMMs) constitute a class of statistical models that has rapidly gained prominence in ecology because they are able to accommodate complex structures that account for changes between unobservable system states \citep[][]{EphraimMerhav2002,CappeEtAl2005,ZucchiniEtAl2016}. By simultaneously modelling two time series --- one consisting of the underlying state dynamics and a second consisting of observations arising from the true state of the system --- HMMs are able to detect state changes in noisy time-dependent phenomena by formally disentangling the state and observation processes. For example, using HMMs and their variants: 
\begin{itemize}
\item historical regime shifts can be identified from reconstructed chronologies;
\item long-term dynamics of populations, species, communities, and ecosystems in changing environments can be inferred from dynamic biodiversity data; 
\item species identity and biodiversity can be determined from environmental DNA (eDNA);
\item hidden evolutionary traits can be accounted for when assessing drivers of diversification;
\item species occurrence can be linked to variation in habitat, population density, land use, host-pathogen dynamics, or predator-prey interactions;
\item survival, dispersal, reproduction, disease status, and habitat use can be inferred from capture-recapture time series;
\item animal movements can be classified into foraging, migrating, or other modes for inferences about behaviour, activity budgets, resource selection, and physiology;
\item trade-offs between dormancy and colonisation can be inferred from standing flora or fungal fruiting bodies.
\end{itemize}

The increasing popularity of HMMs has been fuelled by new and detailed data streams, such as those arising from modern remote sensing and geographic information systems \citep[][]{ViovySaint1994,Gao2002}, eDNA \citep[][]{BalintEtAl2018}, and genetic sequencing \citep[][]{Hudson2008}, as well as advances in computing power and user-friendly software \citep[][]{VisserSpeenkenbrink2010}. However, despite their utility and ubiquity in other fields such as finance \citep[][]{BharShigeyuki2004}, speech recognition \citep[][]{Rabiner1989
}, and bioinformatics \citep[][]{DurbinEtAl1998
}, the vast potential of HMMs for uncovering latent system dynamics from readily available data remains largely unrecognised by the broader ecological community. This is likely attributable to a tendency for the existing ecological literature to characterise HMMs as a subject-specific tool reserved for a particular type of data rather than a general conceptual framework for probabilistic modelling of sequential data. This is also likely exacerbated by a tendency for HMMs to be applied and described quite differently across disciplines. Indeed, many ecologists may not recognise that some of the most well-established inferential frameworks in population, community, and movement ecology are in fact special cases of HMMs.

Catering to ecologists and non-statisticians, we describe the structure and properties of HMMs (Section \ref{sec:HMMs}), establish some common 
terminology (Table $\ref{tab:glossary}$), and review case studies from the biological, ecological, genetics, and statistical literature (Section \ref{sec:ecoapps}). Central to our review and synthesis is a simple but flexible conceptual template that ecologists can use to customise HMMs for their specific systems of interest. In addition to highlighting new areas where HMMs may be particularly promising in ecology, we also demonstrate cases where these models have (perhaps unknowingly) already been used by ecologists for decades. We then identify some practical considerations, including 
implementation, 
software, and potential 
challenges that practitioners may encounter when using HMMs 
(Section \ref{sec:implementation}). Using an illustrative example, we provide a brief tutorial on some of the more technical aspects of HMM implementation in the Supplementary Tutorial. The overall aim of our review is thus to provide a synthesis of the various ways in which HMMs can be used, reveal methodological links between existing applications, and thereby establish HMMs as a fundamental inferential tool for ecologists working with sequential data.

{\small
\setlength{\tabcolsep}{2pt}
\renewcommand{\arraystretch}{0.5}
\begin{table}
\caption{Glossary.\label{tab:glossary}}
\vspace{0em}
\begin{tabular}{|l|l|l|}
   \hline
   {\bf Term}  & {\bf Definition} & {\bf Synonyms} \\ \hline
{\it Conditional indepen-}       &  Assumption made for the state-dependent            &  \\
{\it dence property}             & process: conditional on the state at time $t$,      & \\
                                 &  the observation at time $t$ is independent of      & \\
                                 &   all other observations and states                 & \\[0.5em]
  {\it  Forward algorithm}       &  Recursive scheme for updating the                  &  filtering \\
                                 &  likelihood and state probabilities                 &   \\
                                 &  of an HMM through time                             &   \\[0.5em]
{\it Forward-backward}           &  Recursive scheme for calculating state             & local state decoding;\\
{\it algorithm}                  &   probabilities for any point in time:              & smoothing  \\
                                 &    $\Pr(S_t=i \mid x_1,\ldots,x_T)$                 &   \\[0.5em]
   {\it Hidden Markov}           &  A special class of state-space model               & dependent mixture model;\\
   {\it model (HMM)}             &  with a finite number of hidden states              & latent Markov model;\\
                                 &  that typically assumes some form of                 & Markov-switching model;\\
                                 &  the Markov property and the conditional            & regime-switching model;\\
                                 &  independence property                              & state-switching model; \\
                                 &                                                     & multi-state model\\[0.5em]
{\it Initial distribution} $({\boldsymbol \delta})$& The probability of being in any of the $N$& initial probabilities;\\
                                 &   states at the start of the sequence:              & prior probabilities \\
                                 &$\boldsymbol{\delta}=\bigl( \Pr(S_1=1),\ldots,\Pr(S_1=N) \bigr)$ & \\[0.5em]
   {\it Markov property}         &  Assumption made for the state process:             & memoryless property \\
                                 & $ \Pr(S_{t+1} \, \mid \, S_{t}, S_{t-1} ,\ldots) = 
                                                 \Pr(S_{t+1} \, \mid \, S_{t})  $         & \\ 
                                 & (``conditional on the present, the future           & \\ 
                                 & is independent of the past'')                       & \\[0.5em]
{\it Sojourn time}               &  The amount of time spent in a state before         & dwell time;\\
                                 &  switching to another state                         & occupancy time \\[0.5em]
   {\it State process} $(S_t)$   &  Unobserved, serially correlated sequence of        & hidden/latent process;\\ 
                                 & states describing how the system evolves over       & system process \\
                                 &  time: $S_t \in \{1,\ldots,N\}$ for $t=1,\ldots,T$& \\[0.5em]
{\it State transition}           &  The probability of switching from state $i$ at     & \\
{\it probability}  $(\gamma_{ij})$ & time $t$ to state $j$ at time $t+1$,              & \\
                                 & $\gamma_{ij}=\Pr(S_{t+1}=j \mid S_t=i)$, usually represented & \\
                                 & as a $N \times N$ transition probability matrix $(\boldsymbol \Gamma)$  & \\[0.5em]
{\it State-dependent}            & Probability distribution of an observation $x_t$    & emission distribution;\\
{\it distribution}               & conditional on a particular state being active      & measurement model;\\  
$\bigl(f(x_t \mid S_t=i) \bigr)$     & at time $t$, usually from some parametric class      & observation distribution;\\
                                 & (e.g.\ categorical, Poisson, normal) and            & output distribution; \\     
                                 &   represented as a $N \times N$ diagonal matrix      & response distribution\\
                                 &   ${\mathbf P}(x_t)=\text{diag}\bigl(f(x_t \mid S_t=1), \ldots, f(x_t \mid S_t =N)\bigr)$ & \\[0.5em]
{\it State-dependent}            &  The observed process within an HMM,                 & observation process \\
{\it process} $(X_t)$            &   which is assumed to be driven by the              & \\
                                 &  underlying unobserved state process                &  \\[0.5em]
{\it State-space model}          &  A conditionally specified hierarchical model       & 
\\                               &  consisting of two linked stochastic processes,     &  \\
                                 &  a latent system process model and an               &  \\
                                 &  observation process model                          & \\[0.5em]
{\it Viterbi algorithm}          &  Recursive scheme for finding the sequence of       & global state decoding \\
                                 &  states which is most likely to have given rise     &   \\
                                 &   to the observed sequence                          &   \\[0.5em]
   \hline
\end{tabular}
\end{table}
}

\section{Hidden Markov models} 
\label{sec:HMMs}
We begin by providing a gentle introduction to HMMs, including model formulation, inference, and extensions. Although we have endeavoured to minimise technical material and provide illustrative examples wherever possible, we assume the reader has at least some basic understanding of linear algebra concepts such as matrix multiplication and diagonal matrices \citep[e.g.\ see Appendix A in][]{Caswell2001} and probability theory concepts such as uncertainty, random variables, and probability distributions 
\citep[][Chapters 1--2]{GotelliEllison2013}. 

\subsection{Basic model formulation}

\begin{figure}[h!]
\centering
\includegraphics[]{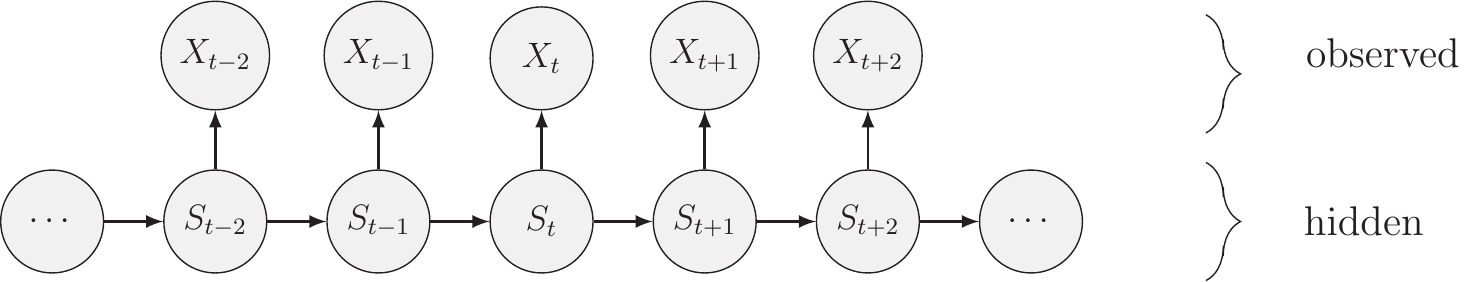}
\caption{Dependence structure of a basic hidden Markov model, with an observed sequence $X_1,\ldots,X_T$ arising from an unobserved sequence of underlying states $S_1,\ldots,S_T$. 
}
\label{fig:HMM}
\end{figure}

Hidden Markov models (HMMs) are a class of statistical models for sequential data, in most instances related to systems evolving over time. The system of interest is modelled using a \textit{state process} (or \textit{system process}; Table \ref{tab:glossary}), which evolves dynamically such that future states depend on the current state. Many ecological phenomena 
can naturally be described by such a process (Fig.\ \ref{fig:EcoStates}). In an HMM, the state process is not directly observed --- it is a ``hidden'' (or ``latent'') variable (see Box \ref{box:latent}). Instead, observations are made of a \textit{state-dependent process} (or \textit{observation process}) that is driven by the underlying state process. As a result, the observations can be regarded as noisy measurements of the system states of interest, but they are typically insufficient to precisely determine the state. Mathematically, an HMM is composed of two sequences:
\begin{itemize}
\item an observed state-dependent process $X_1, X_2, \ldots, X_T$;
\item an unobserved (hidden) state process $S_1,S_2,\ldots,S_T$.
\end{itemize}
In most applications, the indices refer to observations made over time at a regular sampling interval (e.g.\ daily or annual rainfall measurements),
but they can also refer to position \citep[e.g.\ in a sequence of DNA;][]{HendersonEtAl1997,eddy2004hidden} or order \citep[e.g.\ in a sequence of marine mammal dives;][]{DeRuiterEtAl2017}. 
HMMs can also be formulated in continuous time \citep[][]{jackson2003multistate,amoros2019continuous}, but these have 
tended to be 
less frequently applied in ecology \citep[but see ][]{langrock2013markov,ChoquetEtAl2017,OlajosEtAl2018}. 
Among the many HMM formulations of relevance to ecology that we highlight in Section \ref{sec:ecoapps}, some example observation sequences $(X_1,\ldots,X_T)$ and underlying states $(S_1,\ldots,S_T)$ include:
\begin{itemize}
    \item $X_t=$ observation of feeding/not feeding, with underlying state $S_t=$ hungry or sated;
    \item $X_t=$ count of individuals, with underlying state $S_t=$ true population abundance;
    \item $X_t=$ daily rainfall measurement, with underlying state $S_t=$ wet or dry season.
\end{itemize}
Unlike the larger class of {\it state-space models} (see Box \ref{box:latent}), the state process within an HMM can take on only finitely many possible values: $S_t \in \{1, \ldots, N\}$ for $t=1,\ldots,T$. The basic HMM formulation further involves two key dependence assumptions: 1) the probability of a particular state being active at any time $t$ is completely determined by the state active at time $t-1$ (the so-called \textit{Markov property}
); and 2) the probability distribution of an observation at any time $t$ is completely determined by the state active at time $t$ (Fig.\ \ref{fig:HMM}). The latter assumption is called the \textit{conditional independence property}, as this implies that $X_t$ is conditionally independent of past and future observations, given $S_t$. Whether or not these simplifying assumptions can faithfully characterise the underlying dynamics for the system of interest must be carefully considered (see Section \ref{sec:challenges}).
\bigskip

\begin{textBox}[1]{Box \theboxCount. Where do HMMs reside in the taxonomic zoo of latent variable models?}
\label{box:latent}
Latent state (or latent variable) models come in many different forms, with a particular variant often evolving its own nomenclature, notation, and jargon that can be confusing for non-specialists. Here we use broad and non-technical strokes to differentiate the HMM from its close relatives in the taxonomy of latent state models, with the aim to more clearly position HMMs relative to alternative modelling frameworks. Above all, these models are united by assuming \textit{latent states} --- a fundamental property of the system being modelled that is either partially, or completely, unobservable. They also tend to make a clear distinction between an observation process model --- describing noise in the data --- and the hidden state process model --- describing the underlying patterns and dynamics of interest. 

The umbrella terms mixed effects, multilevel, or hierarchical models \citep[e.g.][]{skrondal2004generalized,gelman2006data,RoyleDorazio2008,lee2012basic} typically include the most widely known types of latent variable models \citep[e.g.][]{clogg1995latent}. These often treat latent variables as random effects assumed to arise from a distribution as structural elements of a hierarchical statistical model. There is therefore not only random variation in the observations, but also in the parameters of the model itself. While there are special cases and generalisations that are not so easily classified, a simplified taxonomy for a subset of hierarchical latent variable models can be based on the structural dependence in the hidden state process and whether or not the state space of this hidden process is discrete (i.e.\ taking on finitely many values) or continuous (Table \ref{tab:latent}). 

\begin{boxTable}
\textbf{A simplistic taxonomy for a subset of hierarchical latent variable models based on temporal dependence in the hidden state process and the support of the state space.}
\label{tab:latent} 

\vspace{0.1in}
\hspace{-.65cm}
\begin{tabular}{lcc}
\hline
                               & \multicolumn{2}{c}{State space}\tabularnewline
\cline{2-3}
                               & Continuous     & Discrete \tabularnewline
\hline
Temporal dependence   & \textbf{State-space model}        & \textbf{Hidden Markov model} \tabularnewline
Temporal independence & \textbf{Continuous mixture model} & \textbf{Finite mixture model} \tabularnewline
\hline
\end{tabular}
\vspace{0.3in}  
\end{boxTable}

Latent variable models with a continuous state space and no temporal dependence in the hidden state process fall under the broad class of \textit{continuous mixture models} \citep[e.g.][]{lindsay1995mixture}, with ecological applications including the modelling of closed population abundance \citep[][]{royle2004n}, disease prevalence \citep[][]{calabrese2011partitioning}, and species distribution \citep[][]{OvaskainenEtAl2017}. \textit{State-space models} (SSMs) are a special class of continuous mixture model where the observation process is conditionally specified by a continuous hidden state process with temporal dependence \citep[e.g.][]{durbin2012time,auger2020introduction}, with applications including population dynamics \citep[][]{schnute1994general,wang2007latent,tavecchia2009estimating,newman2014modelling}, disease dynamics \citep[][]{rohani2010never,cooch2012disease}, and animal movement \citep[][]{patterson2008state,HootenEtAl2017,patterson2017statistical}. An HMM is a special class of SSM where the state space is finite (see Section \ref{sec:ecoapps} for many ecological examples). \textit{Finite mixture models} \citep[e.g.][]{fruhwirth2006finite} assume the state space is finite with no temporal dependence in the hidden state process (e.g.\ the latent states are non-Markov or do not change over time), with examples including static species occurrence \citep[][]{MackenzieEtAl2002}, closed population capture-recapture \citep[][]{pledger2000unified}, and species distribution \citep[][]{pledger2014multivariate}  models. HMMs and SSMs can therefore be regarded as specific variations of a hierarchical model with serial dependence, where the random effects vary over time. Furthermore, an HMM can be viewed as a discrete version of a SSM or a time-dependent version of a finite mixture model. 

It is important to note that things may not be as simple as depicted in Table \ref{tab:latent}. For example, an HMM might include continuous random effects on its parameters or a state-dependent observation distribution specified as a finite mixture \citep[][]{Altman2007}. If the number of states becomes very large in an HMM, then it can become a discrete approximation of a SSM \citep[][]{BesbeasMorgan2019}. A SSM with both continuous and discrete latent variables can also encompass features of an HMM \citep[][]{JonsenEtAl2005}. Box \ref{box:toHMM} considers circumstances where application of a standard HMM is not supported and other approaches or extensions might be required. 
\end{textBox}
\vspace{.25in}

As a consequence of these assumptions, HMMs generally facilitate model building and computation that might otherwise be intractable. A basic $N$-state HMM that formally distinguishes the state and observation processes can be fully specified by the following three components: 1) the {\it initial distribution}, $\boldsymbol{\delta}=\bigl( \Pr(S_1=1),\ldots,\Pr(S_1=N) \bigr)$, specifying the probabilities of being in each state at the start of the sequence; 2) the {\it state transition probabilities}, $\gamma_{ij}=\Pr(S_{t+1}=j \mid S_t=i)$, specifying the probability of switching from state $i$ at time $t$ to state $j$ at time $t+1$ and usually represented as a $N \times N$ state transition probability matrix 
\begin{equation*}
\boldsymbol{\Gamma}=   
    \begin{blockarray}{ccccc}
    S_{t+1}=1 & S_{t+1}=2 & \ldots & S_{t+1}=N & \\
    \begin{block}{[cccc]c}
    \gamma_{1,1} & \gamma_{1,2} & \ldots & \gamma_{1,N} & S_t=1 \\
    \gamma_{2,1} & \gamma_{2,2} & \ldots & \gamma_{2,N} & S_t=2 \\
    \vdots       & \vdots       & \ddots & \vdots       & \vdots \\
    \gamma_{N,1} & \gamma_{N,2} & \ldots & \gamma_{N,N} & S_t=N \\
    \end{block}
    \end{blockarray}
\end{equation*}
where $\sum_{j=1}^N \gamma_{ij}=1$; and 3) the {\it state-dependent distributions}, $f(x_t \mid S_t=i)$, specifying the probability distribution of an observation $x_t$ conditional on the state at time $t$ and usually represented as a $N \times N$ diagonal matrix ${\mathbf P}(x_t)=\text{diag}\bigl(f(x_t \mid S_t=1), \ldots, f(x_t \mid S_t =N)\bigr)$ for computational purposes (see Section \ref{sec:inference}). These distributions can pertain to discrete or continuous observations and are generally chosen from an appropriate distributional family. For example, behavioural observation $X_t \in \{\text{feeding},\text{not feeding}\}$ could be modelled using a categorical distribution \citep{macdonald1995hidden}, count $X_t \in \{0,1,2,\ldots\}$ using a non-negative discrete distribution \citep[e.g.\ Poisson;][]{BesbeasMorgan2019}, and measurement $X_t \in [0,\infty)$ using a non-negative continuous distribution \citep[e.g.\ zero-inflated exponential;][]{WoolhiserRoldan1982}. After specifying ${\boldsymbol \delta}$, ${\boldsymbol \Gamma}$, and ${\mathbf P}(x_t)$ in terms of the particular system of interest, one can proceed to drawing inferences about unobservable state dynamics from the observation process.

\subsection{Inference}
\label{sec:inference}

In addition to the ease with which a wide variety of ecological state and observation process models can be specified (see Section \ref{sec:ecoapps}), a key strength of the HMM framework is that efficient recursive algorithms are available for conducting statistical inference. 
Here we will briefly outline some of the most common inferential techniques for HMMs, but motivated readers can find additional technical material and a worked example on model fitting, assessment, and interpretation in the Supplementary Tutorial. Using the \textit{forward algorithm}, the likelihood $\mathcal{L}(\boldsymbol{\theta} \mid x_1,\ldots,x_T)$ as a function 
of the unknown parameters $({\boldsymbol \theta})$ given the observation sequence $(x_1,\ldots,x_T)$ can be calculated at a computational cost that is (only) linear in $T$
. The parameter vector $\boldsymbol{\theta}$, which is to be estimated, contains any unknown parameters embedded in the three model-defining components $\boldsymbol{\delta}$
, $\boldsymbol{\Gamma}$
, and $\mathbf{P}({x}_{t})$. Made possible by the relatively simple dependence structure of an HMM, the forward algorithm traverses along the time series, updating the likelihood step-by-step while retaining information on the probabilities of being in the different states \citep[][pp.\ 37-39]{ZucchiniEtAl2016}. Application of the forward algorithm is equivalent to evaluating the likelihood using a 
simple matrix product expression,
\begin{equation}\label{eq:lik}
\mathcal{L}(\boldsymbol{\theta} \mid x_1,\ldots,x_T) = 
\boldsymbol{\delta} \mathbf{P}({x}_1) \boldsymbol{\Gamma} \mathbf{P}({x}_2) \cdots \boldsymbol{\Gamma} \mathbf{P}({x}_{T-1}) \boldsymbol{\Gamma} \mathbf{P}({x}_T) \mathbf{1} \, ,
\end{equation}
where $\mathbf{1}$ is a column vector of ones (see Supplementary Tutorial for technical derivation). 

In practice, the main challenge when working with HMMs tends to be the estimation of the model parameters. The two main strategies for fitting an HMM are numerical maximisation of the likelihood \citep[ML;][]{Myung2003,ZucchiniEtAl2016} or Bayesian inference \citep[][]{ellison2004bayesian,GelmanEtAl2004} using Markov chain Monte Carlo (MCMC) sampling \citep[][]{BrooksEtAl2011}. The former seeks to identify the parameter values that maximise the likelihood function (i.e.\ the maximum likelihood estimates ${\hat {\boldsymbol \theta}}$), whereas the latter yields a sample from the posterior distribution of the parameters \citep[][]{ellison2004bayesian}. Specifically for the ML approach, the forward algorithm makes it possible to use a standard optimisation routine, e.g.\ Newton-Raphson \citep[][]{Hildebrand1987}, 
to directly numerically maximise the likelihood (eqn \ref{eq:lik}). An alternative ML approach is to employ an expectation-maximisation (EM) algorithm that uses similar recursive techniques to iterate between \textit{state decoding} and updating the parameter vector until convergence \citep[][]{Rabiner1989}. For MCMC, many different strategies can be used \citep[][]{GelmanEtAl2004,BrooksEtAl2011}. However, MCMC samplers that include both the parameter vector $({\boldsymbol \theta})$ and the latent states $(S_1,\ldots,S_T)$ are inherently slow; sampling from the parameter vector only while applying the forward algorithm to evaluate the likelihood will often be preferable \citep[][]{TurekEtAl2016,YackulicEtAl2020}. From a computational point of view, none of these methods is vastly superior to the others, and in practice users will typically adopt the approach they are most comfortable with (cf.\ \citealp{patterson2017statistical,YackulicEtAl2020}).

The forward algorithm and similar recursive techniques can further be used for forecasting and state decoding, as well as to conduct formal model checking using pseudo-residuals  \citep[][Chapters 5 \& 6]{ZucchiniEtAl2016}. The latter task is usually accomplished using the \textit{Viterbi algorithm} or the \textit{forward-backward algorithm}, which respectively identify the most likely sequence of states or the probability of each state at any time $t$, conditional on the observations. Fortunately, practitioners can often use existing software for most aspects of HMM-based data analyses and need not dwell on many of the more technical details of implementation (see Section \ref{sec:implementation} and Supplementary Tutorial).

To illustrate some of the basic mechanics, we use a simple example based on observations of the feeding behaviour of a blue whale (\textit{Balaenoptera musculus}; cf.\ \citealp{DeRuiterEtAl2017})
. Suppose we assume that observations of the number of feeding lunges performed in each of $T=53$ consecutive dives ($X_t \in \{0,1,2,\ldots\}$ for $t=1,\ldots,T$) arise from $N=2$ states of feeding activity. Building on Fig.\ \ref{fig:HMM}, we could for example have:

\bigskip
\begin{center}
\includegraphics[]{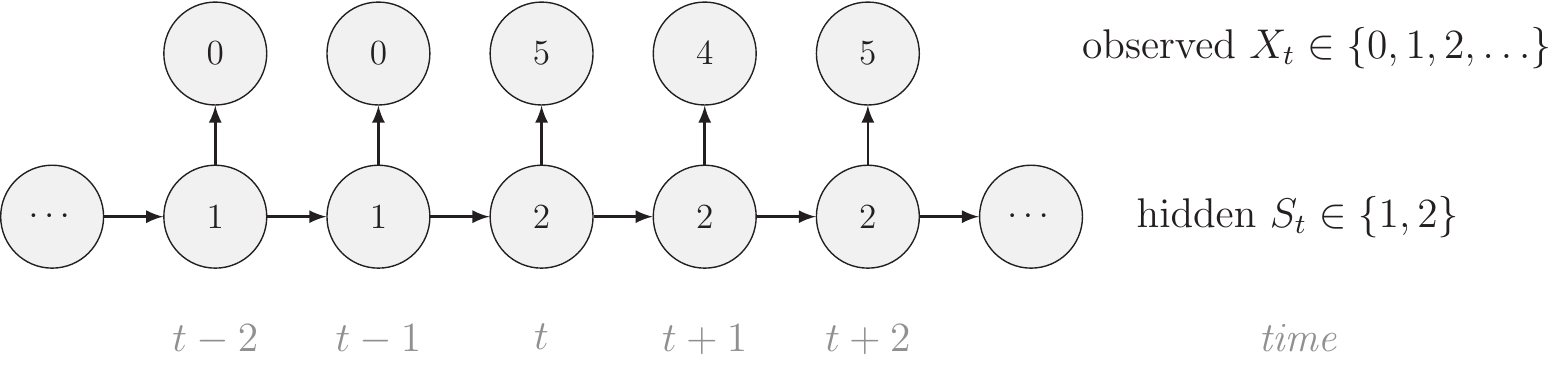}\\
\end{center}
\begin{figure}[h!]
\centering
\includegraphics[width=0.75\textwidth]{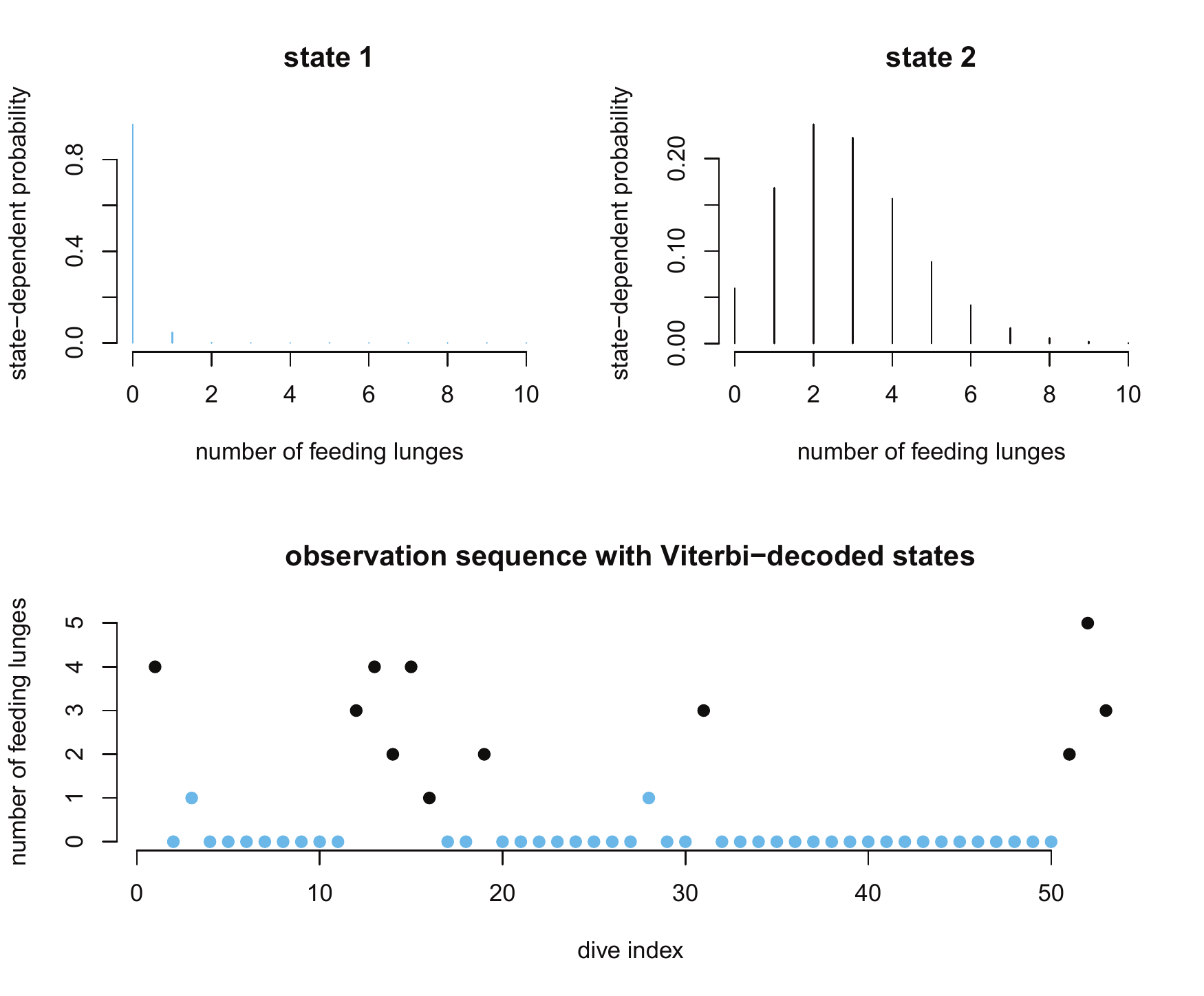}
\caption{Estimated state-dependent distributions (top row) and Viterbi-decoded 
states from a 2-state HMM fitted to counts of feeding lunges performed by a blue whale during a sequence of $T=53$ consecutive dives
. Here the most likely state sequence identifies periods of ``low'' (state 1; blue) and ``high'' (state 2; black) feeding activity.}\label{fig:bluewhale}
\end{figure} 

Fig.\ \ref{fig:bluewhale} displays the results for this simple 2-state HMM assuming Poisson state-dependent (observation) distributions, $X_t \mid S_t=i \sim \text{Poisson}(\lambda_i)$ for $i \in \{1,2\}$, when fitted to the full observation sequence via direct numerical maximisation of eqn \ref{eq:lik}. The rates of the state-dependent distributions were estimated as $\hat{\lambda}_1=0.05$ and $\hat{\lambda}_2=2.82$, suggesting states 1 and 2 correspond to ``low'' and ``high'' feeding activity, respectively. The estimated state transition probability matrix,
\begin{equation*}
    \hat{\boldsymbol{\Gamma}} =
    \begin{blockarray}{ccc}
    S_{t+1}=1 & S_{t+1}=2 & \\
    \begin{block}{[cc]c}
    0.88 & 0.12 & S_t=1 \\
    0.36 & 0.64 & S_t=2 \\
    \end{block}
    \end{blockarray},
\end{equation*}
suggests interspersed bouts of ``low'' and ``high'' feeding activity, but with bouts of ``high'' activity tending to span fewer dives. 
The estimated initial distribution $\hat{\boldsymbol{\delta}}=(0.75,0.25)$ suggests this individual was more likely to have been in the ``low'' activity state at the start of the sequence.
Most ecological applications of HMMs involve more complex inferences related to specific hypotheses about system state dynamics, and a great strength of the HMM framework is 
the relative ease with which the basic model formulation can be modified to describe a wide variety of processes \citep[][Chapters 9-13]{ZucchiniEtAl2016}. 
In Section \ref{sec:extensions}, we highlight some extensions that we consider to be highly relevant in ecological research.

\subsection{Extensions}
\label{sec:extensions}

\begin{figure}[!htb]
    \centering
    \includegraphics[width=\textwidth]{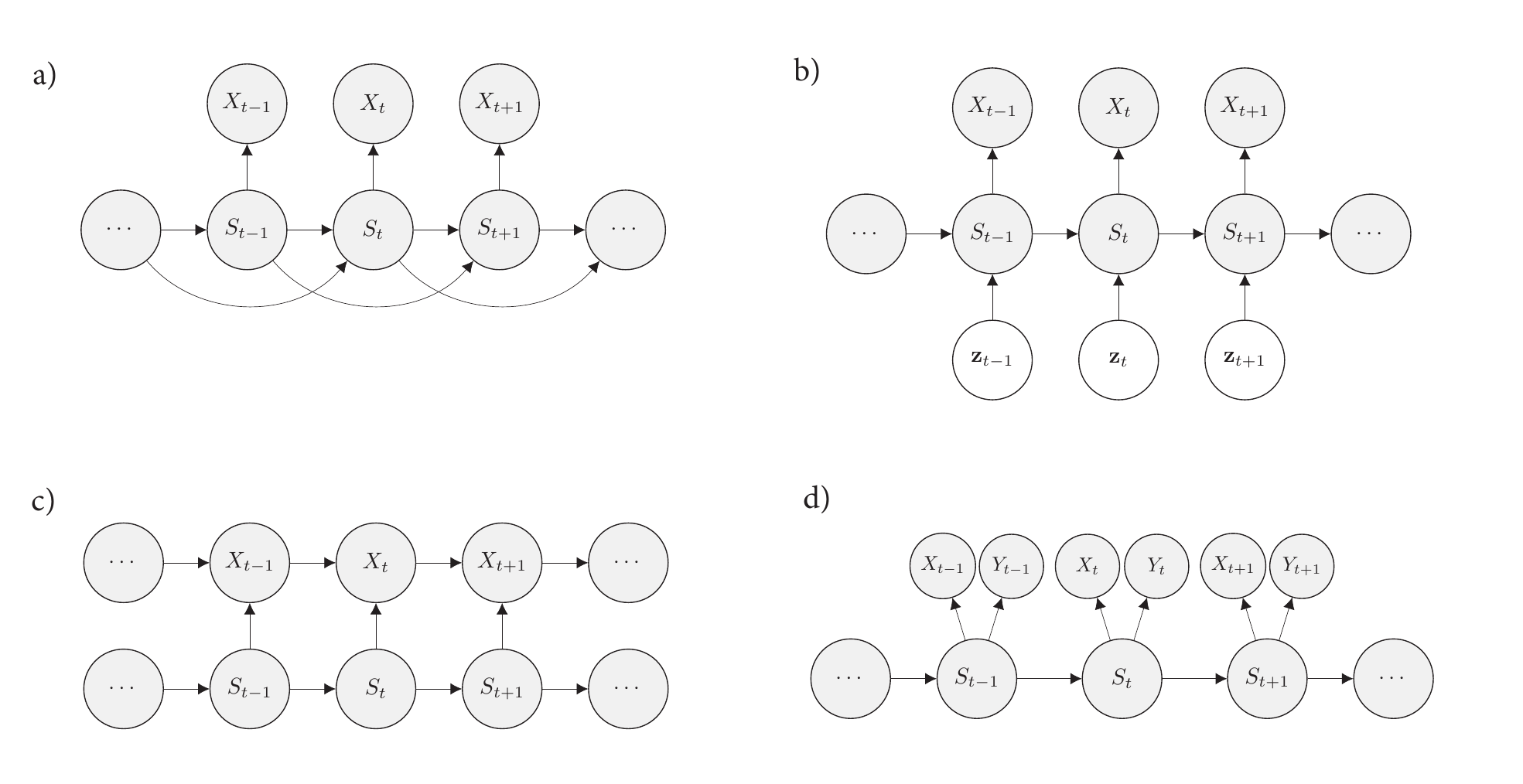}
    \caption{Graphical models associated with different extensions of the basic HMM formulation: a) state sequence with memory order 2; b) influence of covariate vectors $\mathbf{z}_1,\ldots,\mathbf{z}_T$ on state dynamics; c) observations depending on both states and previous observations; d) bivariate observation sequence, conditionally independent given the states.}
    \label{fig:EcoStates4}
\end{figure}

The dependence assumptions made within the basic HMM are mathematically convenient, but not always appropriate. The Markov property implies that the amount of time spent in a state before switching to another state --- the so-called \textit{sojourn time} --- follows a geometric distribution. The most likely length of any given sojourn time hence is one unit, which may not be realistic for certain state processes. The obvious extension is to allow for 
$k$th-order dependencies in the state process (Fig.\ \ref{fig:EcoStates4}a), such that the state at time $t$ depends on the states at times $t-1,t-2,\ldots,t-k$. 
A more parsimonious alternative assumes the state process is ``semi-Markov'' with the sojourn time flexibly modelled using any distribution on the positive integers \citep[][]{ChoquetEtAl2011,van2015hidden,king2016semi}. 

HMMs are often used to infer drivers of ecological state processes by relating the state transition probabilities 
to explanatory covariates (Fig.\ \ref{fig:EcoStates4}b). Indeed, any of the parameters of a basic HMM can be modelled as a function of covariates (e.g.\ sex, age, habitat type, chlorophyll-a) using an appropriate link function \citep[][]{McCullaghNelder1989}. Link functions $(l)$ can relate the natural scale parameters $({\boldsymbol \theta})$ to a $T \times r$ design matrix of covariates $({\mathbf Z})$ and $r$-vector of working scale parameters $({\boldsymbol \beta} \in \mathbb{R}^r)$ such that $l({\boldsymbol \theta})={\mathbf Z}{\boldsymbol \beta}$ \citep[see][for common examples of link functions in HMMs]{WhiteBurnham1999,MackenzieEtAl2002,PattersonEtAl2009}. When simultaneously analysing multiple observation sequences, 
potential heterogeneity across the different sequences can be modelled through explanatory covariates or 
mixed HMMs that include 
random effects \citep{Altman2007,schliehe2012application,TownerEtAl2016}. 

At the level of the observation process, it is relatively straightforward to relax the conditional independence assumption. For example, it can be assumed that the observation at time $t$ depends not only on the state at time $t$ but also the observation at time $t-1$ \citep[Fig.\ \ref{fig:EcoStates4}c;][]{langrock2014modeling,lawler2019conditionally}. 
It is also straightforward to model multivariate observation sequences 
using multivariate state-dependent distributions 
\citep[][]{choquet2013,phillips2015objective,van2019classifying}
, where it is often assumed that the different variables observed are conditionally independent of each other, given the state, 
and a univariate distribution is specified for each of the variables (Fig.\ \ref{fig:EcoStates4}d)
. This {\it contemporaneous conditional independence} assumption does not imply that the individual components are serially independent or mutually independent; the Markov property induces both serial dependence and cross-dependence in the different sequences \citep[][Chapter 9]{ZucchiniEtAl2016}. However, this assumption will not always be appropriate, in which case a multivariate distribution that accounts for any additional dependence between the different variables can be employed.

\section{Ecological applications of hidden Markov models}
\label{sec:ecoapps}
In their classic textbook, \citet{BegonEtAl2006} present the evolutionary foundation of ecology and its superstructure built from individual organisms to populations, communities, and ecosystems. 
At each level of this hierarchy, we will 
illustrate how HMMs can be used for identifying patterns and dynamics of many different types of ecological state variables that would otherwise be difficult or impossible to observe directly. 
For each application, we 
emphasise the two principle components of any HMM --- the observation process and the (hidden) state process --- as a conceptual template for ecologists to formulate HMMs in terms of their particular systems of interest.

The observation process in ecological studies is often driven by many factors, including the system state variable(s) of interest, the biotic and/or abiotic components of the system, and the desired level of inference (Fig.\ \ref{fig:EcoStates}). Among the most common types of observation processes in ecology are capture-recapture \citep[][]{WilliamsEtAl2002
}, DNA sampling \citep[][]{BohmannEtAl2014, RoweEtAl2017, BalintEtAl2018}, animal-borne telemetry \citep[][]{CookeEtAl2004, WhiteGarrott2012, HootenEtAl2017}, count surveys \citep[][]{BucklandEtAl2004, CharmantierEtAl2006, NicholsEtAl2009}, presence-absence surveys \citep[][]{KoleffEtAl2003, MackenzieEtAl2018}, and abiotic measurement (e.g.\ temperature, precipitation, sediment type). These observation processes are not mutually exclusive (e.g.\ capture-recapture or presence-absence time series can be derived from DNA samples), can contribute information at different levels of the hierarchy, and can be pooled for inference 
\citep[][]{Schaub2011, GimenezEtAl2012, EvansEtAl2016}. 

Using Fig.\ \ref{fig:EcoStates} as our expositional roadmap, we begin with 
applications for individual-level state dynamics
. We then work our way up to the population, community, and ecosystem levels. Within each level of the ecological hierarchy, we find it convenient to distinguish ``existential'', ``developmental'', and ``spatial'' states. Although there is inevitably some degree of overlap, particularly at the higher levels of the hierarchy that are inherently spatial, 
we use this distinction in an attempt to separate states of being that in isolation can be viewed as essentially non-spatial 
from state dynamics that are more strictly spatial in nature
. We further delineate the non-spatial states as ``Existental'' based on a fundamental measure of existence at each level of the hierarchy 
and ``Developmental'' based on specific characteristics of this fundamental measure of existence
.

Although typically not referred to as HMMs in the ecological literature, several subfields of ecology have been using HMMs for individual- to community-level inference for decades. 
HMMs have also become standard in biological sequence analysis and molecular ecology \citep{
DurbinEtAl1998,BarbuLimnios2009,Yoon2009
}, and there is much crossover potential for state-of-the-art bioinformatic methods to other applications in ecology \citep{JonesEtAl2006,TuckerDuplisea2012}. HMMs are also used for very specialised tasks of relevance to ecology, such as counting annual layers in ice cores \citep{WinstrupEtAl2012} or characterising plant architectures \citep{DurandEtAl2005}. There are therefore many example HMM applications within some areas of ecology, of which only a handful can be covered in the material that follows. However, in other areas 
the promise of HMMs has only just begun to be recognised.

\subsection{Individual level}
\label{sec:individualLevel}

\subsubsection{Existential state}
\label{sec:individualExist}
At the level of an individual organism, a fundamental measure of existence is to be alive or not (i.e.\ dead or unborn)
. We will therefore begin 
by demonstrating that one of the oldest and most popular inferential tools in wildlife ecology, the Cormack-Jolly-Seber (CJS) model of survival \citep[][]{
WilliamsEtAl2002}, is a special case of an HMM. The CJS model estimates survival probabilities ($\phi$) from capture-recapture data. 
Capture-recapture data consist of $n$ sequences of encounter histories for marked individuals collected through time, where for each individual the observed data are represented as a binary series of ones and zeros. For the CJS model, $X_t=1$ indicates a marked individual was alive and detected at time $t$, while $X_t=0$ indicates non-detection. Marked individuals can either be alive or dead at time $t$, but the ``alive'' state is only partially observable and the ``dead'' state is completely unobservable. Under this observation process, if $X_t=1$ it is known that the individual survived from time $t-1$ to time $t$ (with probability $\phi$) and was detected with probability $p$. However, when $X_t=0$ there are two possibilities: 1) the individual survived to time $t$ (with probability $\phi$) but was not detected (with probability $1-p$); or 2) the individual did not survive from time $t-1$ to time $t$ (with probability $1-\phi$). 

Although not originally described as such, the CJS model is simply a 2-state HMM that conditions on first capture. Framing the observed and hidden processes within the dependence structure of a basic HMM (Fig.\ \ref{fig:HMM}), we could for example have:

\bigskip
\begin{center}
\includegraphics[]{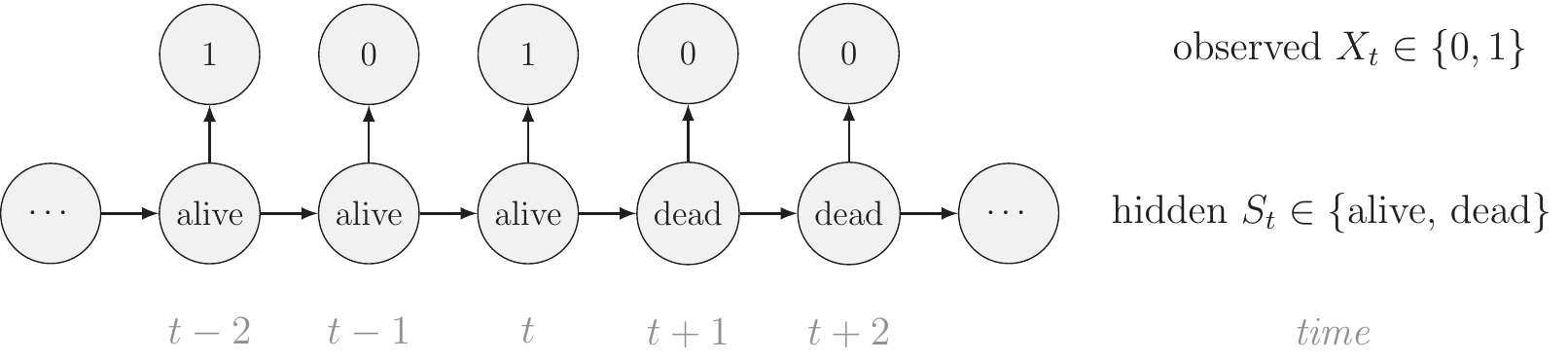}\\
\end{center}
The state-dependent observation distribution for $X_t$ is a simple Bernoulli (i.e.\ a coin flip) with success probability $p$ if alive and success probability $0$ if dead:
\begin{equation*}
    f\left(X_t = x_t \mid S_t=i\right) = 
    \begin{cases}
        p^{x_t}(1-p)^{1-x_t} & \text{if } i = \text{alive}\\
        0^{x_t}(1-0)^{1-x_t} = 1-x_t & \text{if } i = \text{dead}
    \end{cases}
\end{equation*}
We thus have the initial distribution
\begin{equation*}
    \begin{blockarray}{cccl}
    & \text{alive} & \text{dead} & \\
    \begin{block}{c(cc)l}
    \boldsymbol{\delta}= & 1 & 0 & , \\
    \end{block}
    \end{blockarray}
\end{equation*}
state transition probability matrix
\begin{equation*}
    \boldsymbol{\Gamma} =
    \begin{blockarray}{ccc}
    \text{alive} & \text{dead} & \\
    \begin{block}{[cc]c}
    \phi & 1-\phi & \text{alive} \\
    0 & 1 & \text{dead} \\
    \end{block}
    \end{blockarray}
\end{equation*}
and state-dependent observation distribution matrix
\begin{equation*}
    \mathbf{P}(x_t) =
    \begin{blockarray}{cc}
    \text{alive} & \text{dead} \\
    \begin{block}{[cc]}
    p^{x_t}(1-p)^{1-x_t} & 0 \\
    0 & 1-x_t \\ 
    \end{block}
    \end{blockarray} .
\end{equation*}
The CJS model 
is thus a very simple HMM with an absorbing ``dead'' state and only two unknown parameters ($\phi$ and $p$). As an HMM, it can not only be used to estimate survival, but also the point in time when any given individual was most likely to have died (based on local or global state decoding; see Table \ref{tab:glossary}). 

The classic Jolly-Seber capture-recapture model and its various extensions \citep[][]{Pradel1996,WilliamsEtAl2002} 
go a step further by incorporating both birth and death processes. It simply involves extending the 2-state model to an additional ``unborn'' (UB) state. We could for example now have:

\bigskip
\begin{center}
\includegraphics[]{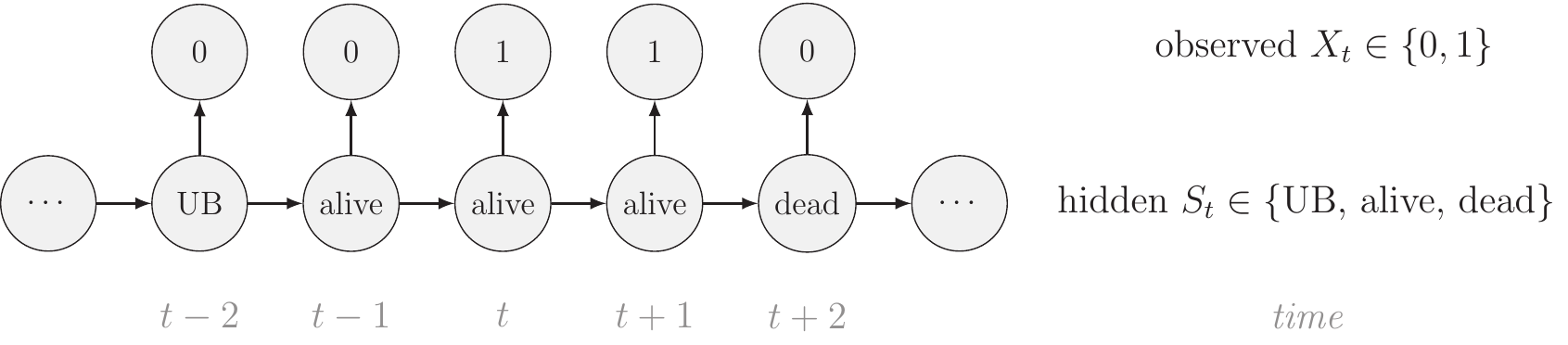}\\
\end{center}
\noindent To formulate a 3-state HMM with an additional ``unborn'' state, we must extend our components for the hidden and observed processes accordingly:
\begin{equation*}
    \begin{blockarray}{ccccl}
    & \text{unborn} & \text{alive} & \text{dead} & \\
    \begin{block}{c(ccc)l}
    \boldsymbol{\delta}= & 1-\alpha_1 & \alpha_1 & 0 & , \\
    \end{block}
    \end{blockarray}
\end{equation*}
\begin{equation*}
    \boldsymbol{\Gamma}^{(t)}=
    \begin{blockarray}{cccc}
    \text{unborn} & \text{alive} & \text{dead} & \\
    \begin{block}{[ccc]c}
    1-\beta_t & \beta_t & 0      & \text{unborn} \\
        0     & \phi    & 1-\phi & \text{alive}  \\
        0     &    0    & 1      & \text{dead}   \\
    \end{block}
    \end{blockarray}
\end{equation*}
and
\begin{equation*}
    \mathbf{P}(x_t) =
    \begin{blockarray}{ccc}
    \text{unborn} & \text{alive} & \text{dead} \\
    \begin{block}{[ccc]}
    1-x_t & 0 &  0 \\
    0 & p^{x_t}(1-p)^{1-x_t} & 0  \\
    0 & 0 & 1-x_t \\
    \end{block}
    \end{blockarray} 
\end{equation*}
where
\begin{equation*}
    \beta_t = \begin{cases}
    \alpha_1 & \text{if } t = 1\\
    \frac{\alpha_t}{\prod_{l=1}^{t-1} (1-\beta_l)} & \text{if } t>1
    \end{cases},
\end{equation*}
$\alpha_1$ is the probability that an individual was already in the population at the beginning of the study, $\alpha_t$ is the probability that any given individual was born at time $t \in \{2,\ldots,T\}$, and $\beta_t$ is the probability that an individual entered the population on occasion $t$ given it had not already entered up to that time. Importantly, note that the 2-state and 3-state HMMs rely on the exact same 
binary data $(X_t \in \{0,1\})$, but we are able to make additional inferences in the 3-state model by re-formulating the observed and hidden processes in terms of both birth and death.
While we have employed these well-known individual-level capture-recapture models to initially demonstrate the key idea of linking observed state-dependent processes to the underlying state dynamics via HMMs, these types of inferences are not limited to traditional capture-recapture observation processes. For example, telemetry and count data can also be utilised in HMMs describing individual-level birth and death processes \citep[][]{SchmidtEtAl2015,CowenEtAl2017}. 

\subsubsection{Developmental state}
\label{sec:individualDevel}

Individual-level data often contain additional information about developmental states 
such as those related to size \citep{NicholsEtAl1992}, reproduction \citep{NicholsEtAl1994}, social groups \citep{MarescotEtAl2018}, or disease \citep{Benhaiem2018}. However, assigning individuals to states can be difficult when traits such as breeding \citep[][]{KendallEtAl2012}, infection \citep[][]{ChambertEtAl2012}, sex \citep[][]{PradelEtAl2008}, or even species \citep[][]{RungeEtAl2007} are ascertained through observations in the field. This difficulty has motivated 
models for individual histories
that can not only account for multiple developmental states \citep{LebretonEtAl2009}, but also uncertainty arising from partially 
or completely unobservable 
states \citep{
Pradel2005}. Such multi-state models can be used  
for testing a broad range of formal biological hypotheses, including host-pathogen dynamics in disease ecology \citep[][]{LachishEtAl2011}, reproductive costs in evolutionary ecology \citep[][]{GarnierEtAl2016}, and social dominance in behavioural ecology \citep{DupontEtAl2015}. For example, it is straightforward to extend the capture-recapture HMM to multiple ``alive'' states parameterised in terms of state-specific survival probabilities $(\phi)$ 
and transition probabilities between these ``alive'' states $(\psi)$. Consider a $3$-state HMM for capture-recapture data that incorporates reproductive status, where $S_t=\text{B}$ indicates ``alive and breeding'' and $S_t=\text{NB}$ indicates ``alive and non-breeding'': 

\bigskip
\begin{center}
\includegraphics[]{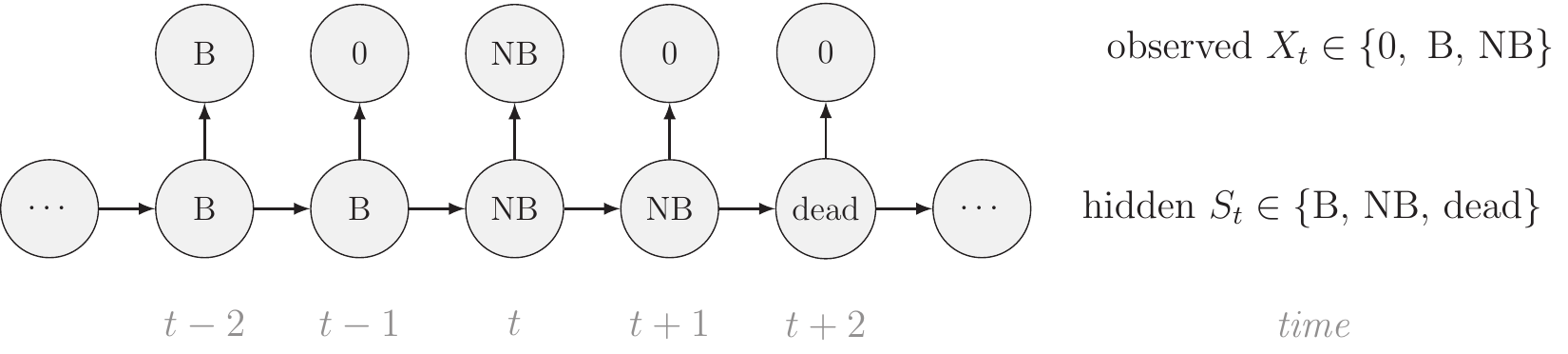}\\
\end{center}
\begin{equation*}
    \begin{blockarray}{ccccl}
    & \text{breeding} & \text{non-breeding} & \text{dead} & \\
    \begin{block}{c(ccc)l}
    \boldsymbol{\delta}= & \delta_B & 1-\delta_B & 0 & , \\
    \end{block}
    \end{blockarray}
\end{equation*}
\begin{equation*}
    \boldsymbol{\Gamma}=
    \begin{blockarray}{cccc}
    \text{breeding} & \text{non-breeding} & \text{dead} & \\
    \begin{block}{[ccc]c}
    \phi_B (1-\psi_{B,NB}) & \phi_B \psi_{B,NB}        &  1-\phi_B     & \text{breeding} \\
    \phi_{NB} \psi_{NB,B}  & \phi_{NB} (1-\psi_{NB,B}) &  1-\phi_{NB}  & \text{non-breeding}  \\
        0                  &    0                      & 1             & \text{dead}   \\
    \end{block}
    \end{blockarray}
\end{equation*}
and
\begin{equation*}
    \mathbf{P}(x_t) =
    \begin{blockarray}{ccc}
    \text{breeding} & \text{non-breeding} & \text{dead} \\
    \begin{block}{[ccc]}
    p_B^{I(x_t=B)}(1-p_B)^{1-I(x_t=B)} & 0                                          &  0 \\
    0                                  & p_{NB}^{I(x_t=NB)}(1-p_{NB})^{1-I(x_t=NB)} & 0  \\
    0                                  & 0                                          & I(x_t=0) \\
    \end{block}
    \end{blockarray} 
\end{equation*}
where $I(x_t=k)$ is an indicator function taking the value 1 when $x_t=k$ and 0 otherwise.
To assess costs of reproduction, a biologist will be interested in the probability of breeding in year $t$, given breeding $(\psi_{B,B}=1-\psi_{B,NB})$ or not $(\psi_{NB,B})$ in year $t-1$, as well as assessing any differences in survival probability between breeders $(\phi_B)$ and non-breeders $(\phi_{NB})$. By simply re-expressing the $\boldsymbol \delta$, $\boldsymbol \Gamma$, and $\mathbf{P}(x_t)$ components in terms of the specific state and observation processes of interest, 
such models can be used to infer the dynamics of conjunctivitis in house finches \citep{ConnCooch2009}, senescence in deer \citep{ChoquetEtAl2011}, reproduction in Florida manatees \citep{KendallEtAl2012}, and life-history trade-offs in 
elephant seals \citep[][]{LloydEtAl2020}. Similar HMMs can also be used to investigate relationships between life-history traits and demographic parameters that are important in determining the fitness of phenotypes or genotypes \citep[][]{StoeltingEtAl2015}. Several measures of individual fitness have been proposed, but one commonly used 
for field studies is lifetime reproductive success \citep[][]{rouan_estimation_2009,gimenez_estimating_2018}. These approaches can be readily adapted to quantify other measures of fitness \citep[][]{mcgraw_estimation_1996,link_model-based_2002,coulson_estimating_2006,MarescotEtAl2018}. 

Inferences about developmental states are of course not limited to traditional capture-recapture data, and significant advancements in animal-borne biotelemetry technology have brought many new and exciting opportunities \citep[][]{CookeEtAl2004,HootenEtAl2017,patterson2017statistical}. For example, telemetry location data can be used to identify migratory phases \citep[][]{weng2007migration}, predation events \citep{FrankeEtAl2006}, or the torpor-arousal cycle of hibernation \citep{hope2012warming}. 
The multi-state (i.e.\ hidden Markov) movement model 
is often used to infer these types of movement behaviour modes from trajectories in two-dimensional space
, where the observations are typically expressed in terms of the bivariate sequence of Euclidean distances (or ``step lengths'') and turning angles between consecutive locations \citep{FrankeEtAl2004,MoralesEtAl2004}. For a model involving $N=2$ states that assumes conditional independence between step length ($X_t$; in meters) and turning angle ($Y_t$; in radians) as in Fig.\ \ref{fig:EcoStates4}d, 
we could for example have:

\bigskip
\begin{center}
\includegraphics[]{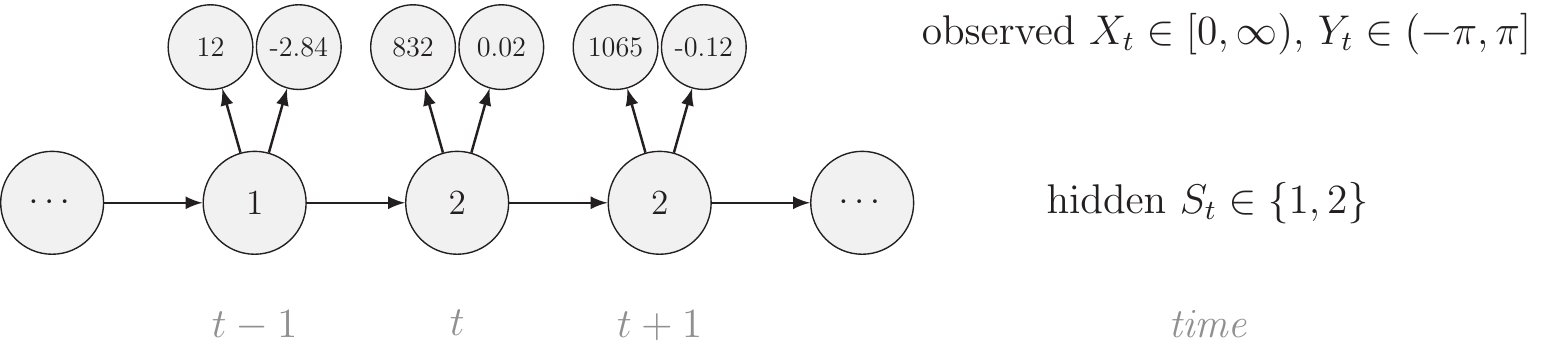}\\
\end{center}
These states could correspond to ``resident'' (state 1) and ``transient'' (state 2) behavioural phases, 
such that within state 2 the movements tend to be longer and directionally persistent (i.e.\ with turning angles concentrated near zero). Under the contemporaneous conditional independence assumption, the bivariate state-dependent observation distribution for $(X_t,Y_t)$ is simply the product of two univariate state-dependent distributions,
\begin{equation*}
    f\left(x_t,y_t \mid S_t=i\right) = f\left(x_t \mid S_t=i\right)   f\left(y_t \mid S_t=i\right). 
\end{equation*}
These univariate distributions are typically assumed to be the gamma or Weibull distribution for step length and the von Mises or wrapped Cauchy distribution for turning angle
.
Unlike 
our previous examples so far, the number of underlying states 
in 
these types of HMMs is generally not clear {\it a priori} and needs to be selected based on 
both biological and statistical criteria \citep[][]{pohle2017selecting}. 
Another difference is that there is often no predetermined structure in the state transition probability matrix, 
\begin{equation*}
    \boldsymbol{\Gamma}=
    \begin{blockarray}{ccc}
    \text{resident} & \text{transient}  & \\
    \begin{block}{[cc]c}
    \gamma_{11} & \gamma_{12}     & \text{resident} \\
    \gamma_{21} & \gamma_{22}     & \text{transient}  \\
    \end{block}
    \end{blockarray},
\end{equation*}
and all entries are freely estimated (but still subject to $\sum_{j=1}^N \gamma_{ij}=1$). As a consequence, the characteristics of the model states as represented by the state-dependent distributions are fully data driven, and hence may not correspond exactly to biologically meaningful entities (see Section \ref{sec:implementation}). 

\begin{figure}[h!]
\centering
\includegraphics[width=0.9\textwidth]{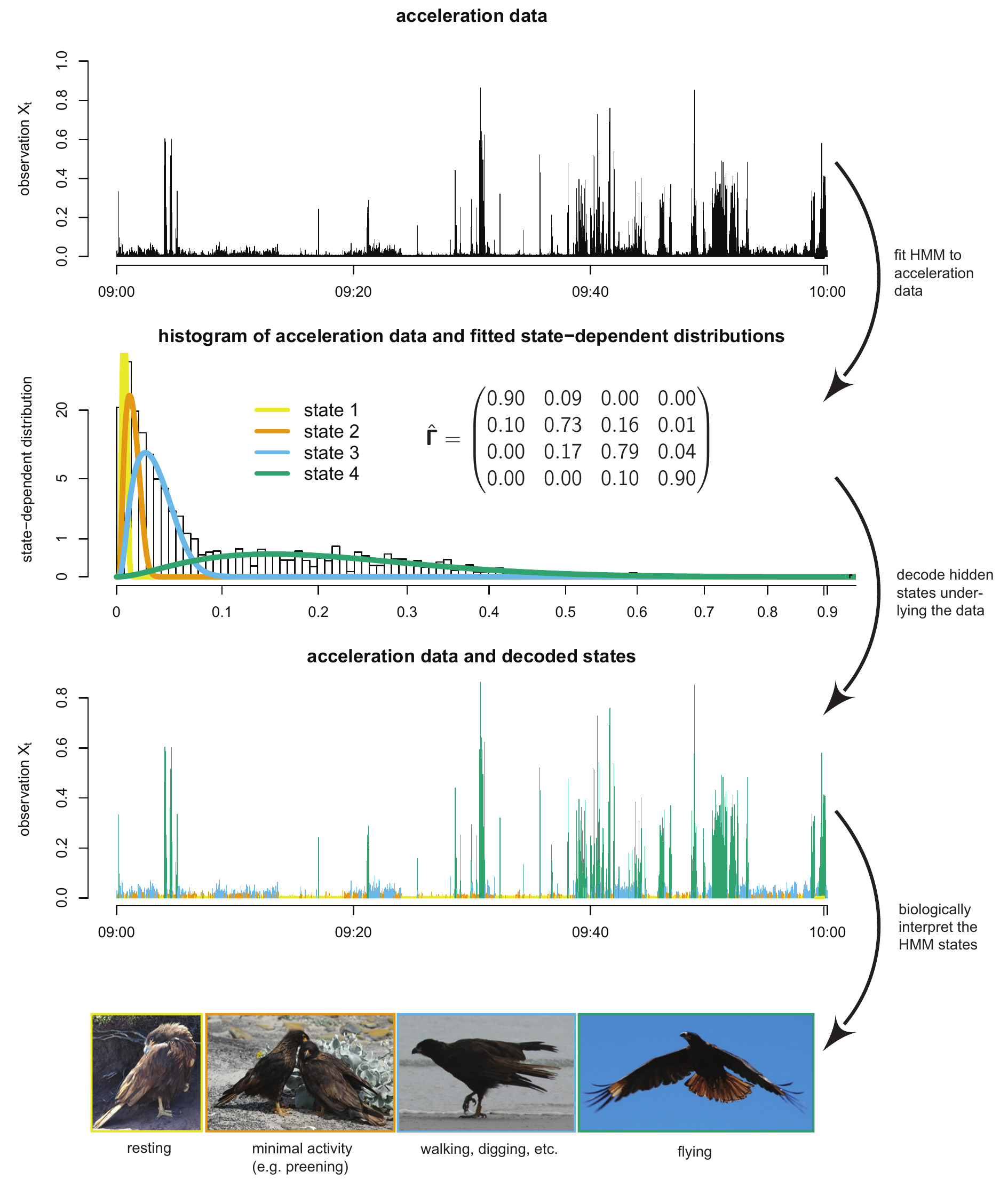}
\caption{Illustration of a possible workflow when using an HMM to infer behavioural modes from the vector of dynamic body acceleration data of a striated caracara (\textit{Phalcoboenus australis}) 
over a period of 60 minutes (see \citealp{fahlbusch2019low}, for data details).   
Four behavioural modes were identified and biologically interpreted to be associated with resting (yellow), minimal activity (orange), moderate activity (blue), and flying (green).}\label{fig:caracara}
\end{figure}

Similar HMMs for animal movement have been used, {\it inter alia}, to identify wolf 
kill-sites \citep{FrankeEtAl2006}, 
the relationship between southern bluefin tuna 
behaviour and ocean temperature \citep{PattersonEtAl2009}, 
activity budgets for 
harbour seals 
\citep{McClintockEtAl2013c}, hunting strategies of white sharks (
\citealp{TownerEtAl2016}), 
the behavioural response of northern gannets 
to frontal activity \citep{grecian2018understanding}, and 
how common noctules 
adjust their space use to the lunar cycle \citep{roeleke2018aerial}. 
Driven by the influx of new biotelemetry sensor technology, 
HMMs have also been used to analyse sequences of dives of marine animals \citep{hart2010behavioural,quick2017hidden,DeRuiterEtAl2017,van2019classifying}. The remote collection of activity data at potentially very high temporal resolutions using accelerometers is another 
emerging application area 
\citep{diosdado2015classification, Leos-BarajasEtAl2017accelerometer, papastamatiou2018optimal, papastamatiou2018activity, adam2019penalized}. These HMM formulations are conceptually very similar to the movement model outlined above, with the state process corresponding to behavioural modes (or at least proxies thereof), and the activity data represented by the state-dependent process. Fig.\ \ref{fig:caracara} illustrates a possible workflow for inferring four behavioural modes from high resolution accelerometer data collected from a striated caracara ({\it Phalcoboenus australis}) over a period of one hour. Here the vector of dynamic body acceleration was used as a univariate summary of the three-dimensional raw acceleration data, and a gamma distribution was used for the state-dependent observation process. In this example, the HMM can be regarded as a clustering scheme which maps observed input data to unobserved underlying classes with biological interpretations roughly corresponding to ``resting'', ``minimal activity'' (e.g.\ preening), ``moderate activity'' (e.g.\ walking, digging), and ``flying''. Complete details of this analysis, including each step of the workflow and example R \citep[][]{RCoreTeam2019} code, can be found in the Supplementary Tutorial.

\subsubsection{Spatial state}
\label{sec:individualSpatial}

HMMs can also be used for inferences about the unobserved spatial location of an individual. For example, capture-recapture data can consist of sequences of 
observations arising from a set of discrete spatial states, where these 
often refer to ecologically important geographic areas, such as wintering and breeding sites for migratory birds \citep[][]{BrownieEtAl1993} or spawning sites for fish \citep[][]{SchwarzEtAl1993}. 
For a $3$-state HMM with two sites (A and B), where $S_t=\text{A}$ indicates ``alive at site A'' and $S_t=\text{B}$ indicates ``alive at site B'', we could for example have:

\bigskip
\begin{center}
\includegraphics[]{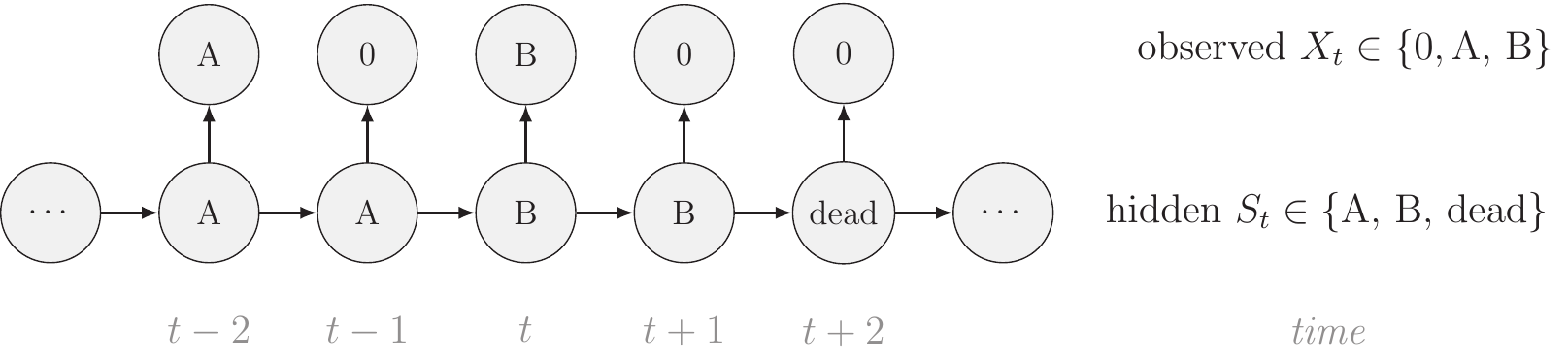}\\
\end{center}
Clearly, this discrete-space HMM is structurally identical to the multi-state capture-recapture HMMs already described in Section \ref{sec:individualDevel}; the only difference is the state transition probability parameters are now interpreted as site-specific survival and movement probabilities between the sites \citep[e.g.\ fidelity or dispersal;][]{LagrangeEtAl2014,CayuelaEtAl2020}. Based on global state decoding, these HMMs can therefore also be used to infer the most likely spatial state for periods when an individual was alive but its location was not observed. 

Another important 
application of HMMs is for geolocation based on indirect measurements that vary with space, such as light, pressure, temperature, and tidal patterns \citep[][]{ThygesenEtAl2009,RakhimberdievEtAl2015}. Although too technical to be described in detail here, geolocation HMMs can be particularly useful for inferring individual location from archival tag data 
\citep[][]{BassonEtAl2016}. These HMMs have even been extended to include state-switching behaviours such as those described in Section \ref{sec:individualDevel} \citep{PedersenEtAl2008,PedersenEtAl2011estimating}. 
Animal movement behaviour HMMs have also been extended to accomodate partially-observed location data common to marine mammal satellite telemetry studies \citep[][]{JonsenEtAl2005,McClintockEtAl2012}.

\subsection{Population level}
\label{sec:populationLevel}

We consider two ways that inference on the population level can arise: 1) an individual-level model, based on data from multiple individuals (e.g.\ capture-recapture), quantitatively connected to a population-level concept through an explicit model; or 2) a population-level model, based on population-level data (e.g.\ counts or presence-absence), with no explicit model for processes at the individual level. 

\subsubsection{Existential state}
\label{sec:populationExist}
A fundamental existential state at the population level is abundance, the number of individuals alive in a population at a particular point in time. 
A common way to infer this using capture-recapture HMMs 
is to formally link abundance to the individual-level processes (e.g.\ survival, recruitment) that drive its dynamics. Intuitively, the abundance model specifies how many individuals go through the life history specified by the HMM. For the abundance component, the key pieces of information are the number of individuals in the population that were detected at least once $(n)$ and the probability of being detected at least once, given an individual was alive at any time during the study $(p^*)$. The former is observed while the latter can be calculated as
\begin{equation*}
    p^*=1-\boldsymbol\delta\mathbf{P}(x_1=0)\boldsymbol\Gamma^{(1)}\mathbf{P}(x_2=0)\boldsymbol\Gamma^{(2)}\cdots\boldsymbol\Gamma^{(T-1)}\mathbf{P}(x_T=0)\mathbf{1}
\end{equation*}
using notation for the Jolly-Seber HMM presented in Section \ref{sec:individualExist}. This HMM formulation is equivalent to the original Jolly-Seber open population model (shown in \citealp{glennie2019open}), 
where population abundance at each time $t$ 
is derived from 
the individual-level process parameters. 

Instead of inducing changes in abundance through individual-level HMMs, abundance itself can be modelled 
as the hidden state within an HMM \citep[][]{SchmidtEtAl2015,CowenEtAl2017,BesbeasMorgan2019}. 
Here population dynamics are 
inferred from population-level surveys \citep{BucklandEtAl2004}, where the observation process can include counts 
or other quantities that are noisy measurements of the true abundance (the hidden state), and the state transition probability matrix $(\boldsymbol\Gamma)$ is naturally formulated in terms of the well-known Leslie matrix for population growth \citep[][]{Caswell2001}.
For example, for imperfect count data $X_t \in \{0,1,2,\ldots\}$ that were collected from a population of true size $S_t \in \{0,1,\ldots,N_{max}\}$ (note the requirement to specify a maximum possible population size $N_{max}$), we could have: 

\bigskip
\begin{center}
\includegraphics[]{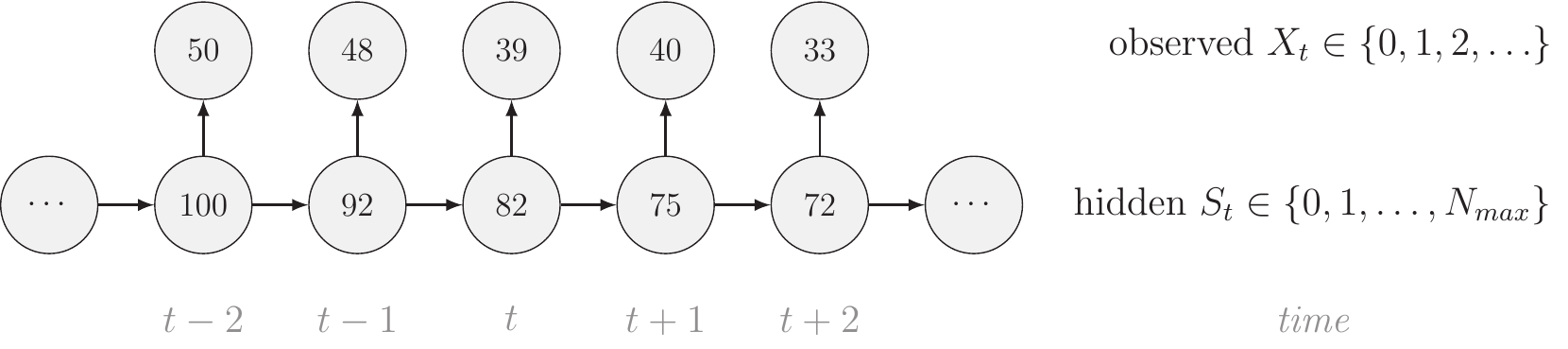}\\
\end{center}

\begin{equation*}
    \begin{blockarray}{cccccl}
    & S_1=0 & S_1=1 & \ldots & S_1=N_{max} & \\
    \begin{block}{c(cccc)l}
    \boldsymbol{\delta}= & \delta_0 & \delta_1 & \ldots & \delta_{N_{max}} & , \\
    \end{block}
    \end{blockarray}
\end{equation*}
\begin{equation*}
    \boldsymbol{\Gamma}^{(t)}=
    \begin{blockarray}{ccccc}
    S_{t+1}=0 & S_{t+1}=1 & \ldots & S_{t+1}=N_{max} & \\
    \begin{block}{[cccc]c}
    \gamma_{0,0} & \gamma_{0,1} & \ldots & \gamma_{0,N_{max}} & S_t=0 \\
    \gamma_{1,0} & \gamma_{1,1} & \ldots & \gamma_{1,N_{max}} & S_t=1 \\
    \vdots                          &   \vdots                         & \ddots &  \vdots                          &   \vdots       \\
    \gamma_{N_{max},0} & \gamma_{N_{max},1} & \ldots & \gamma_{N_{max},N_{max}} & S_t=N_{max} \\
    \end{block}
    \end{blockarray}
\end{equation*}
and
\begin{equation*}
    \mathbf{P}(x_t) =
    \begin{blockarray}{cccc}
    S_t=0 & S_t=1 & \ldots & S_t=N_{max} \\
    \begin{block}{[cccc]}
    f(x_t \mid S_t=0) & 0                 & \ldots & 0 \\
    0                 & f(x_t \mid S_t=1) & \ldots & 0 \\
    \vdots            & \vdots            & \ddots & \vdots \\
    0                 & 0                 & \ldots & f(x_t \mid S_t=N_{max}) \\
    \end{block}
    \end{blockarray} 
\end{equation*}
Each state transition probability $(\gamma_{ij})$ describes the population dynamics from time $t$ to time $t+1$ and can be parameterised in terms of survival, reproduction, emigration, the current population size $S_t$, and any additional population structure (e.g.\ sex or age classes; see Section \ref{sec:populationDevel}). The state-dependent distributions $f(x_t \mid S_t=i)$ can take many different forms depending on the specific observation process, but common choices for count data are binomial or Poisson models \citep[][]{SchmidtEtAl2015,BesbeasMorgan2019}. 
Sometimes count data alone can be insufficient for describing complex population processes, and this has led to integrated population modelling \citep{Schaub2011} that also utilises auxiliary data such as capture-recapture, telemetry, or productivity data \citep{SchmidtEtAl2015,BesbeasMorgan2019}. 

\subsubsection{Developmental state}
\label{sec:populationDevel}

Populations have more structure than simply their overall abundance or density. Sex, age demographics, size of breeding sub-population, fitness of individuals, and behavioural or genetic heterogeneity all have an impact on the development of a population \citep{Seber2019}. Many of these processes can be accounted for within the HMM framework presented in the previous section for individual-level data. As before, the idea is to extend the ``alive" state to a more complex network of states whose state-dependent distributions and transitions match the structure in the population. 
Combinations of these provide the opportunity to build a rich state process to describe the population dynamics. 
This framework is built around the idea that individuals are the singular units that together drive population change, but there has also been increasing use of HMMs from a different viewpoint: that of 
evolutionary processes at lower levels of organisation (e.g.\ genes).

With recent advances in genetic sequencing, the need for interpreting and modelling biological sequences (e.g.\ protein or DNA) has boosted the development of HMMs in molecular ecology \citep[][]{DurbinEtAl1998, BoitardEtAl2009, Yoon2009, GhoshEtAl2012}. Many of these applications use HMMs strictly as a tool for biological sequence analysis \citep[e.g.\ identifying species from DNA barcodes;][]{HebertEtAl2016} and are too technical to delve into detail here, but HMMs for molecular sequence data are commonly formulated in terms of evolutionary state dynamics, including for example speciation and extinction \citep{hobolth_genomic_2007, SoriaEtal2014, crampton_pacing_2018, OlajosEtAl2018}, hybridisation \citep{SchumerEtAl2018,palkopoulou_comprehensive_2018}, mutualism \citep{WernerEtAl2018}, hidden drivers of diversification \citep{caetano_hidden_2018}, and evolutionary rates among sites \citep{FelsensteinChurchill1996}. 

Telemetry locations are another form of individual-level data that, when combined across individuals, can provide population-level inferences about movement, space use, and resource selection \citep[][]{HootenEtAl2017}. As such, telemetry data can be well suited for addressing hypotheses related to intraspecific interactions. While such applications are still relatively rare, HMMs that utilise 
location data have been used to investigate intraspecific competition 
in marine mammals \citep{BreedEtAl2013}, 
herding 
in ungulates \citep{LangrockEtAl2014}, and 
social behaviour in fish \citep{BodeSeitz2018}.

Similar to approaches for inferring population-level developmental states from individual-level data, a rich structure can also be specified within an HMM for population-level data. Multiple states and processes can be represented: age classes/survival, size classes/growth, sex/birth, genotypes, and metapopulations are all states or networks of states with specified connections \citep[][]{newman2014modelling}. Such HMMs can be informed by a wide variety of population-level observations, e.g.\, counts of plants \citep{BorgyEtAl2015} or animals \citep{SchmidtEtAl2015}, as well as auxiliary individual-level observations \citep[][]{BesbeasMorgan2019}. From this general viewpoint, HMMs can be seen as the structure behind open population N-mixture models \citep{SchmidtEtAl2015,CowenEtAl2017}, distance sampling models \citep{Sollmann2015}, and approximate state-space population dynamics models \citep{BesbeasMorgan2019}. 

\subsubsection{Spatial state}
\label{sec:populationSpatial}

The spatial state of a population can be conceived as a surface (or map) quantifying density at each point in space, and population models for individual-level data can be extended to allow density to change over space \citep{Borchers2008}. Inferring density as a spatial population state, however, requires spatial information within the data. Spatial capture-recapture surveys \citep[][]{RoyleEtAl2013book}, an extension of capture-recapture, collect precisely this data. Spatial capture-recapture HMMs can be formulated in terms of survival, recruitment, movement, and population density \citep{RoyleEtAl2018,glennie2019open} and are readily extendable for relating environment and population distribution across space, including how distribution is affected by landscape connectivity, dispersal, resource selection, or environmental impacts such as oil spills \citep[][]{Mcdonald2017,RoyleEtAl2018}.

A different viewpoint is to consider population-level data that are commonly collected over both space and time: presence-absence data. These data provide information on a population's spatial state that is not derived from abundance and arise from the monitoring of spatial units for the (apparent) presence or absence of a species. One of the most popular tools for analysing these data are patch (or site) occupancy models, which can be used to infer patterns and dynamics of species occurrence while accounting for imperfect detection \citep{MackenzieEtAl2018}. As with capture-recapture models, patch occupancy models are also HMMs \citep{royle_bayesian_2007, GimenezEtAl2014} where, instead of the state dynamics of individual organisms, the hidden process describes the state dynamics of sites. Let $S_t=\text{O}$ indicate ``occupied '' and $S_t=\text{U}$ indicate ``unoccupied'', where the species can be detected $(X_{t,k}=1)$ or not $(X_{t,k}=0)$ during multiple visits $k=1,\ldots,K$ to each site, with the following representation:

\bigskip
\begin{center}
\includegraphics[]{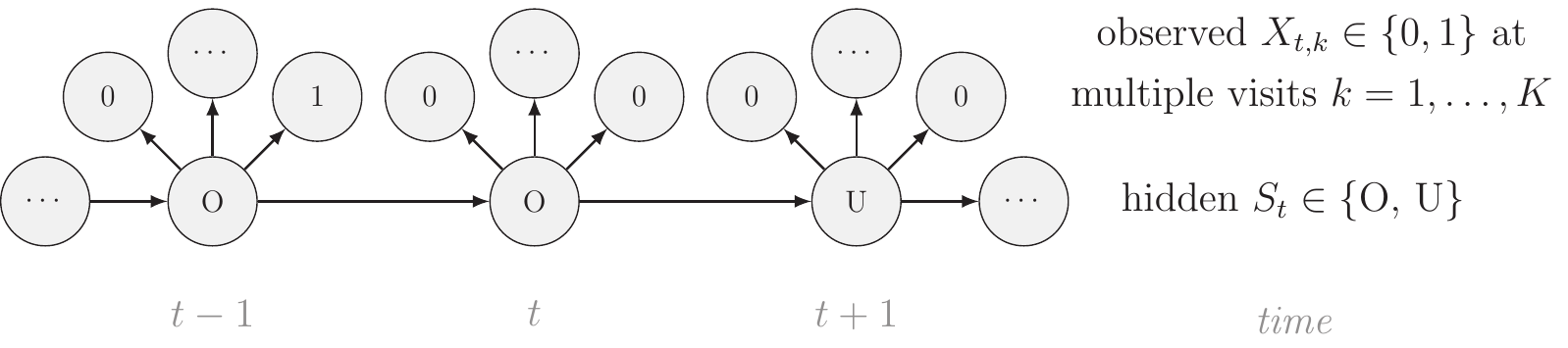}\\
\end{center}
\begin{equation*}
    \begin{blockarray}{cccl}
    & \text{occupied} & \text{unoccupied} \\
    \begin{block}{c(cc)l}
    \boldsymbol{\delta}= & \psi_1 & 1-\psi_1 \\
    \end{block}
    \end{blockarray}
\end{equation*}
\begin{equation*}
    \boldsymbol{\Gamma}=
    \begin{blockarray}{ccc}
    \text{occupied} & \text{unoccupied} \\
    \begin{block}{[cc]c}
    1-\epsilon & \epsilon       & \text{occupied} \\
    \kappa     & 1-\kappa       & \text{unoccupied}  \\
    \end{block}
    \end{blockarray}
\end{equation*}
and
\begin{equation*}
    \mathbf{P}(\mathbf{x}_t) =
    \begin{blockarray}{ccc}
    \text{occupied} & \text{unoccupied} \\
    \begin{block}{[ccc]}
    \prod_{k=1}^K p^{x_{t,k}} (1-p)^{1-x_{t,k}} & 0    \\
    0                     & \prod_{k=1}^K (1-x_{t,k}) \\
    \end{block}
    \end{blockarray} 
\end{equation*}
where $\psi_1$ is the initial patch occupancy probability at time $t=1$, $p$ is the species detection probability at each occupied patch, and 
${\boldsymbol\Gamma}$ is composed of the local colonisation $(\kappa)$ and extinction $(\epsilon)$ probabilities. Single-season (or static) occupancy models \citep[][]{MackenzieEtAl2002} are obtained as a special case with $T=1$ or $\epsilon=\kappa=0$ \citep{GimenezEtAl2014}. This HMM can not only be used to estimate patch occupancy, extinction, and colonisation probabilities, but also the most likely state and times of any colonisation or extinction events within a patch
. The flexibility of the HMM formulation allows patch occupancy to be conveniently extended to cope with site-level heterogeneity in detection using finite mixtures \citep{louvrier_accounting_2018} or a discrete measure of population density \citep[][]{GimenezEtAl2014,VeranEtAl2015} and even false positives due to species misidentification \citep[][]{MillerEtAl2011,louvrier_use_2019}. Just as with multi-state capture-recapture HMMs (Section \ref{sec:individualDevel})
, species occurrence HMMs can be readily extended to multiple ``occupied'' states accommodating reproduction \citep[][]{MackenzieEtAl2009,MartinEtAl2009}, disease \citep[][]{McClintockEtAl2010}, and other (meta-)population dynamics \citep[][]{LamyEtAl2013}.

Inferences from HMMs for presence-absence data are not limited to occupancy models that account for 
imperfect species detection. For example, \citet{PluntzEtAl2018} developed an HMM characterising seed dormancy, colonisation, and germination in annual plant metapopulations based entirely on presence-absence observations of standing flora. In their study, the presence of a completely unobservable soil seed bank was the hidden state of interest, and they modified the dependence structure of a basic HMM such that the seed bank state dynamics at time $t$ depended not only on the seed bank state at time $t-1$, but also on the presence or absence of standing flora at time $t$. Let $S_t=\text{AA}$ indicate ``seed bank absent at time $t-1$, flora absent at time $t$'', $S_t=\text{PA}$ indicate ``seed bank present at time $t-1$, flora absent at time $t$'', and $S_t=\text{PP}$ indicate ``seed bank present at time $t-1$, flora present at time $t$'', where standing flora is present $(X_t=1)$ or not $(X_t=0)$ during visit $t$ to each site and is assumed to be detected without error. We could for example have:

\bigskip
\begin{center}
\includegraphics[]{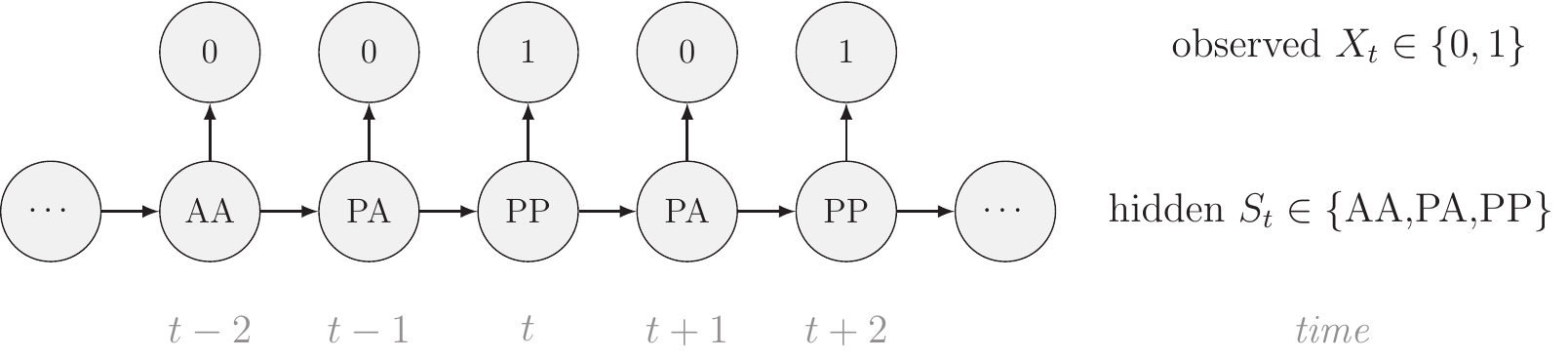}\\
\end{center}
\begin{equation*}
    \begin{blockarray}{ccccl}
    & S_1=\text{AA} & S_1=\text{PA} & S_1=\text{PP} & \\
    \begin{block}{c(ccc)l}
    \boldsymbol{\delta}= & 1-\psi_0 & \psi_0 (1-g) & \psi_0g & , \\
    \end{block}
    \end{blockarray}
\end{equation*}
\begin{equation*}
    \boldsymbol{\Gamma}=
    \begin{blockarray}{cccc}
    S_{t+1}=\text{AA} & S_{t+1}=\text{PA} & S_{t+1}=\text{PP} & \\
    \begin{block}{[ccc]c}
    1-c            & (1-g)c                         & gc                        & S_t=\text{AA} \\
    (1-c)(1-s)     & (1-g)\bigl(1-(1-c)(1-s)\bigr)  & g\bigl(1-(1-c)(1-s)\bigr) & S_t=\text{PA}  \\
        0          &    1-g                         & g                         & S_t=\text{PP}   \\
    \end{block}
    \end{blockarray}
\end{equation*}
where $\psi_0$ is the probability that a seed bank was present the year before the first observation, $g$ is the probability of germination and survival to reproduction, $s$ is the probability of seed bank survival, $c$ is the probability of external colonisation, and ${\mathbf P}(x_t)$ is a $3 \times 3$ diagonal matrix of ones. Similar formulations could be applied to other organisms with dormant life cycles (e.g.\ fungi, crustaceans).

\subsection{Community level}
\label{sec:communityLevel}

Community-level studies often focus on a subset of species based on taxonomy, trophic position, or particular interactions of interest
, and the diversity of topics addressed in community ecology reflects its large scope \citep{vellend_conceptual_2010,vellend_theory_2016}. Here we will only scratch the surface of two study systems 
that can be formulated as HMMs for multi-species presence-absence data commonly collected from field surveys or (e)DNA samples: 1) patch systems composed of (potentially) many species; and 2) patch systems composed of a few (possibly interacting) species.  

\subsubsection{Existential state}

A fundamental measure of biodiversity is the number of species within a community 
(species richness). This community-level state is often unobservable in studies of natural systems \citep[][]{dorazio_estimating_2006}, even for communities composed entirely of sessile organisms \citep[][]{ConwayDoak2011,ChenEtAl2013}. Multi-species occupancy HMMs expand single-species occupancy HMMs (see Section \ref{sec:populationSpatial}) to the community level using presence-absence data for each species that could (potentially) occupy the sampling units within a study area \citep[][Chapter 15]{MackenzieEtAl2018}. By combining single-species 
HMMs, either independently or 
by sharing common parameters among species \citep[][]{EvansEtAl2016,guilleraarroita_modelling_2017}, community-level attributes (e.g.\ species richness) and species-level attributes (e.g.\ patch occupancy) can be integrated within a single modelling framework \citep[][Chapter 12]{RoyleDorazio2008}. By jointly modelling species- and community-level processes, the approach proposed by \citet{dorazio_estimating_2005} and its extensions \citep[reviewed by][Chapter 11]{kery_applied_2015} facilitate the simultaneous testing of formal hypotheses about factors influencing occupancy \citep{rich_using_2016, tenan_quantifying_2017}, 
species richness \citep{sutherland_multiregion_2016}, and their dynamics through time \citep[][]{russell_modeling_2009,DorazioEtAl2010}, with important consequences for conservation and management \citep{zipkin_multi-species_2010}. Although these community dynamics models are typically fitted using hierarchical Bayesian methods and not explicitly referred to as HMMs, they share the same properties and can be similarly decomposed in terms of $\boldsymbol \delta$, $\boldsymbol \Gamma$, and $\mathbf{P}(x_t)$. Viewing the species richness of a community as analogous to the abundance of a population, HMM formulations similar in spirit to those described in Section \ref{sec:populationExist} could account for species that were never detected ({\it sensu} \citealp{dorazio_estimating_2006}).

\subsubsection{Developmental state}

Many community-level attributes can be constructed from ``metacommunity'' HMMs for species richness at both the community and metacommunity level \citep[][Chapter 11]{dorazio_estimating_2005,kery_applied_2015}. 
Species richness at each site is the $\alpha$ diversity metric, and total richness in the whole metacommunity is the $\gamma$ diversity \citep[][Chapter 6]{magurran2004measuring}. A possible metric for the $\beta$ diversity is the similarity Jaccard index: the proportion of species that occur at two sites among the species that occur at either site. Multi-species occupancy models have also been used to address variation in 
community attributes within distinct regions using Hill numbers for species richness, Shannon diversity, and Simpson diversity \citep[][]{BromsEtAl2015, sutherland_multiregion_2016, tenan_quantifying_2017, BoronEtAl2019}. Dynamic multi-species occupancy HMMs can provide inferences about changes in community composition and structure over time, entry (or ``turnover'') probabilities of ``new'' species into the community, and species ``extinction'' probabilities from the community \citep[][]{russell_modeling_2009,DorazioEtAl2010}. Although to our knowledge this has not yet been attempted, community assembly or succession dynamics could naturally be parameterised in terms of such quantities within a multi-state, multi-species HMM describing transitions among different community states (e.g.\ disturbed, climax). Community structure and composition also depend on interspecific interactions, and multi-species occupancy HMMs 
can empirically test for 
any such evidence 
\citep{GimenezEtAl2014,rota_multispecies_2016, DavisEtAl2018, MackenzieEtAl2018, MarescotEtAl2019}. To date these co-occurrence models have mostly been used to infer predator-prey interactions \citep[][]{miller_lions_2018,murphy_using_2019}. Other emerging frameworks for inferences about processes that structure communities could also potentially be formulated as HMMs to account for observation error in presence-absence or count data \citep[][]{OvaskainenEtAl2017}.

\subsubsection{Spatial state}

Understanding geographic variation in the size and structure of communities is one of the major goals in ecology. While we have so far focused on some of the more ``non-spatial'' aspects of community-level inference, all multi-species presence-absence HMMs are of course inherently spatial and describe community distribution as well. Dynamic multi-species occupancy models provide inferences about changes in community distributions over time \citep[][]{russell_modeling_2009,DorazioEtAl2010}, and, when spatio-temporal interactions between species are of primary interest, dynamic co-existence HMMs can incorporate local species extinction and colonisation to investigate interspecific drivers of co-occurrence dynamics and community distribution \citep{fidino_multistate_2019,MarescotEtAl2019}. As a final illustrative example, suppose we have the states $S_t=\text{A}$ (respectively $S_t=\text{B}$ and $S_t=\text{AB}$) for ``site occupied by species A'' (respectively by species B and by both species) and $S_t=\text{U}$ indicates ``unoccupied site''. Define $X_{t,k} \in \{0,1,2,3\}$, where $0$ indicates neither species was detected, $1$ indicates only species A was detected, $2$ indicates only species B was detected, and $3$ indicates both species were detected on the $k$th visit at time $t$. We could for example have:

\bigskip
\begin{center}
\includegraphics[]{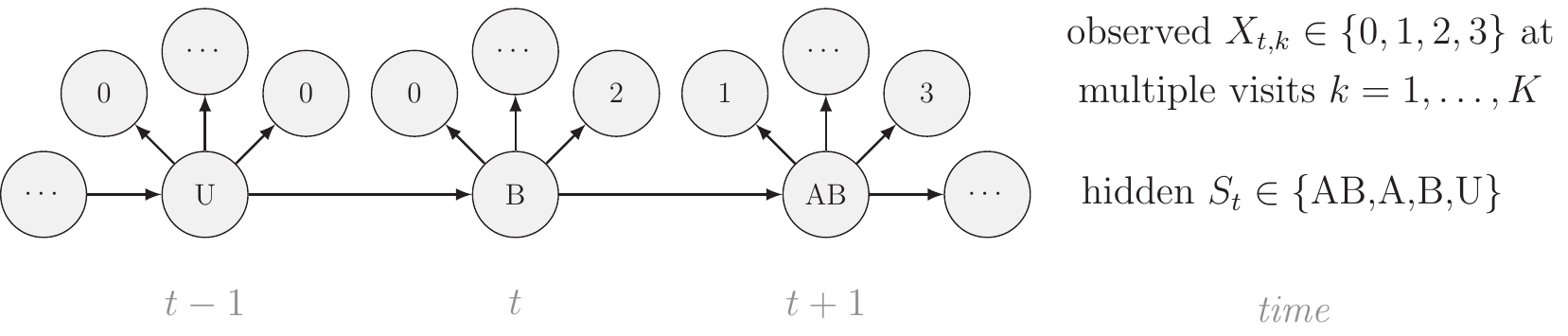}\\
\end{center}

Here the observation and state process models are more complex than previous examples, but they can still be readily expressed in terms of $\boldsymbol \delta$, $\boldsymbol \Gamma$, and ${\mathbf P}(x_t)$ for inferring patterns and drivers of species co-existence distribution dynamics (see Appendix A in Supplementary Material).

\subsection{Ecosystem level}
\label{sec:ecosystemLevel}
Despite the well-recognised need for reliable inferences about broad-scale ecological dynamics in the face of climate change and other challenges \citep[][]{TurnerEtAl1995}, HMMs have thus far seldom been applied at the ecosystem level. This is likely attributable to many factors, including the difficulty of obtaining and integrating observational data at the large spatio-temporal scales required \citep[][]{JonesEtAl2006,BohmannEtAl2014,DietzeEtAl2018,EstesEtAl2018,CompagnoniEtAl2019}. However, Markov models \citep[][]{GrewalEtal2019} are commonly used for inferring community- or ecosystem-level dynamics \citep[][]{WaggonerStephens1970,Wootton2001,TuckerAnand2005,BreiningerEtAl2010} and providing measures of stability, resilience, or persistence \citep[][]{Li1995,PawlowskiMcCord2009,zweig2020applied}, especially in systems composed of sessile organisms such as plant \citep[][but see \citealt{ChenEtAl2013}]{Horn1975,VanHulst1979,Usher1981,TallutoEtAl2017} or 
benthic communities  \citep[][]{TannerEtAl1994,HillEtAl2004,LoweEtAl2011}. 
Ecologists interested in ecosystem state transitions that are less apparent in observable dynamics 
may not recognise that the more widely-used Markov model is just a special case of an HMM \citep[][]{BreiningerEtAl2010,FukayaRoyle2013,GrewalEtal2019hidden}. A Markov model can simply be viewed as an HMM where it is assumed that the state process 
is perfectly observed, i.e., $X_t=S_t$ with ${\mathbf P}(x_t)$ a matrix with entry one in row $s_t$, column $s_t$, and otherwise zeros. For example, patch dynamics HMMs \citep{MackenzieEtAl2003} are simply generalisations of well-known Markov models for patch dynamics \citep[][]{Hanski1994,Moilanen1999} for cases when presence-absence data are subject to imperfect detection. Likewise, any ecosystem-level Markov model can naturally be embedded as the state process within an HMM for less observable phenomena. 

Although there are fewer examples in the literature, HMMs have been used to make ecosystem-level inferences about stability and regime shifts \citep[][]{GalAnderson2010,GennarettiEtAl2014,EconomouMenary2019}, climate-driven community and disease dynamics \citep[][]{MoritzEtAl2008,MartinezEtAl2016,MillerEtAl2018}, the effects of management action on 
habitat dynamics \citep{BreiningerEtAl2010}, climatic niches \citep[][]{TingleyEtAl2009}, and ecosystem health \citep[][]{XiaoEtAl2019}. HMMs are also frequently used by atmospheric scientists, hydrologists, and landscape ecologists to describe regional- to global-scale ecosystem processes such as precipitation \citep[][]{ZucchiniGuttorp1991,SrikanthanMcMahon2001}, streamflow \citep[][]{Jackson1975,BrackenEtAl2014}, wetland dynamics \citep[][]{SiachalouEtAl2014}, and land cover dynamics \citep[][]{AurdalEtAl2005,LazrakEtAl2010,TrierSalberg2011,AbercrombieEtAl2015,SiachalouEtAl2015}. While many of these examples tend to focus on a few specific biotic and/or abiotic components in which to frame ecosystem state dynamics, we can envision future applications adopting a more holistic approach that integrates increasingly more complex ecosystem-level processes with observational data arising from a variety of sources and spatio-temporal scales (see Section \ref{sec:futureDirections}).

\section{Implementation, challenges, and pitfalls}
\label{sec:implementation} 

\subsection{Software}
\label{sec:software}

Recent advances in computing power and 
user-friendly software have made the implementation of HMMs much more feasible for practitioners. However, the features and capabilities of the software are varied, and it can be challenging to determine which software may be most appropriate for a specific objective. We briefly describe some of the HMM software currently available, limiting our treatment to freely available R \citep{RCoreTeam2019} packages and stand-alone programs that we believe are most accessible to ecologists and non-statisticians. While most HMM packages in R include data simulation, parameter estimation, and state decoding for an arbitrary number of system states
, they differ in many key respects (Table \ref{tab:software}). Some of the more general packages provide greater flexibility for specifying state-dependent probability distributions \citep[][]{VisserSpeenkenbrink2010,Jackson2011,Harte2017,McClintockMichelot2018}. One of the earliest and most flexible HMM packages, \verb|depmixS4| \citep{VisserSpeenkenbrink2010}, can accommodate multivariate HMMs, multiple observation sequences, parameter covariates, parameter constraints, and missing observations. Similar to \verb|depmixS4| in terms of features and flexibility, \verb|momentuHMM| \citep{McClintockMichelot2018} can also be used to implement mixed HMMs \citep[][]{DeRuiterEtAl2017}, hierarchical HMMs \citep[][]{Leos-BarajasEtAl2017,AdamEtAl2019}, zero-inflated probability distributions \citep[][]{MartinEtAl2005}, and partially-observed state sequences. In addition to the R packages presented in Table \ref{tab:software}, there are numerous R and stand-alone software packages that are less general and specialise on particular HMM applications in ecology, as well as programs with which these types of models can be relatively easily implemented by users with minimal statistical programming experience (see Appendix B in Supplementary Material).

\begin{landscape}
\begin{table}[!htb]
\caption{Features of HMM packages available in the R environment for statistical computing, including capabilities for multiple observation sequences (``Multiple sequences''), multivariate HMMs (``Multivariate''), mixed HMMs (``Mixed''), hierarchical HMMs (``Hierarchical''), hidden semi-Markov models (``Semi-Markov''), parameter covariate modelling (``Covariates''), parameter constraints (``Constraints''), missing observations (``Missing data''), and state-dependent probability distributions. ``Covariates'' and ``Constraints'' can pertain to initial distribution $(\delta)$, state-dependent probability distribution $(f)$, state transition probability $(\gamma)$, and/or mixture probability $(\pi)$ parameters. Several packages facilitate extensions for user-specified state-dependent probability distributions that require no modifications to the package source code (``custom''). \label{tab:software}}
\centering
\vspace{0.5em}
\resizebox{1.3\textwidth}{!}{
\begin{tabular}{llcccccccccccccccccc}
\hline
\multicolumn{2}{l}{\bf Package} & \multicolumn{2}{c}{\bf Multiple}  & \multicolumn{2}{c}{\bf Multivariate} & \multicolumn{2}{c}{\bf Mixed} & \multicolumn{2}{c}{\bf Hierarchical} & \multicolumn{2}{c}{\bf Semi-Markov} & \multicolumn{2}{c}{\bf Covariates} & \multicolumn{2}{c}{\bf Constraints}    & \multicolumn{2}{c}{\bf Missing data} & \multicolumn{2}{c}{\bf Reference} \\ 
\multicolumn{2}{l}{}            & \multicolumn{2}{c}{\bf sequences} &  \multicolumn{2}{c}{}                &  \multicolumn{2}{c}{}         &   \multicolumn{2}{c}{}               &    \multicolumn{2}{c}{}             &    \multicolumn{2}{c}{}            &    \multicolumn{2}{c}{}                &    \multicolumn{2}{l}{}               & \multicolumn{2}{c}{} \\
\hline
\rowcolor{Gray}\multicolumn{2}{l}{{\texttt aphid}}        &    \multicolumn{2}{c}{\checkmark}              &    \multicolumn{2}{c}{}                &    \multicolumn{2}{c}{}         &        \multicolumn{2}{c}{}            &     \multicolumn{2}{c}{}              &     \multicolumn{2}{c}{}                         &        \multicolumn{2}{c}{}              &         \multicolumn{2}{c}{}               &  \multicolumn{2}{l}{\citet{Wilkinson2019}}\\
\multicolumn{2}{l}{{\texttt depmixS4}}     &   \multicolumn{2}{c}{\checkmark}    &     \multicolumn{2}{c}{\checkmark}     &    \multicolumn{2}{c}{}         &     \multicolumn{2}{c}{}               &     \multicolumn{2}{c}{}              &  \multicolumn{2}{c}{$\delta,f,\gamma$}           & \multicolumn{2}{c}{$\delta,f,\gamma$}    &       \multicolumn{2}{c}{\checkmark}                          &  \multicolumn{2}{l}{\citet{VisserSpeenkenbrink2010}}\\
\rowcolor{Gray}\multicolumn{2}{l}{{\texttt HiddenMarkov}} &    \multicolumn{2}{c}{}             &    \multicolumn{2}{c}{}                &   \multicolumn{2}{c}{}          &    \multicolumn{2}{c}{}                &      \multicolumn{2}{c}{}             &     \multicolumn{2}{c}{$f^\ast$}                 &      \multicolumn{2}{c}{}                &       \multicolumn{2}{c}{}                   &  \multicolumn{2}{l}{\citet{Harte2017}}\\
\multicolumn{2}{l}{{\texttt HMM}}          &   \multicolumn{2}{c}{}              &   \multicolumn{2}{c}{}                 &    \multicolumn{2}{c}{}         &   \multicolumn{2}{c}{}                 &   \multicolumn{2}{c}{}                &      \multicolumn{2}{c}{}                        &         \multicolumn{2}{c}{}             &    \multicolumn{2}{c}{}               &  \multicolumn{2}{l}{\citet{Himmelmann2010}}\\
\rowcolor{Gray}\multicolumn{2}{l}{{\texttt hsmm}}         &    \multicolumn{2}{c}{}             &    \multicolumn{2}{c}{}                &   \multicolumn{2}{c}{}          &     \multicolumn{2}{c}{}               &   \multicolumn{2}{c}{\checkmark}      &        \multicolumn{2}{c}{}                      &     \multicolumn{2}{c}{}                 &  \multicolumn{2}{c}{}  &  \multicolumn{2}{l}{\citet{BullaBulla2013}}\\
\multicolumn{2}{l}{{\texttt LMest}}        &   \multicolumn{2}{c}{\checkmark}    &     \multicolumn{2}{c}{\checkmark}     & \multicolumn{2}{c}{\checkmark}  &     \multicolumn{2}{c}{}               &     \multicolumn{2}{c}{}              &\multicolumn{2}{c}{$f^\dagger$ or $\delta,\gamma$}&         \multicolumn{2}{c}{}             &  \multicolumn{2}{c}{\checkmark} &  \multicolumn{2}{l}{\citet{BartolucciEtAl2017}}\\
\rowcolor{Gray}\multicolumn{2}{l}{{\texttt mhsmm}}        &   \multicolumn{2}{c}{\checkmark}    &     \multicolumn{2}{c}{}               &    \multicolumn{2}{c}{}         &      \multicolumn{2}{c}{}              &  \multicolumn{2}{c}{ \checkmark}      &        \multicolumn{2}{c}{}                      &      \multicolumn{2}{c}{}                &  \multicolumn{2}{c}{\checkmark}  &  \multicolumn{2}{l}{\citet{OConnellHojsgaard2011}}\\
\multicolumn{2}{l}{{\texttt momentuHMM}}   &   \multicolumn{2}{c}{\checkmark}    &     \multicolumn{2}{c}{\checkmark}     & \multicolumn{2}{c}{\checkmark}  &    \multicolumn{2}{c}{\checkmark}      &    \multicolumn{2}{c}{}               &\multicolumn{2}{c}{$\delta,f,\gamma,\pi$}         & \multicolumn{2}{c}{$\delta,f,\gamma,\pi$}&      \multicolumn{2}{c}{\checkmark}             &  \multicolumn{2}{l}{\citet{McClintockMichelot2018}}\\
\rowcolor{Gray}\multicolumn{2}{l}{{\texttt msm}}          &   \multicolumn{2}{c}{\checkmark}    &     \multicolumn{2}{c}{\checkmark}     &     \multicolumn{2}{c}{}        &      \multicolumn{2}{c}{}              &      \multicolumn{2}{c}{}             &      \multicolumn{2}{c}{$f^\ddagger,\gamma$}     &    \multicolumn{2}{c}{$f,\gamma$}        &     \multicolumn{2}{c}{\checkmark}             &  \multicolumn{2}{l}{\citet{Jackson2011}}\\
\multicolumn{2}{l}{{\texttt RcppHMM}}      &     \multicolumn{2}{c}{}            &    \multicolumn{2}{c}{}                &    \multicolumn{2}{c}{}         &      \multicolumn{2}{c}{}              &    \multicolumn{2}{c}{}               &        \multicolumn{2}{c}{}                      &            \multicolumn{2}{c}{}          &       \multicolumn{2}{c}{}         &  \multicolumn{2}{l}{\citet{CardenasOvandoEtAl2017}}\\
\rowcolor{Gray}\multicolumn{2}{l}{{\texttt seqHMM}}       &   \multicolumn{2}{c}{\checkmark}    &     \multicolumn{2}{c}{\checkmark}     & \multicolumn{2}{c}{\checkmark}  &      \multicolumn{2}{c}{}              &     \multicolumn{2}{c}{}              &      \multicolumn{2}{c}{$\pi$}                   & \multicolumn{2}{c}{$\delta,\gamma$}      &  \multicolumn{2}{c}{\checkmark}     &  \multicolumn{2}{l}{\citet{HelskeHelske2019}}\\
\hline
\multicolumn{20}{l}{}\\
\multicolumn{2}{l}{} & \multicolumn{18}{c}{\bf State-dependent probability distributions} \\
\multicolumn{20}{l}{}\\
\cline{3-20}
\multicolumn{2}{l}{}                                  &\bf Bernoulli & \bf beta      &\bf binomial& \bf categorical & \bf custom    & \bf exponential & \bf gamma    & \bf lognormal & \bf logistic & \bf negative & \bf normal  & \bf multivariate  & \bf truncated  & \bf Poisson      & \bf Student's $t$  & \bf Von Mises & \bf Weibull    & \bf wrapped \\
\multicolumn{2}{l}{}                                  & & & & & & & & & & \bf binomial & & \bf normal & \bf normal  & & &  &  & \bf Cauchy \\
\rowcolor{Gray}\multicolumn{2}{l}{{\texttt aphid}}        &              &               &            & \checkmark      &               &                 &              &               &              &                       &             &                         &                      &                  &                    &               &                &                    \\
\multicolumn{2}{l}{{\texttt depmixS4}}                    &              &               & \checkmark & \checkmark      &  \checkmark   &                 & \checkmark   &               &              &                       & \checkmark  &                         &                      &  \checkmark      &                    &               &                &                    \\  
\rowcolor{Gray}\multicolumn{2}{l}{{\texttt HiddenMarkov}} &              & \checkmark    & \checkmark &                 &  \checkmark   &   \checkmark    & \checkmark   &  \checkmark   &  \checkmark  &                       & \checkmark  &                         &                      &  \checkmark      &                    &               &                &                    \\
\multicolumn{2}{l}{{\texttt HMM}}                         &              &               &            & \checkmark      &               &                 &              &               &              &                       &             &                         &                      &                  &                    &               &                &                    \\          
\rowcolor{Gray}\multicolumn{2}{l}{{\texttt hsmm}}         & \checkmark   &               &            &                 &               &                 &              &               &              &                       & \checkmark  &                         &                      &  \checkmark      &   \checkmark       &               &                &                    \\         
\multicolumn{2}{l}{{\texttt LMest}}                       &              &               &            & \checkmark      &               &                 &              &               &              &                       & \checkmark  &      \checkmark         &                      &                  &                    &               &                &                    \\         
\rowcolor{Gray}\multicolumn{2}{l}{{\texttt mhsmm}}        &              &               &            &                 &  \checkmark   &                 &              &               &              &                       & \checkmark  &      \checkmark         &                      &  \checkmark      &                    &               &                &                    \\        
\multicolumn{2}{l}{{\texttt momentuHMM}}                  & \checkmark   & \checkmark    &            & \checkmark      &               &   \checkmark    &  \checkmark  &  \checkmark   &  \checkmark  &      \checkmark       & \checkmark  &      \checkmark         &                      &  \checkmark      & \checkmark         & \checkmark    &  \checkmark    &    \checkmark      \\     
\rowcolor{Gray}\multicolumn{2}{l}{{\texttt msm}}          & \checkmark   & \checkmark    & \checkmark & \checkmark      &               &   \checkmark    &  \checkmark  &  \checkmark   &              &      \checkmark       & \checkmark  &                         &      \checkmark      &  \checkmark      & \checkmark         &               &  \checkmark    &                    \\          
\multicolumn{2}{l}{{\texttt RcppHMM}}                     &              &               &            & \checkmark      &               &                 &              &               &              &                       & \checkmark  &      \checkmark         &                      &  \checkmark      &                    &               &                &                    \\               
\rowcolor{Gray}\multicolumn{2}{l}{{\texttt seqHMM}}       &              &               &            & \checkmark      &               &                 &              &               &              &                       &             &                         &                      &                  &                    &               &                &                    \\ 
\multicolumn{20}{l}{\LARGE{$^\ast$Covariates are only permitted on state-dependent distribution location parameters for the binomial, gamma, normal, and Poisson distributions.}}\\
\multicolumn{20}{l}{\LARGE{$^\dagger$Covariates are only permitted on state-dependent categorical distribution parameters.}}\\
\multicolumn{20}{l}{\LARGE{$^\ddagger$Covariates are only permitted on state-dependent distribution location parameters.}}
\end{tabular}
}
\end{table}
\end{landscape}

\subsection{Challenges and pitfalls}
\label{sec:challenges}

HMMs are natural candidates for conducting inference related to a wide range of ecological phenomena, but they are not a panacea (see Box \ref{box:toHMM}). There are many ecological processes that cannot be faithfully characterised under the simplifying assumptions of HMMs, in which case other latent variable models may be more appropriate (see Box \ref{box:latent}). When HMMs are appropriate, it can be challenging to tailor HMMs to real data, even when using user-friendly software packages. Here we briefly highlight those issues that, based on our experience, constitute the key challenges when using HMMs to analyse ecological data. Other important aspects of statistical practice that are not unique to HMMs, such as model checking and selection \citep[e.g.][Chapter 6]{ZucchiniEtAl2016}, are covered in more detail in the Supplementary Tutorial.

\vspace{.25cm}

\begin{textBox}[1]{Box \theboxCount. To HMM, or not to HMM, that is the question}
\label{box:toHMM}
The structure of a statistical model should be congruent with the data-generating process in question. HMMs are neither a panacea nor a black box --- the appropriateness and feasibility of a particular model will be case-dependent and requires careful consideration. In determining if HMMs are appropriate for describing a particular system, one must consider two questions: 
\begin{enumerate}
\item \textbf{Do the hidden state dynamics display time dependence which can be represented using Markov chains?} If the current system state is not related to the previous state(s), then a latent variable model without time dependence should be considered (see Box \ref{box:latent}). Diagnostics examining temporal patterns in residuals \citep{li2003diagnostic} can help to empirically determine if the assumptions of conditional independence and Markovity are sufficient (see Supplementary Tutorial). When the first-order Markov assumption may not be appropriate for the state process, one can further ask the question: \textit{can system memory be adequately approximated while preserving Markovity?} Faithful representation of system memory may require the inclusion of informative covariates or more complex time dependence structures, and it is possible to expand HMMs to higher-order Markovian or semi-Markovian dependence \citep[][Chapter 12]{ZucchiniEtAl2016}. While modelling this higher-order temporal dependence is sometimes preferable \citep[][]{hestbeck1991estimates}, it is more complex and thus less widely used. General time series modelling often captures complex dependence structures using autoregressive processes \citep[][Chapter 3]{durbin2012time}, and more complicated variations of HMMs can capture some of these features \citep[][]{lawler2019conditionally}. However, other latent variable approaches will often be better suited for more complex temporal dependence structures. There is no foolproof or automatic way to make this determination, and we must typically rely on residual diagnostics \citep[][Chapter 6]{li2003diagnostic,ZucchiniEtAl2016} and expert knowledge of the system dynamics. 

\item \textbf{Can the system be well described by a feasibly finite set of latent states?} Our review highlights a wide range of ecological scenarios where the possible states of the system of interest form (or can be approximated by) a finite set. The number of parameters and the computational burden of an HMM can become large with increases in state dimension, and this can be of particular concern when the finite set of states is a coarser approximation of a finer discrete space (e.g.\ population abundance) or a continuous space (e.g.\ spatial location). Such approximations have strengths and weaknesses. When used as discrete approximations to state-space models (see Box \ref{box:latent}), HMMs can be useful when arbitrary constraints on the state space are required (e.g.\ restricting aquatic organisms to location states off land) or when combining both discrete and continuous state processes
. However, an HMM for a large number of states with a fully parameterised transition probability matrix --- where transitions between any of the states are possible 
--- will be computationally expensive, perhaps prohibitively so. Systems with large state spaces can often be approximated by an HMM when transitions between states are local --- where 
transitions can only occur between neighbouring states --- and the transition probabilities therefore include a relatively small number of parameters that describe this local behaviour. For example, \citet{ThygesenEtAl2009}, \citet{PedersenEtAl2011estimating}, and \citet{glennie2019open} use these properties of sparsity to make an HMM approach computationally efficient for very large state spaces. In short, large numbers of states do not necessarily prohibit application of an HMM; this is dependent on the computer resources available and the properties of the state process. Alternatively, it is possible to reduce the size of an infeasible state space by making a coarser approximation (e.g.\ binning abundance states together into larger states; \citealt[][pp.\ 162--163]{ZucchiniEtAl2016}; \citealt{BesbeasMorgan2019}). Appropriateness will depend on the sensitivity of the inference to the precise value of the state process and is best investigated by varying the coarseness of the approximation. If the set of states is too coarse-grained, approximation might lead to spurious inference about the latent states. For example, coarse-graining could result in masking or misclassification of meaningfully distinct states. The decision of the appropriate number of states can be challenging; there is again no foolproof or automatic way to determine this, and we must usually rely on expert knowledge of the specific system of interest. When the finite state space of an HMM is infeasible or inappropriate, it will often be better to consider other approaches \citep[e.g.][]{patterson2008state,cooch2012disease,patterson2017statistical,auger2020introduction}.      
\end{enumerate}

\end{textBox}
\vspace{.25in}

Depending on the complexity of the state and observation processes, various modelling decisions may need to be made. Among these are the number of states to include, whether to incorporate covariates for the model parameters, and whether the basic dependence structure is sufficient. These decisions tend to be case-dependent and require expert knowledge of the system of interest, so we make no attempt to provide general guidance in this respect. However, in some cases the model structure may be a direct consequence of the ecological process. For example, in the Cormack-Jolly-Seber capture-recapture model, the number of states (namely two: alive or dead) and also the state-dependent (Bernoulli) distributions follow immediately from the capture-recapture process. In situations with more complex data, such as multivariate time series related to animal behaviour \citep{DeRuiterEtAl2017,ngo2019understanding,van2019classifying}, it takes experience and a good intuition both for the data and for the HMM framework to identify an adequate model formulation \citep[][]{pohle2017selecting}. 

Unlike other statistical models such as linear regression, there is no analytical solution for HMM parameter estimation. As a consequence, one needs to resort to numerical procedures, all of which involve technical challenges: local maxima in case of maximum likelihood estimation \citep[][]{Myung2003}, 
or label switching \citep[][]{JasraEtAl2005} and poor mixing \citep[][]{BrooksEtAl2011} when using MCMC sampling. Any increase in model complexity with respect to the number of states or the parameters tends to rapidly exacerbate these problems. When working with HMMs, it is thus important to develop an appreciation for these challenges and the associated risks. For maximum likelihood in particular, the risk of false convergence to a local rather than the global maximum of the likelihood must not be underestimated. In addition to the general advice to avoid overly complex models \citep[][]{Cole2019}, the main strategy to reduce this risk is to try out many initial parameter vectors within the maximisation.

While it is tempting to interpret the states of an HMM fitted to ecological data as biologically meaningful entities, oftentimes this is in fact not justifiable. Outside of standard capture-recapture or species occurrence applications, HMMs are often applied in an unsupervised learning context (see Figs \ref{fig:bluewhale} and \ref{fig:caracara}, Supplementary Tutorial), such that the state characteristics are completely data-driven rather than pre-defined \citep[][]{Leos-BarajasEtAl2017accelerometer}. The model then picks up the \textit{statistically} most relevant modal patterns in the data, and these may or may not correspond closely to \textit{ecologically} meaningful states. It is thus important not to over-interpret the model states, as in some cases they may only be crude proxies for the ecological system states of interest. A classic example is the simple $N=2$ state HMM for animal movement behaviour based on step lengths and turning angles \citep[][]{MoralesEtAl2004}, where evidence of an area-restricted search-type state is often labelled as ``foraging'' activity. Although for many animals area-restricted search is commonly associated with foraging, one usually cannot definitively conclude when and where an individual was actually foraging based solely on location data. Furthermore, while it can be useful to refer to these modalities using descriptive terms such as ``foraging'' (or ``resident'') and ``searching'' (or ``transient'') states, this does not mean that an animal has only two simple modes of behaviour. However, using auxiliary information \citep[][]{AustinEtAl2006,FrankeEtAl2006} and/or partially pre-defining state characteristics based on the system of interest \citep[][]{McClintockEtAl2013c, McClintockEtAl2017} can help mitigate these issues and facilitate ecological interpretation.

\section{Future directions}
\label{sec:futureDirections}

\begin{figure}[!htb]
    \centering
    \includegraphics[width=\textwidth]{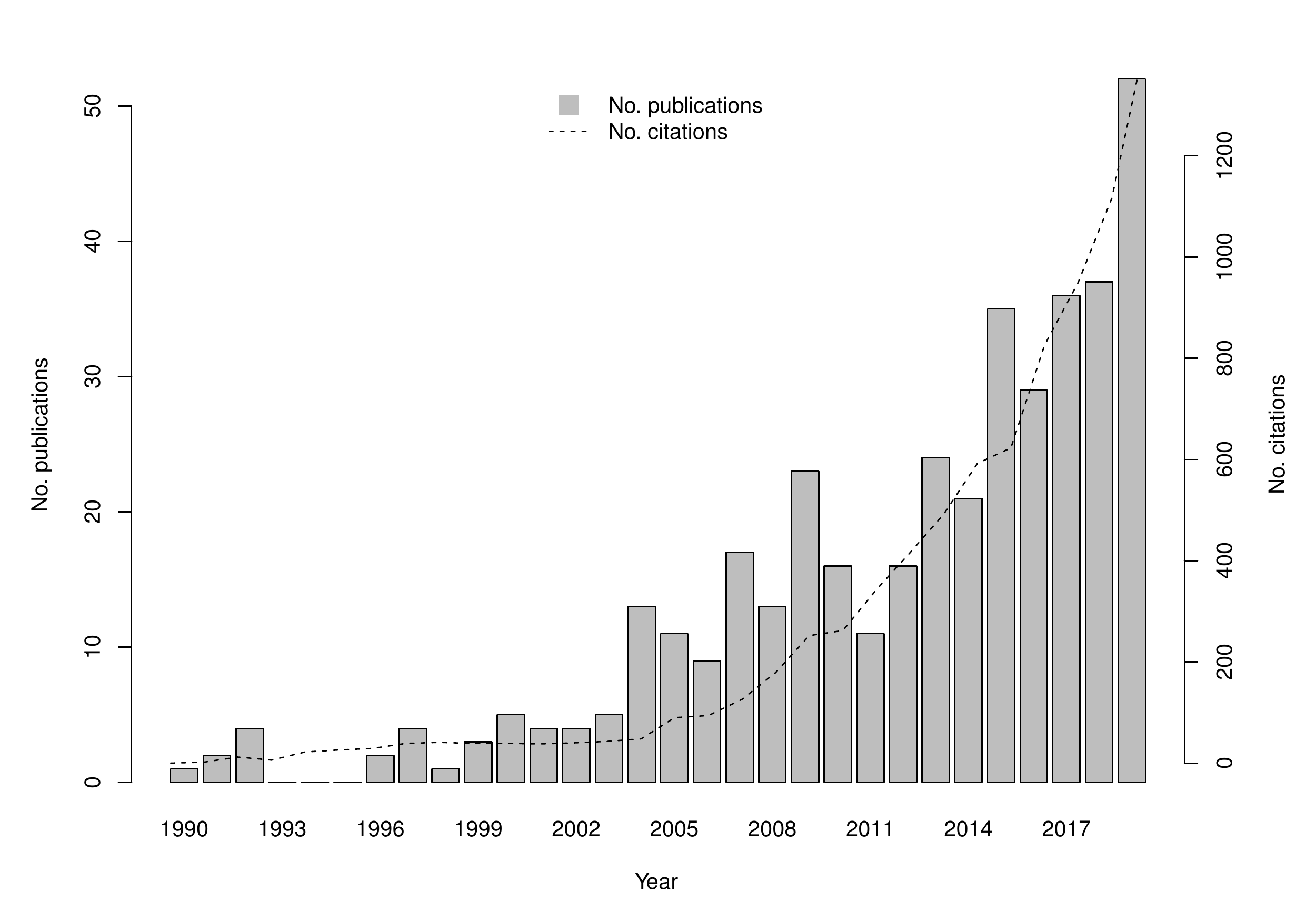}
    \caption{Number of publications (left axis) and total number of times these publications were cited (right axis) per year based on a Web of Science search for ``hidden Markov'' conducted within the categories of ``Biology'', ``Ecology'', ``Marine Freshwater Biology'' and ``Zoology'' on 7 July 2020.}
    \label{fig:citationReport}
\end{figure}

We have highlighted many realised and potential applications of HMMs in ecology.   
We anticipate increased application and development of HMMs as ecologists continue to discover how this relatively simple and flexible class of statistical models can reveal complex state dynamics that are inherently difficult to observe. Indeed, a Web of Science search for ``hidden Markov'' suggests a rapidly increasing awareness of these models within the ecological community over the past two decades (Fig.\ \ref{fig:citationReport}). Given differences in terminology and a tendency for ecologists to use HMMs without explicitly referring to them as such, the use of HMMs is surely becoming even more widespread in our field.

In order for the power and flexibility of HMMs to be harnessed by the broader ecological community, researchers must first be able to recognise the limitations of their data and 
how these can be leveraged by formally linking observable phenomena to the actual ecological processes of interest. Such hierarchical modelling exercises are critical to reliable inference \citep[][]{RoyleDorazio2008,kery_applied_2015}, and it is no coincidence that HMMs have independently ``evolved'' in different ecological contexts over the years. By assuming a discrete state space with basic dependence structures, HMMs can easily capture complex system processes, such as those involving serial correlation, non-linearity, non-normality, and non-stationarity, in a tractable manner that goes well beyond the examples highlighted in Section 3. Instead of viewing these examples as a series of disparate domain-specific applications of HMMs, we view them as a synthesis of the process by which ecologists can begin to critically think about their own sequential data, relate them to their particular system of interest, and formulate an HMM for their specific domain using a simple conceptual template.

We foresee HMMs being more frequently used to integrate biotic and abiotic observations at large spatio-temporal scales to investigate complex ecosystem-level processes. The state process of the HMM could itself be at the ecosystem level (e.g.\ alternative stable states), or it could simply be used to account for unobservable state dynamics at lower levels of the hierarchy as a component of a larger (non-Markovian) ecosystem-level process model. Recent HMM methodological developments such as hierarchical formulations that allow data collection and/or state transitions to occur at multiple temporal resolutions \citep{FineEtAl1998,Leos-BarajasEtAl2017,AdamEtAl2019}, nonparametric approaches avoiding restrictive distributional assumptions \citep{yau2011bayesian,langrock2018spline}, and coupled HMMs for 
interacting state processes associated with different sequences \citep{sherlock2013coupled,touloupou2019scalable} 
extend our capability to incorporate complex data structures and hierarchical relationships scaled from the individual to ecosystem level. 

Despite this great potential, there remain several hurdles to the widespread implementation of HMMs describing long-term, broad-scale ecological dynamics \citep[][]{TurnerEtAl1995,LindenmayerEtAl2012,Haller2014}. First, much like regression and analysis of variance, HMMs must become a familiar and accessible instrument within the ecologist's statistical ``toolbox''. This has been the primary motivation for our review, and we hope our illustrative examples have provided a template by which researchers can begin to formulate HMMs according to their specific state and observation processes of interest. Second, although this challenge is by no means unique to HMMs, ecosystem-level inferences continue to be limited by data availability, accessibility, and compatibility \citep[][]{JonesEtAl2006,DietzeEtAl2018,EstesEtAl2018,CompagnoniEtAl2019,HalbritterEtAl2019}, which can compromise our ability to empirically link observation and state processes operating at different spatio-temporal scales. Third, as with any application of HMMs, such endeavours will require a faithful conceptualisation of ecosystem dynamics that is amenable to this discrete-state modelling framework, as well as the identification and integration of observation processes that can provide information about the underlying system.

\section*{Acknowledgments}
The scientific results and conclusions, as well as any views or opinions expressed herein, are those of the author(s) and do not necessarily reflect those of NOAA or the Department of Commerce.

\bibliographystyle{ecology_letters2}
\bibliography{master}

\begin{thebibliography}{277}
\expandafter\ifx\csname natexlab\endcsname\relax\def\natexlab#1{#1}\fi
\expandafter\ifx\csname url\endcsname\relax
  \def\url#1{\texttt{#1}}\fi
\expandafter\ifx\csname urlprefix\endcsname\relax\def\urlprefix{URL }\fi

\bibitem[{Abercrombie \& Friedl(2015)}]{AbercrombieEtAl2015}
Abercrombie, S.~P. \& Friedl, M.~A. (2015).
\newblock Improving the consistency of multitemporal land cover maps using a
  hidden {M}arkov model.
\newblock \emph{IEEE Transactions on Geoscience and Remote Sensing}, 54,
  703--713.

\bibitem[{Adam \emph{et~al.}(2019{\natexlab{a}})Adam, Griffiths, Leos-Barajas,
  Meese, Lowe, Blackwell, Righton \& Langrock}]{AdamEtAl2019}
Adam, T., Griffiths, C.~A., Leos-Barajas, V., Meese, E.~N., Lowe, C.~G.,
  Blackwell, P.~G., Righton, D. \& Langrock, R. (2019{\natexlab{a}}).
\newblock Joint modelling of multi-scale animal movement data using
  hierarchical hidden {M}arkov models.
\newblock \emph{Methods in Ecology and Evolution}, 10, 1536--1550.

\bibitem[{Adam \emph{et~al.}(2019{\natexlab{b}})Adam, Langrock \&
  Wei{\ss}}]{adam2019penalized}
Adam, T., Langrock, R. \& Wei{\ss}, C.~H. (2019{\natexlab{b}}).
\newblock Penalized estimation of flexible hidden {M}arkov models for time
  series of counts.
\newblock \emph{METRON}, 77, 87--104.

\bibitem[{Altman(2007)}]{Altman2007}
Altman, R.~M. (2007).
\newblock Mixed hidden {M}arkov models: an extension of the hidden {M}arkov
  model to the longitudinal data setting.
\newblock \emph{Journal of the American Statistical Association}, 102,
  201--210.

\bibitem[{Amoros \emph{et~al.}(2019)Amoros, King, Toyoda, Kumada, Johnson \&
  Bird}]{amoros2019continuous}
Amoros, R., King, R., Toyoda, H., Kumada, T., Johnson, P.~J. \& Bird, T.~G.
  (2019).
\newblock A continuous-time hidden {M}arkov model for cancer surveillance using
  serum biomarkers with application to hepatocellular carcinoma.
\newblock \emph{METRON}, 77, 67--86.

\bibitem[{Auger-M{\'e}th{\'e} \emph{et~al.}(2020)Auger-M{\'e}th{\'e}, Newman,
  Cole, Empacher, Gryba, King, Leos-Barajas, Flemming, Nielsen, Petris
  \emph{et~al.}}]{auger2020introduction}
Auger-M{\'e}th{\'e}, M., Newman, K., Cole, D., Empacher, F., Gryba, R., King,
  A.~A., Leos-Barajas, V., Flemming, J.~M., Nielsen, A., Petris, G.
  \emph{et~al.} (2020).
\newblock An introduction to state-space modeling of ecological time series.
\newblock \emph{arXiv preprint arXiv:2002.02001}.

\bibitem[{Aurdal \emph{et~al.}(2005)Aurdal, Huseby, Eikvil, Solberg, Vikhamar
  \& Solberg}]{AurdalEtAl2005}
Aurdal, L., Huseby, R.~B., Eikvil, L., Solberg, R., Vikhamar, D. \& Solberg, A.
  (2005).
\newblock Use of hidden {M}arkov models and phenology for multitemporal
  satellite image classification: Applications to mountain vegetation
  classification.
\newblock In: \emph{Proceedings of the Third International Workshop on the
  Analysis of Multi-temporal Remote Sensing Images, Biloxi, USA} (eds. King,
  R.~L. \& Younan, N.~H.). IEEE, pp. 220--224.

\bibitem[{Austin \emph{et~al.}(2006)Austin, Bowen, McMillan \&
  Boness}]{AustinEtAl2006}
Austin, D., Bowen, W., McMillan, J. \& Boness, D. (2006).
\newblock Stomach temperature telemetry reveals temporal patterns of foraging
  success in a free-ranging marine mammal.
\newblock \emph{Journal of Animal Ecology}, 75, 408--420.

\bibitem[{B\'alint \emph{et~al.}(2018)B\'alint, Pfenninger, Grossart, Taberlet,
  Vellend, Leibold, Englund \& Bowler}]{BalintEtAl2018}
B\'alint, M., Pfenninger, M., Grossart, H.-P., Taberlet, P., Vellend, M.,
  Leibold, M.~A., Englund, G. \& Bowler, D. (2018).
\newblock Environmental {DNA} time series in ecology.
\newblock \emph{Trends in Ecology \& Evolution}, 33, 945--957.

\bibitem[{Barbu \& Limnios(2009)}]{BarbuLimnios2009}
Barbu, V.~S. \& Limnios, N. (2009).
\newblock \emph{Semi-{M}arkov Chains and Hidden Semi-{M}arkov Models Toward
  Applications: Their Use in Reliability and {DNA} Analysis}, vol. 191 of
  \emph{Lecture Notes in Statistics}.
\newblock Springer.

\bibitem[{Bartolucci \emph{et~al.}(2017)Bartolucci, Pandolfi \&
  Pennoni}]{BartolucciEtAl2017}
Bartolucci, F., Pandolfi, S. \& Pennoni, F. (2017).
\newblock {LMest}: An {R} package for latent {M}arkov models for longitudinal
  categorical data.
\newblock \emph{Journal of Statistical Software}, 81, 1--38.

\bibitem[{Basson \emph{et~al.}(2016)Basson, Bravington, Hartog \&
  Patterson}]{BassonEtAl2016}
Basson, M., Bravington, M.~V., Hartog, J.~R. \& Patterson, T.~A. (2016).
\newblock Experimentally derived likelihoods for light-based geolocation.
\newblock \emph{Methods in Ecology and Evolution}, 7, 980--989.

\bibitem[{Begon \emph{et~al.}(2006)Begon, Harper \& Townsend}]{BegonEtAl2006}
Begon, M., Harper, J. \& Townsend, C. (2006).
\newblock \emph{Ecology: From Individuals to Ecosystems}.
\newblock 4th edn. Wiley-Blackwell.

\bibitem[{Beisner \emph{et~al.}(2003)Beisner, Haydon \&
  Cuddington}]{BeisnerEtAl2003}
Beisner, B., Haydon, D. \& Cuddington, K. (2003).
\newblock Alternative stable states in ecology.
\newblock \emph{Frontiers in Ecology and the Environment}, 1, 376--382.

\bibitem[{Benhaiem \emph{et~al.}(2018)Benhaiem, Marescot, Hofer, East,
  Lebreton, Kramer-Schadt \& Gimenez}]{Benhaiem2018}
Benhaiem, S., Marescot, L., Hofer, H., East, M.~L., Lebreton, J.-D.,
  Kramer-Schadt, S. \& Gimenez, O. (2018).
\newblock Robustness of eco-epidemiological capture-recapture parameter
  estimates to variation in infection state uncertainty.
\newblock \emph{Frontiers in Veterinary Science}, 5, 197.

\bibitem[{Besbeas \& Morgan(2019)}]{BesbeasMorgan2019}
Besbeas, P. \& Morgan, B. J.~T. (2019).
\newblock Exact inference for integrated population modelling.
\newblock \emph{Biometrics}, 75, 475--484.

\bibitem[{Bhar \& Hamori(2004)}]{BharShigeyuki2004}
Bhar, R. \& Hamori, S. (2004).
\newblock \emph{Hidden {M}arkov Models: Applications to Financial Economics},
  vol.~40 of \emph{Advanced Studies in Theoretical and Applied Econometrics}.
\newblock Springer.

\bibitem[{Bode \& Seitz(2018)}]{BodeSeitz2018}
Bode, N.~W. \& Seitz, M.~J. (2018).
\newblock Using hidden {M}arkov models to characterise intermittent social
  behaviour in fish shoals.
\newblock \emph{The Science of Nature}, 105, 7.

\bibitem[{Bohmann \emph{et~al.}(2014)Bohmann, Evans, Gilbert, Carvalho, Creer,
  Knapp, Douglas \& De~Bruyn}]{BohmannEtAl2014}
Bohmann, K., Evans, A., Gilbert, M. T.~P., Carvalho, G.~R., Creer, S., Knapp,
  M., Douglas, W.~Y. \& De~Bruyn, M. (2014).
\newblock Environmental {DNA} for wildlife biology and biodiversity monitoring.
\newblock \emph{Trends in Ecology \& Evolution}, 29, 358--367.

\bibitem[{Boitard \emph{et~al.}(2009)Boitard, Schl{\"o}tterer \&
  Futschik}]{BoitardEtAl2009}
Boitard, S., Schl{\"o}tterer, C. \& Futschik, A. (2009).
\newblock Detecting selective sweeps: a new approach based on hidden {M}arkov
  models.
\newblock \emph{Genetics}, 181, 1567--1578.

\bibitem[{Borchers \& Efford(2008)}]{Borchers2008}
Borchers, D.~L. \& Efford, M. (2008).
\newblock Spatially explicit maximum likelihood methods for capture--recapture
  studies.
\newblock \emph{Biometrics}, 64, 377--385.

\bibitem[{Borgy \emph{et~al.}(2015)Borgy, Reboud, Peyrard, Sabbadin \&
  Gaba}]{BorgyEtAl2015}
Borgy, B., Reboud, X., Peyrard, N., Sabbadin, R. \& Gaba, S. (2015).
\newblock Dynamics of weeds in the soil seed bank: a hidden {M}arkov model to
  estimate life history traits from standing plant time series.
\newblock \emph{PLoS ONE}, 10, e0139278.

\bibitem[{Boron \emph{et~al.}(2019)Boron, Deere, Xofis, Link,
  Qui{\~n}ones-Guerrero, Payan \& Tzanopoulos}]{BoronEtAl2019}
Boron, V., Deere, N.~J., Xofis, P., Link, A., Qui{\~n}ones-Guerrero, A., Payan,
  E. \& Tzanopoulos, J. (2019).
\newblock Richness, diversity, and factors influencing occupancy of mammal
  communities across human-modified landscapes in {C}olombia.
\newblock \emph{Biological Conservation}, 232, 108--116.

\bibitem[{Bracken \emph{et~al.}(2014)Bracken, Rajagopalan \&
  Zagona}]{BrackenEtAl2014}
Bracken, C., Rajagopalan, B. \& Zagona, E. (2014).
\newblock A hidden {M}arkov model combined with climate indices for
  multidecadal streamflow simulation.
\newblock \emph{Water Resources Research}, 50, 7836--7846.

\bibitem[{Breed \emph{et~al.}(2013)Breed, Don~Bowen \& Leonard}]{BreedEtAl2013}
Breed, G.~A., Don~Bowen, W. \& Leonard, M.~L. (2013).
\newblock Behavioral signature of intraspecific competition and density
  dependence in colony-breeding marine predators.
\newblock \emph{Ecology and Evolution}, 3, 3838--3854.

\bibitem[{Breininger \emph{et~al.}(2010)Breininger, Nichols, Duncan, Stolen,
  Carter, Hunt \& Drese}]{BreiningerEtAl2010}
Breininger, D.~R., Nichols, J.~D., Duncan, B.~W., Stolen, E.~D., Carter, G.~M.,
  Hunt, D.~K. \& Drese, J.~H. (2010).
\newblock Multistate modeling of habitat dynamics: factors affecting {F}lorida
  scrub transition probabilities.
\newblock \emph{Ecology}, 91, 3354--3364.

\bibitem[{Broms \emph{et~al.}(2015)Broms, Hooten \&
  Fitzpatrick}]{BromsEtAl2015}
Broms, K.~M., Hooten, M.~B. \& Fitzpatrick, R.~M. (2015).
\newblock Accounting for imperfect detection in {H}ill numbers for biodiversity
  studies.
\newblock \emph{Methods in Ecology and Evolution}, 6, 99--108.

\bibitem[{Brooks \emph{et~al.}(2011)Brooks, Gelman, Jones \&
  Meng}]{BrooksEtAl2011}
Brooks, S., Gelman, A., Jones, G. \& Meng, X.-L. (2011).
\newblock \emph{Handbook of Markov Chain Monte Carlo}.
\newblock CRC press.

\bibitem[{Brownie \emph{et~al.}(1993)Brownie, Hines, Nichols, Pollock \&
  Hestbeck}]{BrownieEtAl1993}
Brownie, C., Hines, J.~E., Nichols, J.~D., Pollock, K.~H. \& Hestbeck, J.~B.
  (1993).
\newblock Capture-recapture studies for multiple strata including
  non-{M}arkovian transitions.
\newblock \emph{Biometrics}, 49, 1173--1187.

\bibitem[{Buckland \emph{et~al.}(2004)Buckland, Newman, Thomas \&
  Koester}]{BucklandEtAl2004}
Buckland, S.~T., Newman, K.~B., Thomas, L. \& Koester, N. (2004).
\newblock State-space models for dynamics of wild animal populations.
\newblock \emph{Ecological Modelling}, 171, 157--175.

\bibitem[{Bulla \& Bulla(2013)}]{BullaBulla2013}
Bulla, J. \& Bulla, I. (2013).
\newblock \emph{hsmm: Hidden Semi {M}arkov Models}.
\newblock \urlprefix\url{https://CRAN.R-project.org/package=hsmm}.
\newblock R package version 0.4.

\bibitem[{Caetano \emph{et~al.}(2018)Caetano, O'Meara \&
  Beaulieu}]{caetano_hidden_2018}
Caetano, D.~S., O'Meara, B.~C. \& Beaulieu, J.~M. (2018).
\newblock Hidden state models improve state-dependent diversification
  approaches, including biogeographical models.
\newblock \emph{Evolution}, 72, 2308--2324.

\bibitem[{Calabrese \emph{et~al.}(2011)Calabrese, Brunner \&
  Ostfeld}]{calabrese2011partitioning}
Calabrese, J.~M., Brunner, J.~L. \& Ostfeld, R.~S. (2011).
\newblock Partitioning the aggregation of parasites on hosts into intrinsic and
  extrinsic components via an extended poisson-gamma mixture model.
\newblock \emph{PLoS ONE}, 6, e29215.

\bibitem[{Capp{\'e} \emph{et~al.}(2005)Capp{\'e}, Moulines \&
  Ryd{\'e}n}]{CappeEtAl2005}
Capp{\'e}, O., Moulines, E. \& Ryd{\'e}n, T. (2005).
\newblock \emph{Inference in Hidden {M}arkov Models}.
\newblock New York: Springer.

\bibitem[{Cardenas-Ovando \emph{et~al.}(2017)Cardenas-Ovando, Noguez \&
  Rangel-Escareno}]{CardenasOvandoEtAl2017}
Cardenas-Ovando, R.~A., Noguez, J. \& Rangel-Escareno, C. (2017).
\newblock \emph{RcppHMM: Rcpp Hidden {M}arkov Model}.
\newblock \urlprefix\url{https://CRAN.R-project.org/package=RcppHMM}.
\newblock R package version 1.2.2.

\bibitem[{Caswell(2001)}]{Caswell2001}
Caswell, H. (2001).
\newblock \emph{Matrix Population Models, 2nd Edition}.
\newblock Sinauer, Sunderland, MA.

\bibitem[{Cayuela \emph{et~al.}(2020)Cayuela, Besnard, Cote, Laporte, Bonnaire,
  Pichenot, Schtickzelle, Bellec, Joly \& L{\'e}na}]{CayuelaEtAl2020}
Cayuela, H., Besnard, A., Cote, J., Laporte, M., Bonnaire, E., Pichenot, J.,
  Schtickzelle, N., Bellec, A., Joly, P. \& L{\'e}na, J.-P. (2020).
\newblock Anthropogenic disturbance drives dispersal syndromes, demography, and
  gene flow in amphibian populations.
\newblock \emph{Ecological Monographs}, 90, e01406.

\bibitem[{Chambert \emph{et~al.}(2012)Chambert, Staszewski, Lobato, Choquet,
  Carrie, McCoy, Tveraa \& Boulinier}]{ChambertEtAl2012}
Chambert, T., Staszewski, V., Lobato, E., Choquet, R., Carrie, C., McCoy,
  K.~D., Tveraa, T. \& Boulinier, T. (2012).
\newblock Exposure of black-legged kittiwakes to {L}yme disease spirochetes:
  dynamics of the immune status of adult hosts and effects on their survival.
\newblock \emph{Journal of Animal Ecology}, 81, 986--995.

\bibitem[{Charmantier \emph{et~al.}(2006)Charmantier, Perrins, McCleery \&
  Sheldon}]{CharmantierEtAl2006}
Charmantier, A., Perrins, C., McCleery, R.~H. \& Sheldon, B.~C. (2006).
\newblock Evolutionary response to selection on clutch size in a long-term
  study of the mute swan.
\newblock \emph{The American Naturalist}, 167, 453--465.

\bibitem[{Chen \emph{et~al.}(2013)Chen, K{\'e}ry, Plattner, Ma \&
  Gardner}]{ChenEtAl2013}
Chen, G., K{\'e}ry, M., Plattner, M., Ma, K. \& Gardner, B. (2013).
\newblock Imperfect detection is the rule rather than the exception in plant
  distribution studies.
\newblock \emph{Journal of Ecology}, 101, 183--191.

\bibitem[{Choquet \emph{et~al.}(2013)Choquet, Carri\'e, Chambert \&
  Boulinier}]{choquet2013}
Choquet, R., Carri\'e, C., Chambert, T. \& Boulinier, T. (2013).
\newblock Estimating transitions between states using measurements with
  imperfect detection: application to serological data.
\newblock \emph{Ecology}, 94, 2160--2165.

\bibitem[{Choquet \emph{et~al.}(2017)Choquet, Garnier, Awuve \&
  Besnard}]{ChoquetEtAl2017}
Choquet, R., Garnier, A., Awuve, E. \& Besnard, A. (2017).
\newblock Transient state estimation using continuous-time processes applied to
  opportunistic capture--recapture data.
\newblock \emph{Ecological Modelling}, 361, 157--163.

\bibitem[{Choquet \emph{et~al.}(2011)Choquet, Viallefont, Rouan, Gaanoun \&
  Gaillard}]{ChoquetEtAl2011}
Choquet, R., Viallefont, A., Rouan, L., Gaanoun, K. \& Gaillard, J.-M. (2011).
\newblock A semi-{M}arkov model to assess reliably survival patterns from birth
  to death in free-ranging populations.
\newblock \emph{Methods in Ecology and Evolution}, 2, 383--389.

\bibitem[{Clogg(1995)}]{clogg1995latent}
Clogg, C.~C. (1995).
\newblock Latent class models.
\newblock In: \emph{Handbook of Statistical Modeling for the Social and
  Behavioral Sciences} (eds. Arminger, G., Clogg, C.~C. \& Sobel, M.~E.).
  Springer, pp. 311--359.

\bibitem[{Cole(2019)}]{Cole2019}
Cole, D.~J. (2019).
\newblock Parameter redundancy and identifiability in hidden {M}arkov models.
\newblock \emph{METRON}, 77, 105--118.

\bibitem[{Compagnoni \emph{et~al.}(2019)Compagnoni, Bibian, Ochocki, Levin, Zhu
  \& Miller}]{CompagnoniEtAl2019}
Compagnoni, A., Bibian, A.~J., Ochocki, B.~M., Levin, S., Zhu, K. \& Miller,
  T.~E. (2019).
\newblock popler: an {R} package for extraction and synthesis of population
  time series from the long-term ecological research ({LTER}) network.
\newblock \emph{Methods in Ecology and Evolution}, 11, 258--264.

\bibitem[{Conn \& Cooch(2009)}]{ConnCooch2009}
Conn, P.~B. \& Cooch, E.~G. (2009).
\newblock Multistate capture--recapture analysis under imperfect state
  observation: an application to disease models.
\newblock \emph{Journal of Applied Ecology}, 46, 486--492.

\bibitem[{Conway-Cranos \& Doak(2011)}]{ConwayDoak2011}
Conway-Cranos, L.~L. \& Doak, D.~F. (2011).
\newblock Sampling errors create bias in {M}arkov models for community
  dynamics: the problem and a method for its solution.
\newblock \emph{Oecologia}, 167, 199--207.

\bibitem[{Cooch \emph{et~al.}(2012)Cooch, Conn, Ellner, Dobson \&
  Pollock}]{cooch2012disease}
Cooch, E.~G., Conn, P.~B., Ellner, S.~P., Dobson, A.~P. \& Pollock, K.~H.
  (2012).
\newblock Disease dynamics in wild populations: modeling and estimation: a
  review.
\newblock \emph{Journal of Ornithology}, 152, 485--509.

\bibitem[{Cooke \emph{et~al.}(2004)Cooke, Hinch, Wikelski, Andrews, Kuchel,
  Wolcott \& Butler}]{CookeEtAl2004}
Cooke, S.~J., Hinch, S.~G., Wikelski, M., Andrews, R.~D., Kuchel, L.~J.,
  Wolcott, T.~G. \& Butler, P.~J. (2004).
\newblock Biotelemetry: a mechanistic approach to ecology.
\newblock \emph{Trends in Ecology \& Evolution}, 19, 334--343.

\bibitem[{Coulson \emph{et~al.}(2006)Coulson, Benton, Lundberg, Dall, Kendall
  \& Gaillard}]{coulson_estimating_2006}
Coulson, T., Benton, T.~G., Lundberg, P., Dall, S. R.~X., Kendall, B.~E. \&
  Gaillard, J.~M. (2006).
\newblock Estimating individual contributions to population growth:
  evolutionary fitness in ecological time.
\newblock \emph{Proceedings of the Royal Society B Biological Sciences}, 273,
  547--555.

\bibitem[{Cowen \emph{et~al.}(2017)Cowen, Besbeas, Morgan \&
  Schwarz}]{CowenEtAl2017}
Cowen, L. L.~E., Besbeas, P., Morgan, B. J.~T. \& Schwarz, C.~J. (2017).
\newblock Hidden {M}arkov models for extended batch data.
\newblock \emph{Biometrics}, 73, 1321--1331.

\bibitem[{Crampton \emph{et~al.}(2018)Crampton, Meyers, Cooper, Sadler, Foote
  \& Harte}]{crampton_pacing_2018}
Crampton, J.~S., Meyers, S.~R., Cooper, R.~A., Sadler, P.~M., Foote, M. \&
  Harte, D. (2018).
\newblock Pacing of {Paleozoic} macroevolutionary rates by {Milankovitch} grand
  cycles.
\newblock \emph{Proceedings of the National Academy of Sciences}, 115,
  5686--5691.

\bibitem[{Davis \emph{et~al.}(2018)Davis, Rich, Farris, Kelly, Di~Bitetti,
  Blanco, Albanesi, Farhadinia, Gholikhani, Hamel
  \emph{et~al.}}]{DavisEtAl2018}
Davis, C.~L., Rich, L.~N., Farris, Z.~J., Kelly, M.~J., Di~Bitetti, M.~S.,
  Blanco, Y.~D., Albanesi, S., Farhadinia, M.~S., Gholikhani, N., Hamel, S.
  \emph{et~al.} (2018).
\newblock Ecological correlates of the spatial co-occurrence of sympatric
  mammalian carnivores worldwide.
\newblock \emph{Ecology Letters}, 21, 1401--1412.

\bibitem[{DeRuiter \emph{et~al.}(2017)DeRuiter, Langrock, Skirbutas, Goldbogen,
  Calambokidis, Friedlaender \& Southall}]{DeRuiterEtAl2017}
DeRuiter, S.~L., Langrock, R., Skirbutas, T., Goldbogen, J.~A., Calambokidis,
  J., Friedlaender, A.~S. \& Southall, B.~L. (2017).
\newblock A multivariate mixed hidden {M}arkov model to analyze blue whale
  diving behaviour during controlled sound exposures.
\newblock \emph{The Annals of Applied Statistics}, 11, 362--392.

\bibitem[{Dietze \emph{et~al.}(2018)Dietze, Fox, Beck-Johnson, Betancourt,
  Hooten, Jarnevich, Keitt, Kenney, Laney, Larsen
  \emph{et~al.}}]{DietzeEtAl2018}
Dietze, M.~C., Fox, A., Beck-Johnson, L.~M., Betancourt, J.~L., Hooten, M.~B.,
  Jarnevich, C.~S., Keitt, T.~H., Kenney, M.~A., Laney, C.~M., Larsen, L.~G.
  \emph{et~al.} (2018).
\newblock Iterative near-term ecological forecasting: Needs, opportunities, and
  challenges.
\newblock \emph{Proceedings of the National Academy of Sciences}, 115,
  1424--1432.

\bibitem[{Diosdado \emph{et~al.}(2015)Diosdado, Barker, Hodges, Amory, Croft,
  Bell \& Codling}]{diosdado2015classification}
Diosdado, J. A.~V., Barker, Z.~E., Hodges, H.~R., Amory, J.~R., Croft, D.~P.,
  Bell, N.~J. \& Codling, E.~A. (2015).
\newblock Classification of behaviour in housed dairy cows using an
  accelerometer-based activity monitoring system.
\newblock \emph{Animal Biotelemetry}, 3, 15.

\bibitem[{Dorazio \emph{et~al.}(2010)Dorazio, Kery, Royle \&
  Plattner}]{DorazioEtAl2010}
Dorazio, R.~M., Kery, M., Royle, J.~A. \& Plattner, M. (2010).
\newblock Models for inference in dynamic metacommunity systems.
\newblock \emph{Ecology}, 91, 2466--2475.

\bibitem[{Dorazio \& Royle(2005)}]{dorazio_estimating_2005}
Dorazio, R.~M. \& Royle, J.~A. (2005).
\newblock Estimating size and composition of biological communities by modeling
  the occurrence of species.
\newblock \emph{Journal of the American Statistical Association}, 100,
  389--398.

\bibitem[{Dorazio \emph{et~al.}(2006)Dorazio, Royle, S{\"o}derstr{\"o}m \&
  Glimsk{\"a}r}]{dorazio_estimating_2006}
Dorazio, R.~M., Royle, J.~A., S{\"o}derstr{\"o}m, B. \& Glimsk{\"a}r, A.
  (2006).
\newblock Estimating species richness and accumulation by modeling species
  occurrence and detectability.
\newblock \emph{Ecology}, 87, 842--854.

\bibitem[{Dupont \emph{et~al.}(2015)Dupont, Pradel, Lardy, Allain{\'e} \&
  Cohas}]{DupontEtAl2015}
Dupont, P., Pradel, R., Lardy, S., Allain{\'e}, D. \& Cohas, A. (2015).
\newblock Litter sex composition influences dominance status of {A}lpine
  marmots ({M}armota marmota).
\newblock \emph{Oecologia}, 179, 753--763.

\bibitem[{Durand \emph{et~al.}(2005)Durand, Gu{\'e}don, Caraglio \&
  Costes}]{DurandEtAl2005}
Durand, J.-B., Gu{\'e}don, Y., Caraglio, Y. \& Costes, E. (2005).
\newblock Analysis of the plant architecture via tree-structured statistical
  models: the hidden {M}arkov tree models.
\newblock \emph{New Phytologist}, 166, 813--825.

\bibitem[{Durbin \& Koopman(2012)}]{durbin2012time}
Durbin, J. \& Koopman, S.~J. (2012).
\newblock \emph{Time Series Analysis by State Space Methods}.
\newblock 2nd edn. Oxford University Press.

\bibitem[{Durbin \emph{et~al.}(1998)Durbin, Eddy, Krogh \&
  Mitchison}]{DurbinEtAl1998}
Durbin, R., Eddy, S.~R., Krogh, A. \& Mitchison, G. (1998).
\newblock \emph{Biological Sequence Analysis: Probabilistic Models of Proteins
  and Nucleic Acids}.
\newblock Cambridge University Press.

\bibitem[{Economou \& Menary(2019)}]{EconomouMenary2019}
Economou, T. \& Menary, M.~B. (2019).
\newblock A hidden semi-{M}arkov model for characterizing regime shifts in
  ocean density variability.
\newblock \emph{Journal of the Royal Statistical Society: Series C (Applied
  Statistics)}, 68, 1529--1553.

\bibitem[{Eddy(2004)}]{eddy2004hidden}
Eddy, S.~R. (2004).
\newblock What is a hidden {M}arkov model?
\newblock \emph{Nature Biotechnology}, 22, 1315.

\bibitem[{Ellison(2004)}]{ellison2004bayesian}
Ellison, A.~M. (2004).
\newblock Bayesian inference in ecology.
\newblock \emph{Ecology Letters}, 7, 509--520.

\bibitem[{Ephraim \& Merhav(2002)}]{EphraimMerhav2002}
Ephraim, Y. \& Merhav, N. (2002).
\newblock Hidden {M}arkov processes.
\newblock \emph{IEEE Transactions on Information Theory}, 48, 1518--1569.

\bibitem[{Estes \emph{et~al.}(2018)Estes, Elsen, Treuer, Ahmed, Caylor, Chang,
  Choi \& Ellis}]{EstesEtAl2018}
Estes, L., Elsen, P.~R., Treuer, T., Ahmed, L., Caylor, K., Chang, J., Choi,
  J.~J. \& Ellis, E.~C. (2018).
\newblock The spatial and temporal domains of modern ecology.
\newblock \emph{Nature Ecology \& Evolution}, 2, 819.

\bibitem[{Evans \emph{et~al.}(2016)Evans, Merow, Record, McMahon \&
  Enquist}]{EvansEtAl2016}
Evans, M.~E., Merow, C., Record, S., McMahon, S.~M. \& Enquist, B.~J. (2016).
\newblock Towards process-based range modeling of many species.
\newblock \emph{Trends in Ecology \& Evolution}, 31, 860--871.

\bibitem[{Fahlbusch \& Harrington(2019)}]{fahlbusch2019low}
Fahlbusch, J.~A. \& Harrington, K.~J. (2019).
\newblock A low-cost, open-source inertial movement {GPS} logger for
  eco-physiology applications.
\newblock \emph{Journal of Experimental Biology}, 222, jeb211136.

\bibitem[{Fasiolo \emph{et~al.}(2016)Fasiolo, Pya \& Wood}]{FasioloEtAl2016}
Fasiolo, M., Pya, N. \& Wood, S.~N. (2016).
\newblock A comparison of inferential methods for highly nonlinear state space
  models in ecology and epidemiology.
\newblock \emph{Statistical Science}, 31, 96--118.

\bibitem[{Felsenstein \& Churchill(1996)}]{FelsensteinChurchill1996}
Felsenstein, J. \& Churchill, G.~A. (1996).
\newblock A hidden {M}arkov model approach to variation among sites in rate of
  evolution.
\newblock \emph{Molecular Biology and Evolution}, 13, 93--104.

\bibitem[{Fidino \emph{et~al.}(2019)Fidino, Simonis \&
  Magle}]{fidino_multistate_2019}
Fidino, M., Simonis, J.~L. \& Magle, S.~B. (2019).
\newblock A multistate dynamic occupancy model to estimate local
  colonization--extinction rates and patterns of co-occurrence between two or
  more interacting species.
\newblock \emph{Methods in Ecology and Evolution}, 10, 233--244.

\bibitem[{Fine \emph{et~al.}(1998)Fine, Singer \& Tishby}]{FineEtAl1998}
Fine, S., Singer, Y. \& Tishby, N. (1998).
\newblock The hierarchical hidden {M}arkov model: Analysis and applications.
\newblock \emph{Machine Learning}, 32, 41--62.

\bibitem[{Folke \emph{et~al.}(2004)Folke, Carpenter, Walker, Scheffer,
  Elmqvist, Gunderson \& Holling}]{FolkeEtAl2004}
Folke, C., Carpenter, S., Walker, B., Scheffer, M., Elmqvist, T., Gunderson, L.
  \& Holling, C.~S. (2004).
\newblock Regime shifts, resilience, and biodiversity in ecosystem management.
\newblock \emph{Annu. Rev. Ecol. Evol. Syst.}, 35, 557--581.

\bibitem[{Franke \emph{et~al.}(2004)Franke, Caelli \& Hudson}]{FrankeEtAl2004}
Franke, A., Caelli, T. \& Hudson, R.~J. (2004).
\newblock Analysis of movements and behavior of caribou ({R}angifer tarandus)
  using hidden {M}arkov models.
\newblock \emph{Ecological Modelling}, 173, 259--270.

\bibitem[{Franke \emph{et~al.}(2006)Franke, Caelli, Kuzyk \&
  Hudson}]{FrankeEtAl2006}
Franke, A., Caelli, T., Kuzyk, G. \& Hudson, R.~J. (2006).
\newblock Prediction of wolf ({C}anis lupus) kill-sites using hidden {M}arkov
  models.
\newblock \emph{Ecological Modelling}, 197, 237--246.

\bibitem[{Fr{\"u}hwirth-Schnatter(2006)}]{fruhwirth2006finite}
Fr{\"u}hwirth-Schnatter, S. (2006).
\newblock \emph{Finite Mixture and {M}arkov Switching Models}.
\newblock Springer Science \& Business Media.

\bibitem[{Fukaya \& Royle(2013)}]{FukayaRoyle2013}
Fukaya, K. \& Royle, J.~A. (2013).
\newblock Markov models for community dynamics allowing for observation error.
\newblock \emph{Ecology}, 94, 2670--2677.

\bibitem[{Gal \& Anderson(2010)}]{GalAnderson2010}
Gal, G. \& Anderson, W. (2010).
\newblock A novel approach to detecting a regime shift in a lake ecosystem.
\newblock \emph{Methods in Ecology and Evolution}, 1, 45--52.

\bibitem[{Gao(2002)}]{Gao2002}
Gao, J. (2002).
\newblock Integration of {GPS} with remote sensing and {GIS}: reality and
  prospect.
\newblock \emph{Photogrammetric Engineering and Remote Sensing}, 68, 447--454.

\bibitem[{Garnier \emph{et~al.}(2016)Garnier, Gaillard, Gauthier \&
  Besnard}]{GarnierEtAl2016}
Garnier, A., Gaillard, J.-M., Gauthier, D. \& Besnard, A. (2016).
\newblock What shapes fitness costs of reproduction in long-lived iteroparous
  species? a case study on the {A}lpine ibex.
\newblock \emph{Ecology}, 97, 205--214.

\bibitem[{Gelman \emph{et~al.}(2004)Gelman, Carlin, Stern \&
  Rubin}]{GelmanEtAl2004}
Gelman, A., Carlin, J.~B., Stern, H.~S. \& Rubin, D.~B. (2004).
\newblock \emph{Bayesian Data Analysis, 2nd Edition}.
\newblock Chapman and Hall, Boca Raton.

\bibitem[{Gelman \& Hill(2006)}]{gelman2006data}
Gelman, A. \& Hill, J. (2006).
\newblock \emph{Data Analysis Using Regression and Multilevel/Hierarchical
  Models}.
\newblock Cambridge University Press.

\bibitem[{Gennaretti \emph{et~al.}(2014)Gennaretti, Arseneault, Nicault,
  Perreault \& B{\'e}gin}]{GennarettiEtAl2014}
Gennaretti, F., Arseneault, D., Nicault, A., Perreault, L. \& B{\'e}gin, Y.
  (2014).
\newblock Volcano-induced regime shifts in millennial tree-ring chronologies
  from northeastern {N}orth {A}merica.
\newblock \emph{Proceedings of the National Academy of Sciences}, 111,
  10077--10082.

\bibitem[{Ghosh \emph{et~al.}(2012)Ghosh, Gajjalla, Mohammed \&
  Mande}]{GhoshEtAl2012}
Ghosh, T.~S., Gajjalla, P., Mohammed, M.~H. \& Mande, S.~S. (2012).
\newblock {C16S} --- a hidden {M}arkov model based algorithm for taxonomic
  classification of {16S} {rRNA} gene sequences.
\newblock \emph{Genomics}, 99, 195--201.

\bibitem[{Gimenez \emph{et~al.}(2014)Gimenez, Blanc, Besnard, Pradel,
  Doherty~Jr, Marboutin \& Choquet}]{GimenezEtAl2014}
Gimenez, O., Blanc, L., Besnard, A., Pradel, R., Doherty~Jr, P.~F., Marboutin,
  E. \& Choquet, R. (2014).
\newblock Fitting occupancy models with {E-SURGE}: hidden {M}arkov modelling of
  presence--absence data.
\newblock \emph{Methods in Ecology and Evolution}, 5, 592--597.

\bibitem[{Gimenez \& Gaillard(2018)}]{gimenez_estimating_2018}
Gimenez, O. \& Gaillard, J.-M. (2018).
\newblock Estimating individual fitness in the wild using capture-recapture
  data.
\newblock \emph{Population Ecology}, 60, 101--109.

\bibitem[{Gimenez \emph{et~al.}(2012)Gimenez, Lebreton, Gaillard, Choquet \&
  Pradel}]{GimenezEtAl2012}
Gimenez, O., Lebreton, J.-D., Gaillard, J.-M., Choquet, R. \& Pradel, R.
  (2012).
\newblock Estimating demographic parameters using hidden process dynamic
  models.
\newblock \emph{Theoretical Population Biology}, 82, 307--316.

\bibitem[{Glennie \emph{et~al.}(2019)Glennie, Borchers, Murchie, Harmsen \&
  Foster}]{glennie2019open}
Glennie, R., Borchers, D.~L., Murchie, M., Harmsen, B.~J. \& Foster, R.~J.
  (2019).
\newblock Open population maximum likelihood spatial capture-recapture.
\newblock \emph{Biometrics}, 75, 1345--1355.

\bibitem[{Gotelli \& Ellison(2013)}]{GotelliEllison2013}
Gotelli, N.~J. \& Ellison, A.~M. (2013).
\newblock \emph{A Primer of Ecological Statistics}.
\newblock 2nd edn. Sinauer Associates Publishers.

\bibitem[{Grecian \emph{et~al.}(2018)Grecian, Lane, Michelot, Wade \&
  Hamer}]{grecian2018understanding}
Grecian, W.~J., Lane, J.~V., Michelot, T., Wade, H.~M. \& Hamer, K.~C. (2018).
\newblock Understanding the ontogeny of foraging behaviour: insights from
  combining marine predator bio-logging with satellite-derived oceanography in
  hidden {M}arkov models.
\newblock \emph{Journal of the Royal Society Interface}, 15, 20180084.

\bibitem[{Grewal \emph{et~al.}(2019{\natexlab{a}})Grewal, Krzywinski \&
  Altman}]{GrewalEtal2019hidden}
Grewal, J.~K., Krzywinski, M. \& Altman, N. (2019{\natexlab{a}}).
\newblock {M}arkov models ---hidden {M}arkov models.
\newblock \emph{Nature Methods}, 16, 795--796.

\bibitem[{Grewal \emph{et~al.}(2019{\natexlab{b}})Grewal, Krzywinski \&
  Altman}]{GrewalEtal2019}
Grewal, J.~K., Krzywinski, M. \& Altman, N.~S. (2019{\natexlab{b}}).
\newblock {M}arkov models ---{M}arkov chains.
\newblock \emph{Nature Methods}, 16, 663--664.

\bibitem[{Guillera-Arroita(2017)}]{guilleraarroita_modelling_2017}
Guillera-Arroita, G. (2017).
\newblock Modelling of species distributions, range dynamics and communities
  under imperfect detection: advances, challenges and opportunities.
\newblock \emph{Ecography}, 40, 281--295.

\bibitem[{Halbritter \emph{et~al.}(2020)Halbritter, De~Boeck, Eycott, Reinsch,
  Robinson, Vicca, Berauer, Christiansen, Estiarte, Gr{\"u}nzweig
  \emph{et~al.}}]{HalbritterEtAl2019}
Halbritter, A.~H., De~Boeck, H.~J., Eycott, A.~E., Reinsch, S., Robinson,
  D.~A., Vicca, S., Berauer, B., Christiansen, C.~T., Estiarte, M.,
  Gr{\"u}nzweig, J.~M. \emph{et~al.} (2020).
\newblock The handbook for standardised field and laboratory measurements in
  terrestrial climate-change experiments and observational studies
  ({C}lim{E}x).
\newblock \emph{Methods in Ecology and Evolution}, 11, 22--37.

\bibitem[{Haller(2014)}]{Haller2014}
Haller, B.~C. (2014).
\newblock Theoretical and empirical perspectives in ecology and evolution: a
  survey.
\newblock \emph{Bioscience}, 64, 907--916.

\bibitem[{Hanski(1994)}]{Hanski1994}
Hanski, I. (1994).
\newblock A practical model of metapopulation dynamics.
\newblock \emph{Journal of Animal Ecology}, 63, 151--162.

\bibitem[{Hart \emph{et~al.}(2010)Hart, Mann, Coulson, Pettorelli \&
  Trathan}]{hart2010behavioural}
Hart, T., Mann, R., Coulson, T., Pettorelli, N. \& Trathan, P. (2010).
\newblock Behavioural switching in a central place forager: patterns of diving
  behaviour in the macaroni penguin ({E}udyptes chrysolophus).
\newblock \emph{Marine Biology}, 157, 1543--1553.

\bibitem[{Harte(2017)}]{Harte2017}
Harte, D. (2017).
\newblock \emph{Hidden{M}arkov: Hidden {M}arkov Models}.
\newblock Statistics Research Associates, Wellington.
\newblock \urlprefix\url{http://www.statsresearch.co.nz/dsh/sslib/}.
\newblock R package version 1.8-11.

\bibitem[{Hebert \emph{et~al.}(2016)Hebert, Ratnasingham, Zakharov, Telfer,
  Levesque-Beaudin, Milton, Pedersen, Jannetta \& deWaard}]{HebertEtAl2016}
Hebert, P.~D., Ratnasingham, S., Zakharov, E.~V., Telfer, A.~C.,
  Levesque-Beaudin, V., Milton, M.~A., Pedersen, S., Jannetta, P. \& deWaard,
  J.~R. (2016).
\newblock Counting animal species with {DNA} barcodes: {C}anadian insects.
\newblock \emph{Philosophical Transactions of the Royal Society B: Biological
  Sciences}, 371, 20150333.

\bibitem[{Helske \& Helske(2019)}]{HelskeHelske2019}
Helske, S. \& Helske, J. (2019).
\newblock Mixture hidden {M}arkov models for sequence data: The seq{HMM}
  package in {R}.
\newblock \emph{Journal of Statistical Software}, 88, 1--32.

\bibitem[{Henderson \emph{et~al.}(1997)Henderson, Salzberg \&
  Fasman}]{HendersonEtAl1997}
Henderson, J., Salzberg, S. \& Fasman, K.~H. (1997).
\newblock Finding genes in {DNA} with a hidden {M}arkov model.
\newblock \emph{Journal of Computational Biology}, 4, 127--141.

\bibitem[{Hestbeck \emph{et~al.}(1991)Hestbeck, Nichols \&
  Malecki}]{hestbeck1991estimates}
Hestbeck, J.~B., Nichols, J.~D. \& Malecki, R.~A. (1991).
\newblock Estimates of movement and site fidelity using mark-resight data of
  wintering canada geese.
\newblock \emph{Ecology}, 72, 523--533.

\bibitem[{Hildebrand(1987)}]{Hildebrand1987}
Hildebrand, F.~B. (1987).
\newblock \emph{Introduction to Numerical Analysis}.
\newblock Courier Corporation.

\bibitem[{Hill \emph{et~al.}(2004)Hill, Witman \& Caswell}]{HillEtAl2004}
Hill, M.~F., Witman, J.~D. \& Caswell, H. (2004).
\newblock {M}arkov chain analysis of succession in a rocky subtidal community.
\newblock \emph{The American Naturalist}, 164, E46--E61.

\bibitem[{Himmelmann(2010)}]{Himmelmann2010}
Himmelmann, L. (2010).
\newblock \emph{HMM: Hidden {M}arkov Models}.
\newblock \urlprefix\url{https://CRAN.R-project.org/package=HMM}.
\newblock R package version 1.0.

\bibitem[{Hobolth \emph{et~al.}(2007)Hobolth, Christensen, Mailund \&
  Schierup}]{hobolth_genomic_2007}
Hobolth, A., Christensen, O.~F., Mailund, T. \& Schierup, M.~H. (2007).
\newblock Genomic relationships and speciation times of human, chimpanzee, and
  gorilla inferred from a coalescent hidden {Markov} model.
\newblock \emph{PLoS Genetics}, 3, e7.

\bibitem[{Hooten \emph{et~al.}(2017)Hooten, Johnson, McClintock \&
  Morales}]{HootenEtAl2017}
Hooten, M.~B., Johnson, D.~S., McClintock, B.~T. \& Morales, J.~M. (2017).
\newblock \emph{Animal Movement: Statistical Models for Telemetry Data}.
\newblock CRC Press.

\bibitem[{Hope \& Jones(2012)}]{hope2012warming}
Hope, P.~R. \& Jones, G. (2012).
\newblock Warming up for dinner: torpor and arousal in hibernating {N}atterer's
  bats ({M}yotis nattereri) studied by radio telemetry.
\newblock \emph{Journal of Comparative Physiology B}, 182, 569--578.

\bibitem[{Horn(1975)}]{Horn1975}
Horn, H.~S. (1975).
\newblock {M}arkovian properties of forest succession.
\newblock In: \emph{Ecology and Evolution of Communities} (eds. Cody, M. \&
  Diamond, J.). Harvard University Press, pp. 196--211.

\bibitem[{Hudson(2008)}]{Hudson2008}
Hudson, M.~E. (2008).
\newblock Sequencing breakthroughs for genomic ecology and evolutionary
  biology.
\newblock \emph{Molecular Ecology Resources}, 8, 3--17.

\bibitem[{Jackson(1975)}]{Jackson1975}
Jackson, B.~B. (1975).
\newblock {M}arkov mixture models for drought lengths.
\newblock \emph{Water Resources Research}, 11, 64--74.

\bibitem[{Jackson(2011)}]{Jackson2011}
Jackson, C.~H. (2011).
\newblock Multi-state models for panel data: the msm package for {R}.
\newblock \emph{Journal of statistical software}, 38, 1--29.

\bibitem[{Jackson \emph{et~al.}(2003)Jackson, Sharples, Thompson, Duffy \&
  Couto}]{jackson2003multistate}
Jackson, C.~H., Sharples, L.~D., Thompson, S.~G., Duffy, S.~W. \& Couto, E.
  (2003).
\newblock Multistate {M}arkov models for disease progression with
  classification error.
\newblock \emph{Journal of the Royal Statistical Society: Series D (The
  Statistician)}, 52, 193--209.

\bibitem[{Jasra \emph{et~al.}(2005)Jasra, Holmes \& Stephens}]{JasraEtAl2005}
Jasra, A., Holmes, C.~C. \& Stephens, D.~A. (2005).
\newblock Markov chain {M}onte {C}arlo methods and the label switching problem
  in {B}ayesian mixture modeling.
\newblock \emph{Statistical Science}, 20, 50--67.

\bibitem[{Jones \emph{et~al.}(2006)Jones, Schildhauer, Reichman \&
  Bowers}]{JonesEtAl2006}
Jones, M.~B., Schildhauer, M.~P., Reichman, O. \& Bowers, S. (2006).
\newblock The new bioinformatics: integrating ecological data from the gene to
  the biosphere.
\newblock \emph{Annual Review of Ecology, Evolution, and Systematics}, 37,
  519--544.

\bibitem[{Jonsen \emph{et~al.}(2005)Jonsen, Flemming \& Myers}]{JonsenEtAl2005}
Jonsen, I.~D., Flemming, J.~M. \& Myers, R.~A. (2005).
\newblock Robust state--space modeling of animal movement data.
\newblock \emph{Ecology}, 86, 2874--2880.

\bibitem[{Kellner \& Swihart(2014)}]{KellnerSwihart2014}
Kellner, K.~F. \& Swihart, R.~K. (2014).
\newblock Accounting for imperfect detection in ecology: a quantitative review.
\newblock \emph{PLoS ONE}, 9, e111436.

\bibitem[{Kendall \emph{et~al.}(2012)Kendall, White, Hines, Langtimm \&
  Yoshizaki}]{KendallEtAl2012}
Kendall, W.~L., White, G.~C., Hines, J.~E., Langtimm, C.~A. \& Yoshizaki, J.
  (2012).
\newblock Estimating parameters of hidden {M}arkov models based on marked
  individuals: use of robust design data.
\newblock \emph{Ecology}, 93, 913--920.

\bibitem[{Kery \& Royle(2015)}]{kery_applied_2015}
Kery, M. \& Royle, J.~A. (2015).
\newblock \emph{Applied {Hierarchical} {Modeling} in {Ecology}: {Analysis} of
  distribution, abundance and species richness in {R} and {BUGS}}.
\newblock Academic Press Inc, Amsterdam.

\bibitem[{K{\'e}ry \& Schmidt(2008)}]{KerySchmidt2008}
K{\'e}ry, M. \& Schmidt, B. (2008).
\newblock Imperfect detection and its consequences for monitoring for
  conservation.
\newblock \emph{Community Ecology}, 9, 207--216.

\bibitem[{King \& Langrock(2016)}]{king2016semi}
King, R. \& Langrock, R. (2016).
\newblock Semi-{M}arkov {A}rnason--{S}chwarz models.
\newblock \emph{Biometrics}, 72, 619--628.

\bibitem[{Koleff \emph{et~al.}(2003)Koleff, Gaston \& Lennon}]{KoleffEtAl2003}
Koleff, P., Gaston, K.~J. \& Lennon, J.~J. (2003).
\newblock Measuring beta diversity for presence--absence data.
\newblock \emph{Journal of Animal Ecology}, 72, 367--382.

\bibitem[{Lachish \emph{et~al.}(2011)Lachish, Knowles, Alves, Wood \&
  Sheldon}]{LachishEtAl2011}
Lachish, S., Knowles, S.~C., Alves, R., Wood, M.~J. \& Sheldon, B.~C. (2011).
\newblock Infection dynamics of endemic malaria in a wild bird population:
  parasite species-dependent drivers of spatial and temporal variation in
  transmission rates.
\newblock \emph{Journal of Animal Ecology}, 80, 1207--1216.

\bibitem[{Lagrange \emph{et~al.}(2014)Lagrange, Pradel, B{\'e}lisle \&
  Gimenez}]{LagrangeEtAl2014}
Lagrange, P., Pradel, R., B{\'e}lisle, M. \& Gimenez, O. (2014).
\newblock Estimating dispersal among numerous sites using capture--recapture
  data.
\newblock \emph{Ecology}, 95, 2316--2323.

\bibitem[{Lamy \emph{et~al.}(2013)Lamy, Gimenez, Pointier, Jarne \&
  David}]{LamyEtAl2013}
Lamy, T., Gimenez, O., Pointier, J.-P., Jarne, P. \& David, P. (2013).
\newblock Metapopulation dynamics of species with cryptic life stages.
\newblock \emph{The American Naturalist}, 181, 479--491.

\bibitem[{Langrock \emph{et~al.}(2018)Langrock, Adam, Leos-Barajas, Mews,
  Miller \& Papastamatiou}]{langrock2018spline}
Langrock, R., Adam, T., Leos-Barajas, V., Mews, S., Miller, D.~L. \&
  Papastamatiou, Y.~P. (2018).
\newblock Spline-based nonparametric inference in general state-switching
  models.
\newblock \emph{Statistica Neerlandica}, 72, 179--200.

\bibitem[{Langrock \emph{et~al.}(2013)Langrock, Borchers \&
  Skaug}]{langrock2013markov}
Langrock, R., Borchers, D.~L. \& Skaug, H.~J. (2013).
\newblock Markov-modulated nonhomogeneous {P}oisson processes for modeling
  detections in surveys of marine mammal abundance.
\newblock \emph{Journal of the American Statistical Association}, 108,
  840--851.

\bibitem[{Langrock \emph{et~al.}(2014{\natexlab{a}})Langrock, Hopcraft,
  Blackwell, Goodall, King, Niu, Patterson, Pedersen, Skarin \&
  Schick}]{LangrockEtAl2014}
Langrock, R., Hopcraft, G., Blackwell, P., Goodall, V., King, R., Niu, M.,
  Patterson, T., Pedersen, M., Skarin, A. \& Schick, R. (2014{\natexlab{a}}).
\newblock Modelling group dynamic animal movement.
\newblock \emph{Methods in Ecology and Evolution}, 5, 190--199.

\bibitem[{Langrock \emph{et~al.}(2014{\natexlab{b}})Langrock, Marques, Baird \&
  Thomas}]{langrock2014modeling}
Langrock, R., Marques, T.~A., Baird, R.~W. \& Thomas, L. (2014{\natexlab{b}}).
\newblock Modeling the diving behavior of whales: a latent-variable approach
  with feedback and semi-{M}arkovian components.
\newblock \emph{Journal of Agricultural, Biological, and Environmental
  Statistics}, 19, 82--100.

\bibitem[{Lawler \emph{et~al.}(2019)Lawler, Whoriskey, Aeberhard, Field \&
  Flemming}]{lawler2019conditionally}
Lawler, E., Whoriskey, K., Aeberhard, W.~H., Field, C. \& Flemming, J.~M.
  (2019).
\newblock The conditionally autoregressive hidden {M}arkov model (carhmm):
  Inferring behavioural states from animal tracking data exhibiting conditional
  autocorrelation.
\newblock \emph{Journal of Agricultural, Biological and Environmental
  Statistics}, 1--18.

\bibitem[{Lazrak \emph{et~al.}(2010)Lazrak, Mari, Beno{\^\i}t
  \emph{et~al.}}]{LazrakEtAl2010}
Lazrak, E., Mari, J., Beno{\^\i}t, M. \emph{et~al.} (2010).
\newblock Landscape regularity modelling for environmental challenges in
  agriculture.
\newblock \emph{Landscape Ecology}, 25, 169--183.

\bibitem[{Lebreton \emph{et~al.}(2009)Lebreton, Nichols, Barker, Pradel \&
  Spendelow}]{LebretonEtAl2009}
Lebreton, J.-D., Nichols, J.~D., Barker, R.~J., Pradel, R. \& Spendelow, J.~A.
  (2009).
\newblock Modeling individual animal histories with multistate
  capture--recapture models.
\newblock \emph{Advances in Ecological Research}, 41, 87--173.

\bibitem[{Lee \& Song(2012)}]{lee2012basic}
Lee, S.-Y. \& Song, X.-Y. (2012).
\newblock \emph{Basic and advanced Bayesian structural equation modeling: With
  applications in the medical and behavioral sciences}.
\newblock John Wiley \& Sons.

\bibitem[{Leos-Barajas \emph{et~al.}(2017{\natexlab{a}})Leos-Barajas, Gangloff,
  Adam, Langrock, Van~Beest, Nabe-Nielsen \& Morales}]{Leos-BarajasEtAl2017}
Leos-Barajas, V., Gangloff, E.~J., Adam, T., Langrock, R., Van~Beest, F.~M.,
  Nabe-Nielsen, J. \& Morales, J.~M. (2017{\natexlab{a}}).
\newblock Multi-scale modeling of animal movement and general behavior data
  using hidden {M}arkov models with hierarchical structures.
\newblock \emph{Journal of Agricultural, Biological and Environmental
  Statistics}, 22, 232--248.

\bibitem[{Leos-Barajas \emph{et~al.}(2017{\natexlab{b}})Leos-Barajas,
  Photopoulou, Langrock, Patterson, Watanabe, Murgatroyd \&
  Papastamatiou}]{Leos-BarajasEtAl2017accelerometer}
Leos-Barajas, V., Photopoulou, T., Langrock, R., Patterson, T.~A., Watanabe,
  Y.~Y., Murgatroyd, M. \& Papastamatiou, Y.~P. (2017{\natexlab{b}}).
\newblock Analysis of animal accelerometer data using hidden {M}arkov models.
\newblock \emph{Methods in Ecology and Evolution}, 8, 161--173.

\bibitem[{Li(1995)}]{Li1995}
Li, B.-L. (1995).
\newblock Stability analysis of a nonhomogeneous {M}arkovian landscape model.
\newblock \emph{Ecological Modelling}, 82, 247--256.

\bibitem[{Li(2003)}]{li2003diagnostic}
Li, W.~K. (2003).
\newblock \emph{Diagnostic Checks in Time Series}.
\newblock CRC Press.

\bibitem[{Lindenmayer \emph{et~al.}(2012)Lindenmayer, Likens, Andersen, Bowman,
  Bull, Burns, Dickman, Hoffmann, Keith, Liddell
  \emph{et~al.}}]{LindenmayerEtAl2012}
Lindenmayer, D.~B., Likens, G.~E., Andersen, A., Bowman, D., Bull, C.~M.,
  Burns, E., Dickman, C.~R., Hoffmann, A.~A., Keith, D.~A., Liddell, M.~J.
  \emph{et~al.} (2012).
\newblock Value of long-term ecological studies.
\newblock \emph{Austral Ecology}, 37, 745--757.

\bibitem[{Lindsay(1995)}]{lindsay1995mixture}
Lindsay, B.~G. (1995).
\newblock Mixture models: Theory, geometry and applications.
\newblock \emph{NSF-CBMS Regional Conference Series in Probability and
  Statistics}, 5, i--163.

\bibitem[{Link \emph{et~al.}(2002)Link, Cooch \& Cam}]{link_model-based_2002}
Link, W.~A., Cooch, E. \& Cam, E. (2002).
\newblock Model-based estimation of individual fitness.
\newblock \emph{Journal of Applied Statistics}, 29, 207--224.

\bibitem[{Lloyd \emph{et~al.}(2020)Lloyd, Oosthuizen, Bester \&
  de~Bruyn}]{LloydEtAl2020}
Lloyd, K.~J., Oosthuizen, W.~C., Bester, M.~N. \& de~Bruyn, P.~N. (2020).
\newblock Trade-offs between age-related breeding improvement and survival
  senescence in highly polygynous elephant seals: Dominant males always do
  better.
\newblock \emph{Journal of Animal Ecology}, 89, 897--909.

\bibitem[{Louvrier \emph{et~al.}(2018)Louvrier, Chambert, Marboutin \&
  Gimenez}]{louvrier_accounting_2018}
Louvrier, J., Chambert, T., Marboutin, E. \& Gimenez, O. (2018).
\newblock Accounting for misidentification and heterogeneity in occupancy
  studies using hidden {Markov} models.
\newblock \emph{Ecological Modelling}, 387, 61--69.

\bibitem[{Louvrier \emph{et~al.}(2019)Louvrier, Molinari-Jobin, K{\'e}ry,
  Chambert, Miller, Zimmermann, Marboutin, Molinari, M{\"u}eller, {\v{C}}erne
  \emph{et~al.}}]{louvrier_use_2019}
Louvrier, J., Molinari-Jobin, A., K{\'e}ry, M., Chambert, T., Miller, D.,
  Zimmermann, F., Marboutin, E., Molinari, P., M{\"u}eller, O., {\v{C}}erne, R.
  \emph{et~al.} (2019).
\newblock Use of ambiguous detections to improve estimates from species
  distribution models.
\newblock \emph{Conservation Biology}, 33, 185--195.

\bibitem[{Lowe \emph{et~al.}(2011)Lowe, Bruno, Selig \& Spencer}]{LoweEtAl2011}
Lowe, P.~K., Bruno, J.~F., Selig, E.~R. \& Spencer, M. (2011).
\newblock Empirical models of transitions between coral reef states: effects of
  region, protection, and environmental change.
\newblock \emph{PLoS ONE}, 6, e26339.

\bibitem[{MacDonald \& Raubenheimer(1995)}]{macdonald1995hidden}
MacDonald, I.~L. \& Raubenheimer, D. (1995).
\newblock Hidden {M}arkov models and animal behaviour.
\newblock \emph{Biometrical Journal}, 37, 701--712.

\bibitem[{MacKenzie \emph{et~al.}(2003)MacKenzie, Nichols, Hines, Knutson \&
  Franklin}]{MackenzieEtAl2003}
MacKenzie, D.~I., Nichols, J.~D., Hines, J.~E., Knutson, M.~G. \& Franklin,
  A.~B. (2003).
\newblock Estimating site occupancy, colonization, and local extinction when a
  species is detected imperfectly.
\newblock \emph{Ecology}, 84, 2200--2207.

\bibitem[{MacKenzie \emph{et~al.}(2002)MacKenzie, Nichols, Lachman, Droege,
  Andrew~Royle \& Langtimm}]{MackenzieEtAl2002}
MacKenzie, D.~I., Nichols, J.~D., Lachman, G.~B., Droege, S., Andrew~Royle, J.
  \& Langtimm, C.~A. (2002).
\newblock Estimating site occupancy rates when detection probabilities are less
  than one.
\newblock \emph{Ecology}, 83, 2248--2255.

\bibitem[{MacKenzie \emph{et~al.}(2018)MacKenzie, Nichols, Royle, Pollock,
  Bailey \& Hines}]{MackenzieEtAl2018}
MacKenzie, D.~I., Nichols, J.~D., Royle, J.~A., Pollock, K.~H., Bailey, L. \&
  Hines, J.~E. (2018).
\newblock \emph{Occupancy Estimation and Modeling: Inferring Patterns and
  Dynamics of Species Occurrence}.
\newblock 2nd edn. Elsevier.

\bibitem[{MacKenzie \emph{et~al.}(2009)MacKenzie, Nichols, Seamans \&
  Guti{\'e}rrez}]{MackenzieEtAl2009}
MacKenzie, D.~I., Nichols, J.~D., Seamans, M.~E. \& Guti{\'e}rrez, R. (2009).
\newblock Modeling species occurrence dynamics with multiple states and
  imperfect detection.
\newblock \emph{Ecology}, 90, 823--835.

\bibitem[{Magurran(2004)}]{magurran2004measuring}
Magurran, A.~E. (2004).
\newblock \emph{Measuring biological diversity}.
\newblock Blackwell Science, Oxford.

\bibitem[{Marescot \emph{et~al.}(2018)Marescot, Benhaiem, Gimenez, Hofer,
  Lebreton, Olarte-Castillo, Kramer-Schadt \& East}]{MarescotEtAl2018}
Marescot, L., Benhaiem, S., Gimenez, O., Hofer, H., Lebreton, J.-D.,
  Olarte-Castillo, X.~A., Kramer-Schadt, S. \& East, M.~L. (2018).
\newblock Social status mediates the fitness costs of infection with canine
  distemper virus in {Serengeti} spotted hyenas.
\newblock \emph{Functional Ecology}, 32, 1237--1250.

\bibitem[{Marescot \emph{et~al.}(2019)Marescot, Lyet, Singh, Carter \&
  Gimenez}]{MarescotEtAl2019}
Marescot, L., Lyet, A., Singh, R., Carter, N. \& Gimenez, O. (2019).
\newblock Inferring wildlife poaching in southeast {A}sia with multispecies
  dynamic occupancy models.
\newblock \emph{Ecography}.

\bibitem[{Martin \emph{et~al.}(2009)Martin, McIntyre, Hines, Nichols, Schmutz
  \& MacCluskie}]{MartinEtAl2009}
Martin, J., McIntyre, C.~L., Hines, J.~E., Nichols, J.~D., Schmutz, J.~A. \&
  MacCluskie, M.~C. (2009).
\newblock Dynamic multistate site occupancy models to evaluate hypotheses
  relevant to conservation of golden eagles in {D}enali {N}ational {P}ark,
  {A}laska.
\newblock \emph{Biological Conservation}, 142, 2726--2731.

\bibitem[{Martin \emph{et~al.}(2005)Martin, Wintle, Rhodes, Kuhnert, Field,
  Low-Choy, Tyre \& Possingham}]{MartinEtAl2005}
Martin, T.~G., Wintle, B.~A., Rhodes, J.~R., Kuhnert, P.~M., Field, S.~A.,
  Low-Choy, S.~J., Tyre, A.~J. \& Possingham, H.~P. (2005).
\newblock Zero tolerance ecology: improving ecological inference by modelling
  the source of zero observations.
\newblock \emph{Ecology Letters}, 8, 1235--1246.

\bibitem[{Martinez \emph{et~al.}(2016)Martinez, King, Yunus, Faruque \&
  Pascual}]{MartinezEtAl2016}
Martinez, P.~P., King, A.~A., Yunus, M., Faruque, A. \& Pascual, M. (2016).
\newblock Differential and enhanced response to climate forcing in diarrheal
  disease due to rotavirus across a megacity of the developing world.
\newblock \emph{Proceedings of the National Academy of Sciences}, 113,
  4092--4097.

\bibitem[{McClintock \emph{et~al.}(2012)McClintock, King, Thomas,
  Matthiopoulos, McConnell \& Morales}]{McClintockEtAl2012}
McClintock, B.~T., King, R., Thomas, L., Matthiopoulos, J., McConnell, B.~J. \&
  Morales, J.~M. (2012).
\newblock A general discrete-time modeling framework for animal movement using
  multistate random walks.
\newblock \emph{Ecological Monographs}, 82, 335--349.

\bibitem[{McClintock \emph{et~al.}(2017)McClintock, London, Cameron \&
  Boveng}]{McClintockEtAl2017}
McClintock, B.~T., London, J.~M., Cameron, M.~F. \& Boveng, P.~L. (2017).
\newblock Bridging the gaps in animal movement: hidden behaviors and ecological
  relationships revealed by integrated data streams.
\newblock \emph{Ecosphere}, 8, e01751.

\bibitem[{McClintock \& Michelot(2018)}]{McClintockMichelot2018}
McClintock, B.~T. \& Michelot, T. (2018).
\newblock momentu{HMM}: {R} package for generalized hidden {M}arkov models of
  animal movement.
\newblock \emph{Methods in Ecology and Evolution}, 9, 1518--1530.

\bibitem[{McClintock \emph{et~al.}(2010)McClintock, Nichols, Bailey, MacKenzie,
  Kendall \& Franklin}]{McClintockEtAl2010}
McClintock, B.~T., Nichols, J.~D., Bailey, L.~L., MacKenzie, D.~I., Kendall,
  W.~L. \& Franklin, A.~B. (2010).
\newblock Seeking a second opinion: uncertainty in disease ecology.
\newblock \emph{Ecology Letters}, 13, 659--674.

\bibitem[{McClintock \emph{et~al.}(2013)McClintock, Russell, Matthiopoulos \&
  King}]{McClintockEtAl2013c}
McClintock, B.~T., Russell, D.~J., Matthiopoulos, J. \& King, R. (2013).
\newblock Combining individual animal movement and ancillary biotelemetry data
  to investigate population-level activity budgets.
\newblock \emph{Ecology}, 94, 838--849.

\bibitem[{McCullagh \& Nelder(1989)}]{McCullaghNelder1989}
McCullagh, P. \& Nelder, J.~A. (1989).
\newblock \emph{Generalized Linear Models}.
\newblock Chapman and Hall, New York.

\bibitem[{McDonald \emph{et~al.}(2017)McDonald, Hornsby, Speakman, Zolman,
  Mullin, Sinclair, Rosel, Thomas \& Schwacke}]{Mcdonald2017}
McDonald, T.~L., Hornsby, F.~E., Speakman, T.~R., Zolman, E.~S., Mullin, K.~D.,
  Sinclair, C., Rosel, P.~E., Thomas, L. \& Schwacke, L.~H. (2017).
\newblock Survival, density, and abundance of common bottlenose dolphins in
  {B}arataria {B}ay ({USA}) following the {D}eepwater {H}orizon oil spill.
\newblock \emph{Endangered Species Research}, 33, 193--209.

\bibitem[{McGraw \& Caswell(1996)}]{mcgraw_estimation_1996}
McGraw, J.~B. \& Caswell, H. (1996).
\newblock Estimation of individual fitness from life-history data.
\newblock \emph{The American Naturalist}, 147, 47--64.

\bibitem[{Miller \emph{et~al.}(2018{\natexlab{a}})Miller, Grant, Muths,
  Amburgey, Adams, Joseph, Waddle, Johnson, Ryan, Schmidt
  \emph{et~al.}}]{MillerEtAl2018}
Miller, D.~A., Grant, E. H.~C., Muths, E., Amburgey, S.~M., Adams, M.~J.,
  Joseph, M.~B., Waddle, J.~H., Johnson, P.~T., Ryan, M.~E., Schmidt, B.~R.
  \emph{et~al.} (2018{\natexlab{a}}).
\newblock Quantifying climate sensitivity and climate-driven change in {N}orth
  {A}merican amphibian communities.
\newblock \emph{Nature communications}, 9, 3926.

\bibitem[{Miller \emph{et~al.}(2011)Miller, Nichols, McClintock, Grant, Bailey
  \& Weir}]{MillerEtAl2011}
Miller, D.~A., Nichols, J.~D., McClintock, B.~T., Grant, E. H.~C., Bailey,
  L.~L. \& Weir, L.~A. (2011).
\newblock Improving occupancy estimation when two types of observational error
  occur: non-detection and species misidentification.
\newblock \emph{Ecology}, 92, 1422--1428.

\bibitem[{Miller \emph{et~al.}(2018{\natexlab{b}})Miller, Pitman, Mann, Fuller
  \& Balme}]{miller_lions_2018}
Miller, J. R.~B., Pitman, R.~T., Mann, G. K.~H., Fuller, A.~K. \& Balme, G.~A.
  (2018{\natexlab{b}}).
\newblock Lions and leopards coexist without spatial, temporal or demographic
  effects of interspecific competition.
\newblock \emph{Journal of Animal Ecology}, 87, 1709--1726.

\bibitem[{Moilanen(1999)}]{Moilanen1999}
Moilanen, A. (1999).
\newblock Patch occupancy models of metapopulation dynamics: efficient
  parameter estimation using implicit statistical inference.
\newblock \emph{Ecology}, 80, 1031--1043.

\bibitem[{Morales \emph{et~al.}(2004)Morales, Haydon, Frair, Holsinger \&
  Fryxell}]{MoralesEtAl2004}
Morales, J.~M., Haydon, D.~T., Frair, J., Holsinger, K.~E. \& Fryxell, J.~M.
  (2004).
\newblock Extracting more out of relocation data: building movement models as
  mixtures of random walks.
\newblock \emph{Ecology}, 85, 2436--2445.

\bibitem[{Moritz \emph{et~al.}(2008)Moritz, Patton, Conroy, Parra, White \&
  Beissinger}]{MoritzEtAl2008}
Moritz, C., Patton, J.~L., Conroy, C.~J., Parra, J.~L., White, G.~C. \&
  Beissinger, S.~R. (2008).
\newblock Impact of a century of climate change on small-mammal communities in
  {Y}osemite {N}ational {P}ark, {USA}.
\newblock \emph{Science}, 322, 261--264.

\bibitem[{Murphy \emph{et~al.}(2019)Murphy, Kelly, Karpanty, Andrianjakarivelo
  \& Farris}]{murphy_using_2019}
Murphy, A., Kelly, M.~J., Karpanty, S.~M., Andrianjakarivelo, V. \& Farris,
  Z.~J. (2019).
\newblock Using camera traps to investigate spatial co-occurrence between
  exotic predators and native prey species: a case study from northeastern
  {Madagascar}.
\newblock \emph{Journal of Zoology}, 307, 264--273.

\bibitem[{Myung(2003)}]{Myung2003}
Myung, I.~J. (2003).
\newblock Tutorial on maximum likelihood estimation.
\newblock \emph{Journal of Mathematical Psychology}, 47, 90--100.

\bibitem[{Newman \emph{et~al.}(2014)Newman, Buckland, Morgan, King, Borchers,
  Cole, Besbeas, Gimenez \& Thomas}]{newman2014modelling}
Newman, K.~B., Buckland, S.~T., Morgan, B. J.~T., King, R., Borchers, D.~L.,
  Cole, D.~J., Besbeas, P., Gimenez, O. \& Thomas, L. (2014).
\newblock \emph{Modelling population dynamics: model formulation, fitting and
  assessment using state-space methods}.
\newblock Springer.

\bibitem[{Ng{\^o} \emph{et~al.}(2019)Ng{\^o}, Heide-J{\o}rgensen \&
  Ditlevsen}]{ngo2019understanding}
Ng{\^o}, M.~C., Heide-J{\o}rgensen, M.~P. \& Ditlevsen, S. (2019).
\newblock Understanding narwhal diving behaviour using hidden {M}arkov models
  with dependent state distributions and long range dependence.
\newblock \emph{PLoS Computational Biology}, 15, e1006425.

\bibitem[{Nichols \emph{et~al.}(1992)Nichols, Sauer, Pollock \&
  Hestbeck}]{NicholsEtAl1992}
Nichols, J., Sauer, J.~R., Pollock, K.~H. \& Hestbeck, J.~B. (1992).
\newblock Estimating transition probabilities for stage-based population
  projection matrices using capture--recapture data.
\newblock \emph{Ecology}, 73, 306--312.

\bibitem[{Nichols \emph{et~al.}(1994)Nichols, Hines, Pollock, Hinz \&
  Link}]{NicholsEtAl1994}
Nichols, J.~D., Hines, J.~E., Pollock, K.~H., Hinz, R.~L. \& Link, W.~A.
  (1994).
\newblock Estimating breeding proportions and testing hypotheses about costs of
  reproduction with capture-recapture data.
\newblock \emph{Ecology}, 75, 2052--2065.

\bibitem[{Nichols \emph{et~al.}(2009)Nichols, Thomas \& Conn}]{NicholsEtAl2009}
Nichols, J.~D., Thomas, L. \& Conn, P.~B. (2009).
\newblock Inferences about landbird abundance from count data: recent advances
  and future directions.
\newblock In: \emph{Modeling Demographic Processes in Marked Populations}.
  Springer, pp. 201--235.

\bibitem[{O'Connell \& H\o{}jsgaard(2011)}]{OConnellHojsgaard2011}
O'Connell, J. \& H\o{}jsgaard, S. (2011).
\newblock Hidden semi {M}arkov models for multiple observation sequences: The
  {mhsmm} package for {R}.
\newblock \emph{Journal of Statistical Software}, 39, 1--22.

\bibitem[{Olajos \emph{et~al.}(2018)Olajos, Bokma, Bartels, Myrstener, Rydberg,
  {\"O}hlund, Bindler, Wang, Zale \& Englund}]{OlajosEtAl2018}
Olajos, F., Bokma, F., Bartels, P., Myrstener, E., Rydberg, J., {\"O}hlund, G.,
  Bindler, R., Wang, X.-R., Zale, R. \& Englund, G. (2018).
\newblock Estimating species colonization dates using {DNA} in lake sediment.
\newblock \emph{Methods in Ecology and Evolution}, 9, 535--543.

\bibitem[{Ovaskainen \emph{et~al.}(2017)Ovaskainen, Tikhonov, Norberg,
  Guillaume~Blanchet, Duan, Dunson, Roslin \& Abrego}]{OvaskainenEtAl2017}
Ovaskainen, O., Tikhonov, G., Norberg, A., Guillaume~Blanchet, F., Duan, L.,
  Dunson, D., Roslin, T. \& Abrego, N. (2017).
\newblock How to make more out of community data? a conceptual framework and
  its implementation as models and software.
\newblock \emph{Ecology Letters}, 20, 561--576.

\bibitem[{Palkopoulou \emph{et~al.}(2018)Palkopoulou, Lipson, Mallick, Nielsen,
  Rohland, Baleka, Karpinski, Ivancevic, To, Kortschak
  \emph{et~al.}}]{palkopoulou_comprehensive_2018}
Palkopoulou, E., Lipson, M., Mallick, S., Nielsen, S., Rohland, N., Baleka, S.,
  Karpinski, E., Ivancevic, A.~M., To, T.-H., Kortschak, R.~D. \emph{et~al.}
  (2018).
\newblock A comprehensive genomic history of extinct and living elephants.
\newblock \emph{Proceedings of the National Academy of Sciences}, 115,
  E2566--E2574.

\bibitem[{Papastamatiou \emph{et~al.}(2018{\natexlab{a}})Papastamatiou,
  Iosilevskii, Leos-Barajas, Brooks, Howey, Chapman \&
  Watanabe}]{papastamatiou2018optimal}
Papastamatiou, Y.~P., Iosilevskii, G., Leos-Barajas, V., Brooks, E.~J., Howey,
  L.~A., Chapman, D.~D. \& Watanabe, Y.~Y. (2018{\natexlab{a}}).
\newblock Optimal swimming strategies and behavioral plasticity of oceanic
  whitetip sharks.
\newblock \emph{Scientific Reports}, 8, 1--12.

\bibitem[{Papastamatiou \emph{et~al.}(2018{\natexlab{b}})Papastamatiou,
  Watanabe, Dem{\v{s}}ar, Leos-Barajas, Bradley, Langrock, Weng, Lowe,
  Friedlander \& Caselle}]{papastamatiou2018activity}
Papastamatiou, Y.~P., Watanabe, Y.~Y., Dem{\v{s}}ar, U., Leos-Barajas, V.,
  Bradley, D., Langrock, R., Weng, K., Lowe, C.~G., Friedlander, A.~M. \&
  Caselle, J.~E. (2018{\natexlab{b}}).
\newblock Activity seascapes highlight central place foraging strategies in
  marine predators that never stop swimming.
\newblock \emph{Movement Ecology}, 6, 9.

\bibitem[{Patterson \emph{et~al.}(2009)Patterson, Basson, Bravington \&
  Gunn}]{PattersonEtAl2009}
Patterson, T.~A., Basson, M., Bravington, M.~V. \& Gunn, J.~S. (2009).
\newblock Classifying movement behaviour in relation to environmental
  conditions using hidden {M}arkov models.
\newblock \emph{Journal of Animal Ecology}, 78, 1113--1123.

\bibitem[{Patterson \emph{et~al.}(2017)Patterson, Parton, Langrock, Blackwell,
  Thomas \& King}]{patterson2017statistical}
Patterson, T.~A., Parton, A., Langrock, R., Blackwell, P.~G., Thomas, L. \&
  King, R. (2017).
\newblock Statistical modelling of individual animal movement: an overview of
  key methods and a discussion of practical challenges.
\newblock \emph{AStA Advances in Statistical Analysis}, 101, 399--438.

\bibitem[{Patterson \emph{et~al.}(2008)Patterson, Thomas, Wilcox, Ovaskainen \&
  Matthiopoulos}]{patterson2008state}
Patterson, T.~A., Thomas, L., Wilcox, C., Ovaskainen, O. \& Matthiopoulos, J.
  (2008).
\newblock State--space models of individual animal movement.
\newblock \emph{Trends in Ecology \& Evolution}, 23, 87--94.

\bibitem[{Pawlowski \& McCord(2009)}]{PawlowskiMcCord2009}
Pawlowski, C.~W. \& McCord, C. (2009).
\newblock A {M}arkov model for assessing ecological stability properties.
\newblock \emph{Ecological Modelling}, 220, 86--95.

\bibitem[{Pedersen \emph{et~al.}(2011{\natexlab{a}})Pedersen, Berg, Thygesen,
  Nielsen \& Madsen}]{PedersenEtAl2011}
Pedersen, M.~W., Berg, C.~W., Thygesen, U.~H., Nielsen, A. \& Madsen, H.
  (2011{\natexlab{a}}).
\newblock Estimation methods for nonlinear state-space models in ecology.
\newblock \emph{Ecological Modelling}, 222, 1394--1400.

\bibitem[{Pedersen \emph{et~al.}(2011{\natexlab{b}})Pedersen, Patterson,
  Thygesen \& Madsen}]{PedersenEtAl2011estimating}
Pedersen, M.~W., Patterson, T.~A., Thygesen, U.~H. \& Madsen, H.
  (2011{\natexlab{b}}).
\newblock Estimating animal behavior and residency from movement data.
\newblock \emph{Oikos}, 120, 1281--1290.

\bibitem[{Pedersen \emph{et~al.}(2008)Pedersen, Righton, Thygesen, Andersen \&
  Madsen}]{PedersenEtAl2008}
Pedersen, M.~W., Righton, D., Thygesen, U.~H., Andersen, K.~H. \& Madsen, H.
  (2008).
\newblock Geolocation of {N}orth {S}ea cod ({G}adus morhua) using hidden
  {M}arkov models and behavioural switching.
\newblock \emph{Canadian Journal of Fisheries and Aquatic Sciences}, 65,
  2367--2377.

\bibitem[{Phillips \emph{et~al.}(2015)Phillips, Patterson, Leroy, Pilling \&
  Nicol}]{phillips2015objective}
Phillips, J.~S., Patterson, T.~A., Leroy, B., Pilling, G.~M. \& Nicol, S.~J.
  (2015).
\newblock Objective classification of latent behavioral states in bio-logging
  data using multivariate-normal hidden {M}arkov models.
\newblock \emph{Ecological Applications}, 25, 1244--1258.

\bibitem[{Pledger(2000)}]{pledger2000unified}
Pledger, S. (2000).
\newblock Unified maximum likelihood estimates for closed capture--recapture
  models using mixtures.
\newblock \emph{Biometrics}, 56, 434--442.

\bibitem[{Pledger \& Arnold(2014)}]{pledger2014multivariate}
Pledger, S. \& Arnold, R. (2014).
\newblock Multivariate methods using mixtures: Correspondence analysis, scaling
  and pattern-detection.
\newblock \emph{Computational Statistics \& Data Analysis}, 71, 241--261.

\bibitem[{Pluntz \emph{et~al.}(2018)Pluntz, Coz, Peyrard, Pradel, Choquet \&
  Cheptou}]{PluntzEtAl2018}
Pluntz, M., Coz, S.~L., Peyrard, N., Pradel, R., Choquet, R. \& Cheptou, P.-O.
  (2018).
\newblock A general method for estimating seed dormancy and colonisation in
  annual plants from the observation of existing flora.
\newblock \emph{Ecology Letters}, 21, 1311--1318.

\bibitem[{Pohle \emph{et~al.}(2017)Pohle, Langrock, van Beest \&
  Schmidt}]{pohle2017selecting}
Pohle, J., Langrock, R., van Beest, F.~M. \& Schmidt, N.~M. (2017).
\newblock Selecting the number of states in hidden {M}arkov models: pragmatic
  solutions illustrated using animal movement.
\newblock \emph{Journal of Agricultural, Biological and Environmental
  Statistics}, 22, 270--293.

\bibitem[{Pradel(1996)}]{Pradel1996}
Pradel, R. (1996).
\newblock Utilization of capture-mark-recapture for the study of recruitment
  and population growth rate.
\newblock \emph{Biometrics}, 52, 703--709.

\bibitem[{Pradel(2005)}]{Pradel2005}
Pradel, R. (2005).
\newblock Multievent: An extension of multistate capture-recapture models to
  uncertain states.
\newblock \emph{Biometrics}, 61, 442--447.

\bibitem[{Pradel \emph{et~al.}(2008)Pradel, Maurin-Bernier, Gimenez, Genovart,
  Choquet \& Oro}]{PradelEtAl2008}
Pradel, R., Maurin-Bernier, L., Gimenez, O., Genovart, M., Choquet, R. \& Oro,
  D. (2008).
\newblock Estimation of sex-specific survival with uncertainty in sex
  assessment.
\newblock \emph{Canadian Journal of Statistics}, 36, 29--42.

\bibitem[{Quick \emph{et~al.}(2017)Quick, Isojunno, Sadykova, Bowers, Nowacek
  \& Read}]{quick2017hidden}
Quick, N.~J., Isojunno, S., Sadykova, D., Bowers, M., Nowacek, D.~P. \& Read,
  A.~J. (2017).
\newblock Hidden {M}arkov models reveal complexity in the diving behaviour of
  short-finned pilot whales.
\newblock \emph{Scientific Reports}, 7, 45765.

\bibitem[{{R Core Team}(2019)}]{RCoreTeam2019}
{R Core Team} (2019).
\newblock \emph{R: A Language and Environment for Statistical Computing}.
\newblock R Foundation for Statistical Computing, Vienna, Austria.
\newblock \urlprefix\url{https://www.R-project.org/}.

\bibitem[{Rabiner(1989)}]{Rabiner1989}
Rabiner, L.~R. (1989).
\newblock A tutorial on hidden {M}arkov models and selected applications in
  speech recognition.
\newblock \emph{Proceedings of the IEEE}, 77, 257--286.

\bibitem[{Rakhimberdiev \emph{et~al.}(2015)Rakhimberdiev, Winkler, Bridge,
  Seavy, Sheldon, Piersma \& Saveliev}]{RakhimberdievEtAl2015}
Rakhimberdiev, E., Winkler, D.~W., Bridge, E., Seavy, N.~E., Sheldon, D.,
  Piersma, T. \& Saveliev, A. (2015).
\newblock A hidden {M}arkov model for reconstructing animal paths from solar
  geolocation loggers using templates for light intensity.
\newblock \emph{Movement Ecology}, 3, 25.

\bibitem[{Rich \emph{et~al.}(2016)Rich, Miller, Robinson, McNutt \&
  Kelly}]{rich_using_2016}
Rich, L.~N., Miller, D. A.~W., Robinson, H.~S., McNutt, J.~W. \& Kelly, M.~J.
  (2016).
\newblock Using camera trapping and hierarchical occupancy modelling to
  evaluate the spatial ecology of an {African} mammal community.
\newblock \emph{Journal of Applied Ecology}, 53, 1225--1235.

\bibitem[{Roeleke \emph{et~al.}(2018)Roeleke, Teige, Hoffmeister, Klingler \&
  Voigt}]{roeleke2018aerial}
Roeleke, M., Teige, T., Hoffmeister, U., Klingler, F. \& Voigt, C.~C. (2018).
\newblock Aerial-hawking bats adjust their use of space to the lunar cycle.
\newblock \emph{Movement Ecology}, 6, 11.

\bibitem[{Rohani \& King(2010)}]{rohani2010never}
Rohani, P. \& King, A.~A. (2010).
\newblock Never mind the length, feel the quality: the impact of long-term
  epidemiological data sets on theory, application and policy.
\newblock \emph{Trends in Ecology \& Evolution}, 25, 611--618.

\bibitem[{Rota \emph{et~al.}(2016)Rota, Ferreira, Kays, Forrester, Kalies,
  McShea, Parsons \& Millspaugh}]{rota_multispecies_2016}
Rota, C.~T., Ferreira, M. A.~R., Kays, R.~W., Forrester, T.~D., Kalies, E.~L.,
  McShea, W.~J., Parsons, A.~W. \& Millspaugh, J.~J. (2016).
\newblock A multispecies occupancy model for two or more interacting species.
\newblock \emph{Methods in Ecology and Evolution}, 7, 1164--1173.

\bibitem[{Rouan \emph{et~al.}(2009)Rouan, Gaillard, Gu\'edon \&
  Pradel}]{rouan_estimation_2009}
Rouan, L., Gaillard, J.-M., Gu\'edon, Y. \& Pradel, R. (2009).
\newblock Estimation of lifetime reproductive success when reproductive status
  cannot always be assessed.
\newblock In: \emph{Modeling {Demographic} {Processes} {In} {Marked}
  {Populations}} (eds. Thomson, D.~L., Cooch, E.~G. \& Conroy, M.~J.).
  Springer, Boston, pp. 867--879.

\bibitem[{Rowe \emph{et~al.}(2017)Rowe, Sweet \& Beebee}]{RoweEtAl2017}
Rowe, G., Sweet, M. \& Beebee, T. J.~C. (2017).
\newblock \emph{An Introduction to Molecular Ecology}.
\newblock Oxford University Press.

\bibitem[{Royle(2004)}]{royle2004n}
Royle, J.~A. (2004).
\newblock N-mixture models for estimating population size from spatially
  replicated counts.
\newblock \emph{Biometrics}, 60, 108--115.

\bibitem[{Royle \emph{et~al.}(2013)Royle, Chandler, Sollmann \&
  Gardner}]{RoyleEtAl2013book}
Royle, J.~A., Chandler, R.~B., Sollmann, R. \& Gardner, B. (2013).
\newblock \emph{Spatial Capture-Recapture}.
\newblock Academic Press.

\bibitem[{Royle \& Dorazio(2008)}]{RoyleDorazio2008}
Royle, J.~A. \& Dorazio, R.~M. (2008).
\newblock \emph{Hierarchical Modeling and Inference in Ecology: the analysis of
  data from populations, metapopulations and communities}.
\newblock Elsevier.

\bibitem[{Royle \emph{et~al.}(2018)Royle, Fuller \& Sutherland}]{RoyleEtAl2018}
Royle, J.~A., Fuller, A.~K. \& Sutherland, C. (2018).
\newblock Unifying population and landscape ecology with spatial
  capture--recapture.
\newblock \emph{Ecography}, 41, 444--456.

\bibitem[{Royle \& K\'ery(2007)}]{royle_bayesian_2007}
Royle, J.~A. \& K\'ery, M. (2007).
\newblock A {Bayesian} state-space formulation of dynamic occupancy models.
\newblock \emph{Ecology}, 88, 1813--23.

\bibitem[{Runge \emph{et~al.}(2007)Runge, Hines \& Nichols}]{RungeEtAl2007}
Runge, J.~P., Hines, J.~E. \& Nichols, J.~D. (2007).
\newblock Estimating species-specific survival and movement when species
  identification is uncertain.
\newblock \emph{Ecology}, 88, 282--288.

\bibitem[{Russell \emph{et~al.}(2009)Russell, Royle, Saab, Lehmkuhl, Block \&
  Sauer}]{russell_modeling_2009}
Russell, R.~E., Royle, J.~A., Saab, V.~A., Lehmkuhl, J.~F., Block, W.~M. \&
  Sauer, J.~R. (2009).
\newblock Modeling the effects of environmental disturbance on wildlife
  communities: avian responses to prescribed fire.
\newblock \emph{Ecological Applications}, 19, 1253--1263.

\bibitem[{Schaub \& Abadi(2011)}]{Schaub2011}
Schaub, M. \& Abadi, F. (2011).
\newblock Integrated population models: a novel analysis framework for deeper
  insights into population dynamics.
\newblock \emph{Journal of Ornithology}, 152, 227--237.

\bibitem[{Scheffer \emph{et~al.}(2001)Scheffer, Carpenter, Foley, Folke \&
  Walker}]{SchefferEtAl2001}
Scheffer, M., Carpenter, S., Foley, J.~A., Folke, C. \& Walker, B. (2001).
\newblock Catastrophic shifts in ecosystems.
\newblock \emph{Nature}, 413, 591.

\bibitem[{Scheffer \& Carpenter(2003)}]{SchefferEtAl2003}
Scheffer, M. \& Carpenter, S.~R. (2003).
\newblock Catastrophic regime shifts in ecosystems: linking theory to
  observation.
\newblock \emph{Trends in Ecology \& Evolution}, 18, 648--656.

\bibitem[{Schliehe-Diecks \emph{et~al.}(2012)Schliehe-Diecks, Kappeler \&
  Langrock}]{schliehe2012application}
Schliehe-Diecks, S., Kappeler, P. \& Langrock, R. (2012).
\newblock On the application of mixed hidden {M}arkov models to multiple
  behavioural time series.
\newblock \emph{Interface Focus}, 2, 180--189.

\bibitem[{Schmidt \emph{et~al.}(2015)Schmidt, Johnson, Lindberg \&
  Adams}]{SchmidtEtAl2015}
Schmidt, J.~H., Johnson, D.~S., Lindberg, M.~S. \& Adams, L.~G. (2015).
\newblock Estimating demographic parameters using a combination of known-fate
  and open {N}-mixture models.
\newblock \emph{Ecology}, 96, 2583--2589.

\bibitem[{Schnute(1994)}]{schnute1994general}
Schnute, J.~T. (1994).
\newblock A general framework for developing sequential fisheries models.
\newblock \emph{Canadian Journal of Fisheries and Aquatic Sciences}, 51,
  1676--1688.

\bibitem[{Schumer \emph{et~al.}(2018)Schumer, Xu, Powell, Durvasula, Skov,
  Holland, Blazier, Sankararaman, Andolfatto, Rosenthal
  \emph{et~al.}}]{SchumerEtAl2018}
Schumer, M., Xu, C., Powell, D.~L., Durvasula, A., Skov, L., Holland, C.,
  Blazier, J.~C., Sankararaman, S., Andolfatto, P., Rosenthal, G.~G.
  \emph{et~al.} (2018).
\newblock Natural selection interacts with recombination to shape the evolution
  of hybrid genomes.
\newblock \emph{Science}, 360, 656--660.

\bibitem[{Schwarz \emph{et~al.}(1993)Schwarz, Schweigert \&
  Arnason}]{SchwarzEtAl1993}
Schwarz, C.~J., Schweigert, J.~F. \& Arnason, A.~N. (1993).
\newblock Estimating migration rates using tag-recovery data.
\newblock \emph{Biometrics}, 49, 177--193.

\bibitem[{Seber \& Schofield(2019)}]{Seber2019}
Seber, G.~A. \& Schofield, M.~R. (2019).
\newblock \emph{Capture-Recapture: Parameter Estimation for Open Animal
  Populations}.
\newblock Springer.

\bibitem[{Sherlock \emph{et~al.}(2013)Sherlock, Xifara, Telfer \&
  Begon}]{sherlock2013coupled}
Sherlock, C., Xifara, T., Telfer, S. \& Begon, M. (2013).
\newblock A coupled hidden {M}arkov model for disease interactions.
\newblock \emph{Journal of the Royal Statistical Society: Series C (Applied
  Statistics)}, 62, 609--627.

\bibitem[{Siachalou \emph{et~al.}(2014)Siachalou, Doxani \&
  Tsakiri-Strati}]{SiachalouEtAl2014}
Siachalou, S., Doxani, G. \& Tsakiri-Strati, M. (2014).
\newblock Time-series analysis of high temporal remote sensing data to model
  wetland dynamics: A hidden {M}arkov model approach.
\newblock In: \emph{Proceedings of the SENTINEL-2 for Science
  Workshop--ESA--ESRIN, Frascati, Italy, 20--22 May 2014}.

\bibitem[{Siachalou \emph{et~al.}(2015)Siachalou, Mallinis \&
  Tsakiri-Strati}]{SiachalouEtAl2015}
Siachalou, S., Mallinis, G. \& Tsakiri-Strati, M. (2015).
\newblock A hidden {M}arkov models approach for crop classification: Linking
  crop phenology to time series of multi-sensor remote sensing data.
\newblock \emph{Remote Sensing}, 7, 3633--3650.

\bibitem[{Skrondal \& Rabe-Hesketh(2004)}]{skrondal2004generalized}
Skrondal, A. \& Rabe-Hesketh, S. (2004).
\newblock \emph{Generalized Latent Variable Modeling: Multilevel, Longitudinal,
  and Structural Equation Models}.
\newblock CRC Press.

\bibitem[{Sollmann \emph{et~al.}(2015)Sollmann, Gardner, Chandler, Royle \&
  Sillett}]{Sollmann2015}
Sollmann, R., Gardner, B., Chandler, R.~B., Royle, J.~A. \& Sillett, T.~S.
  (2015).
\newblock An open-population hierarchical distance sampling model.
\newblock \emph{Ecology}, 96, 325--331.

\bibitem[{Soria-Carrasco \emph{et~al.}(2014)Soria-Carrasco, Gompert, Comeault,
  Farkas, Parchman, Johnston, Buerkle, Feder, Bast, Schwander
  \emph{et~al.}}]{SoriaEtal2014}
Soria-Carrasco, V., Gompert, Z., Comeault, A.~A., Farkas, T.~E., Parchman,
  T.~L., Johnston, J.~S., Buerkle, C.~A., Feder, J.~L., Bast, J., Schwander, T.
  \emph{et~al.} (2014).
\newblock Stick insect genomes reveal natural selection's role in parallel
  speciation.
\newblock \emph{Science}, 344, 738--742.

\bibitem[{Srikanthan \& McMahon(2001)}]{SrikanthanMcMahon2001}
Srikanthan, R. \& McMahon, T. (2001).
\newblock Stochastic generation of annual, monthly and daily climate data: A
  review.
\newblock \emph{Hydrology and Earth System Sciences Discussions}, 5, 653--670.

\bibitem[{Stoelting \emph{et~al.}(2015)Stoelting, Gutierrez, Kendall \&
  Peery}]{StoeltingEtAl2015}
Stoelting, R.~E., Gutierrez, R.~J., Kendall, W.~L. \& Peery, M.~Z. (2015).
\newblock Life-history tradeoffs and reproductive cycles in spotted owls.
\newblock \emph{The Auk}, 132, 46--64.

\bibitem[{Sutherland \emph{et~al.}(2016)Sutherland, Brambilla, Pedrini \&
  Tenan}]{sutherland_multiregion_2016}
Sutherland, C., Brambilla, M., Pedrini, P. \& Tenan, S. (2016).
\newblock A multiregion community model for inference about geographic
  variation in species richness.
\newblock \emph{Methods in Ecology and Evolution}, 7, 783--791.

\bibitem[{Talluto \emph{et~al.}(2017)Talluto, Boulangeat, Vissault, Thuiller \&
  Gravel}]{TallutoEtAl2017}
Talluto, M.~V., Boulangeat, I., Vissault, S., Thuiller, W. \& Gravel, D.
  (2017).
\newblock Extinction debt and colonization credit delay range shifts of eastern
  {N}orth {A}merican trees.
\newblock \emph{Nature Ecology \& Evolution}, 1, 0182.

\bibitem[{Tanner \emph{et~al.}(1994)Tanner, Hughes \& Connell}]{TannerEtAl1994}
Tanner, J.~E., Hughes, T.~P. \& Connell, J.~H. (1994).
\newblock Species coexistence, keystone species, and succession: a sensitivity
  analysis.
\newblock \emph{Ecology}, 75, 2204--2219.

\bibitem[{Tavecchia \emph{et~al.}(2009)Tavecchia, Besbeas, Coulson, Morgan \&
  Clutton-Brock}]{tavecchia2009estimating}
Tavecchia, G., Besbeas, P., Coulson, T., Morgan, B. J.~T. \& Clutton-Brock,
  T.~H. (2009).
\newblock Estimating population size and hidden demographic parameters with
  state-space modeling.
\newblock \emph{The American Naturalist}, 173, 722--733.

\bibitem[{Tenan \emph{et~al.}(2017)Tenan, Brambilla, Pedrini \&
  Sutherland}]{tenan_quantifying_2017}
Tenan, S., Brambilla, M., Pedrini, P. \& Sutherland, C. (2017).
\newblock Quantifying spatial variation in the size and structure of
  ecologically stratified communities.
\newblock \emph{Methods in Ecology and Evolution}, 8, 976--984.

\bibitem[{Thygesen \emph{et~al.}(2009)Thygesen, Pedersen \&
  Madsen}]{ThygesenEtAl2009}
Thygesen, U.~H., Pedersen, M.~W. \& Madsen, H. (2009).
\newblock Geolocating fish using hidden {M}arkov models and data storage tags.
\newblock In: \emph{Tagging and Tracking of Marine Animals with Electronic
  Devices} (eds. Nielsen, J.~L., Arrizabalaga, H., Fragoso, N., Hobday, A.,
  Lutcavage, M. \& Sibert, J.). Springer, pp. 277--293.

\bibitem[{Tingley \emph{et~al.}(2009)Tingley, Monahan, Beissinger \&
  Moritz}]{TingleyEtAl2009}
Tingley, M.~W., Monahan, W.~B., Beissinger, S.~R. \& Moritz, C. (2009).
\newblock Birds track their {G}rinnellian niche through a century of climate
  change.
\newblock \emph{Proceedings of the National Academy of Sciences}, 106,
  19637--19643.

\bibitem[{Touloupou \emph{et~al.}(2020)Touloupou, Finkenst{\"a}dt \&
  Spencer}]{touloupou2019scalable}
Touloupou, P., Finkenst{\"a}dt, B. \& Spencer, S.~E. (2020).
\newblock Scalable {B}ayesian inference for coupled hidden {M}arkov and
  semi-{M}arkov models.
\newblock \emph{Journal of Computational and Graphical Statistics}, 29,
  238--249.

\bibitem[{Towner \emph{et~al.}(2016)Towner, Leos-Barajas, Langrock, Schick,
  Smale, Kaschke, Jewell \& Papastamatiou}]{TownerEtAl2016}
Towner, A.~V., Leos-Barajas, V., Langrock, R., Schick, R.~S., Smale, M.~J.,
  Kaschke, T., Jewell, O. J.~D. \& Papastamatiou, Y.~P. (2016).
\newblock Sex-specific and individual preferences for hunting strategies in
  white sharks.
\newblock \emph{Functional Ecology}, 30, 1397--1407.

\bibitem[{Trier \& Salberg(2011)}]{TrierSalberg2011}
Trier, {\O}.~D. \& Salberg, A.-B. (2011).
\newblock Time-series analysis of satellite images for forest cover change
  monitoring in {T}anzania.
\newblock In: \emph{1st EARSeL Workshop on Operational Remote Sensing in Forest
  Management}.

\bibitem[{Tucker \& Duplisea(2012)}]{TuckerDuplisea2012}
Tucker, A. \& Duplisea, D. (2012).
\newblock Bioinformatics tools in predictive ecology: applications to
  fisheries.
\newblock \emph{Philosophical Transactions of the Royal Society B: Biological
  Sciences}, 367, 279--290.

\bibitem[{Tucker \& Anand(2005)}]{TuckerAnand2005}
Tucker, B.~C. \& Anand, M. (2005).
\newblock On the use of stationary versus hidden {M}arkov models to detect
  simple versus complex ecological dynamics.
\newblock \emph{Ecological Modelling}, 185, 177--193.

\bibitem[{Turek \emph{et~al.}(2016)Turek, de~Valpine \&
  Paciorek}]{TurekEtAl2016}
Turek, D., de~Valpine, P. \& Paciorek, C.~J. (2016).
\newblock Efficient {Markov chain Monte Carlo} sampling for hierarchical hidden
  {M}arkov models.
\newblock \emph{Environmental and Ecological Statistics}, 23, 549--564.

\bibitem[{Turner \emph{et~al.}(1995)Turner, Gardner \&
  O'neill}]{TurnerEtAl1995}
Turner, M.~G., Gardner, R.~H. \& O'neill, R.~V. (1995).
\newblock Ecological dynamics at broad scales.
\newblock \emph{BioScience}, 45, S29--S35.

\bibitem[{Usher(1981)}]{Usher1981}
Usher, M. (1981).
\newblock Modelling ecological succession, with particular reference to
  {M}arkovian models.
\newblock In: \emph{Vegetation Dynamics in Grasslands, Healthlands and
  Mediterranean Ligneous Formations}. Springer, pp. 11--18.

\bibitem[{van Beest \emph{et~al.}(2019)van Beest, Mews, Elkenkamp, Schuhmann,
  Tsolak, Wobbe, Bartolino, Bastardie, Dietz, von Dorrien
  \emph{et~al.}}]{van2019classifying}
van Beest, F.~M., Mews, S., Elkenkamp, S., Schuhmann, P., Tsolak, D., Wobbe,
  T., Bartolino, V., Bastardie, F., Dietz, R., von Dorrien, C. \emph{et~al.}
  (2019).
\newblock Classifying grey seal behaviour in relation to environmental
  variability and commercial fishing activity -- a multivariate hidden {M}arkov
  model.
\newblock \emph{Scientific Reports}, 9, 5642.

\bibitem[{van~de Kerk \emph{et~al.}(2015)van~de Kerk, Onorato, Criffield,
  Bolker, Augustine, McKinley \& Oli}]{van2015hidden}
van~de Kerk, M., Onorato, D.~P., Criffield, M.~A., Bolker, B.~M., Augustine,
  B.~C., McKinley, S.~A. \& Oli, M.~K. (2015).
\newblock Hidden semi-{M}arkov models reveal multiphasic movement of the
  endangered {F}lorida panther.
\newblock \emph{Journal of Animal Ecology}, 84, 576--585.

\bibitem[{van Hulst(1979)}]{VanHulst1979}
van Hulst, R. (1979).
\newblock On the dynamics of vegetation: {M}arkov chains as models of
  succession.
\newblock \emph{Vegetatio}, 40, 3--14.

\bibitem[{Vellend(2010)}]{vellend_conceptual_2010}
Vellend, M. (2010).
\newblock Conceptual synthesis in community ecology.
\newblock \emph{The Quarterly Review of Biology}, 85, 183--206.

\bibitem[{Vellend(2016)}]{vellend_theory_2016}
Vellend, M. (2016).
\newblock \emph{The {Theory} of {Ecological} {Communities}}.
\newblock Princeton University Press, Princeton.

\bibitem[{Veran \emph{et~al.}(2015)Veran, Simpson, Sword, Deveson, Piry, Hines
  \& Berthier}]{VeranEtAl2015}
Veran, S., Simpson, S.~J., Sword, G.~A., Deveson, E., Piry, S., Hines, J.~E. \&
  Berthier, K. (2015).
\newblock Modeling spatiotemporal dynamics of outbreaking species: influence of
  environment and migration in a locust.
\newblock \emph{Ecology}, 96, 737--748.

\bibitem[{Viovy \& Saint(1994)}]{ViovySaint1994}
Viovy, N. \& Saint, G. (1994).
\newblock Hidden {M}arkov models applied to vegetation dynamics analysis using
  satellite remote sensing.
\newblock \emph{IEEE Transactions on Geoscience and Remote Sensing}, 32,
  906--917.

\bibitem[{Visser \& Speekenbrink(2010)}]{VisserSpeenkenbrink2010}
Visser, I. \& Speekenbrink, M. (2010).
\newblock depmix{S}4: an {R} package for hidden {M}arkov models.
\newblock \emph{Journal of Statistical Software}, 36, 1--21.

\bibitem[{Waggoner \& Stephens(1970)}]{WaggonerStephens1970}
Waggoner, P.~E. \& Stephens, G.~R. (1970).
\newblock Transition probabilities for a forest.
\newblock \emph{Nature}, 225, 1160.

\bibitem[{Wang(2007)}]{wang2007latent}
Wang, G. (2007).
\newblock On the latent state estimation of nonlinear population dynamics using
  {B}ayesian and non-{B}ayesian state-space models.
\newblock \emph{Ecological Modelling}, 200, 521--528.

\bibitem[{Weng \emph{et~al.}(2007)Weng, Boustany, Pyle, Anderson, Brown \&
  Block}]{weng2007migration}
Weng, K.~C., Boustany, A.~M., Pyle, P., Anderson, S.~D., Brown, A. \& Block,
  B.~A. (2007).
\newblock Migration and habitat of white sharks ({C}archarodon carcharias) in
  the eastern {P}acific {O}cean.
\newblock \emph{Marine Biology}, 152, 877--894.

\bibitem[{Werner \emph{et~al.}(2018)Werner, Cornelissen, Cornwell,
  Soudzilovskaia, Kattge, West \& Kiers}]{WernerEtAl2018}
Werner, G.~D., Cornelissen, J.~H., Cornwell, W.~K., Soudzilovskaia, N.~A.,
  Kattge, J., West, S.~A. \& Kiers, E.~T. (2018).
\newblock Symbiont switching and alternative resource acquisition strategies
  drive mutualism breakdown.
\newblock \emph{Proceedings of the National Academy of Sciences}, 115,
  5229--5234.

\bibitem[{White \& Burnham(1999)}]{WhiteBurnham1999}
White, G.~C. \& Burnham, K.~P. (1999).
\newblock Program {MARK}: Survival estimation from populations of marked
  animals.
\newblock \emph{Bird Study}, 46, S120--S138.

\bibitem[{White \& Garrott(2012)}]{WhiteGarrott2012}
White, G.~C. \& Garrott, R.~A. (2012).
\newblock \emph{Analysis of Wildlife Radio-Tracking Data}.
\newblock Elsevier.

\bibitem[{Wilkinson(2019)}]{Wilkinson2019}
Wilkinson, S. (2019).
\newblock aphid: an {R} package for analysis with profile hidden {M}arkov
  models.
\newblock \emph{Bioinformatics}, 35, 3829--3830.

\bibitem[{Williams \emph{et~al.}(2002)Williams, Nichols \&
  Conroy}]{WilliamsEtAl2002}
Williams, B.~K., Nichols, J.~D. \& Conroy, M.~J. (2002).
\newblock \emph{Analysis and Management of Animal Populations}.
\newblock Academic Press, San Diego, CA, USA.

\bibitem[{Winstrup \emph{et~al.}(2012)Winstrup, Svensson, Rasmussen, Winther,
  Steig \& Axelrod}]{WinstrupEtAl2012}
Winstrup, M., Svensson, A., Rasmussen, S.~O., Winther, O., Steig, E. \&
  Axelrod, A. (2012).
\newblock An automated approach for annual layer counting in ice cores.
\newblock \emph{Climate of the Past Discussions}, 8, 1881--1895.

\bibitem[{Wood(2010)}]{Wood2010}
Wood, S.~N. (2010).
\newblock Statistical inference for noisy nonlinear ecological dynamic systems.
\newblock \emph{Nature}, 466, 1102.

\bibitem[{Woolhiser \& Roldan(1982)}]{WoolhiserRoldan1982}
Woolhiser, D.~A. \& Roldan, J. (1982).
\newblock Stochastic daily precipitation models: 2. a comparison of
  distributions of amounts.
\newblock \emph{Water Resources Research}, 18, 1461--1468.

\bibitem[{Wootton(2001)}]{Wootton2001}
Wootton, J.~T. (2001).
\newblock Prediction in complex communities: analysis of empirically derived
  {M}arkov models.
\newblock \emph{Ecology}, 82, 580--598.

\bibitem[{Xiao \emph{et~al.}(2019)Xiao, Yu, Shi, Zhang, Yu, Li, Chen \&
  Gao}]{XiaoEtAl2019}
Xiao, R., Yu, X., Shi, R., Zhang, Z., Yu, W., Li, Y., Chen, G. \& Gao, J.
  (2019).
\newblock Ecosystem health monitoring in the {Shanghai-Hangzhou Bay
  Metropolitan Area}: A hidden {M}arkov modeling approach.
\newblock \emph{Environment International}, 133, 105170.

\bibitem[{Yackulic \emph{et~al.}(2020)Yackulic, Dodrill, Dzul, Sanderlin \&
  Reid}]{YackulicEtAl2020}
Yackulic, C.~B., Dodrill, M., Dzul, M., Sanderlin, J.~S. \& Reid, J.~A. (2020).
\newblock A need for speed in {B}ayesian population models: a practical guide
  to marginalizing and recovering discrete latent states.
\newblock \emph{Ecological Applications}, 30, e02112.

\bibitem[{Yau \emph{et~al.}(2011)Yau, Papaspiliopoulos, Roberts \&
  Holmes}]{yau2011bayesian}
Yau, C., Papaspiliopoulos, O., Roberts, G.~O. \& Holmes, C. (2011).
\newblock Bayesian non-parametric hidden {M}arkov models with applications in
  genomics.
\newblock \emph{Journal of the Royal Statistical Society: Series B (Statistical
  Methodology)}, 73, 37--57.

\bibitem[{Yoon(2009)}]{Yoon2009}
Yoon, B.-J. (2009).
\newblock Hidden {M}arkov models and their applications in biological sequence
  analysis.
\newblock \emph{Current Genomics}, 10, 402--415.

\bibitem[{Zipkin \emph{et~al.}(2010)Zipkin, Andrew~Royle, Dawson \&
  Bates}]{zipkin_multi-species_2010}
Zipkin, E.~F., Andrew~Royle, J., Dawson, D.~K. \& Bates, S. (2010).
\newblock Multi-species occurrence models to evaluate the effects of
  conservation and management actions.
\newblock \emph{Biological Conservation}, 143, 479--484.

\bibitem[{Zucchini \& Guttorp(1991)}]{ZucchiniGuttorp1991}
Zucchini, W. \& Guttorp, P. (1991).
\newblock A hidden {M}arkov model for space-time precipitation.
\newblock \emph{Water Resources Research}, 27, 1917--1923.

\bibitem[{Zucchini \emph{et~al.}(2016)Zucchini, MacDonald \&
  Langrock}]{ZucchiniEtAl2016}
Zucchini, W., MacDonald, I.~L. \& Langrock, R. (2016).
\newblock \emph{Hidden {M}arkov Models for Time Series: An Introduction Using
  R}.
\newblock 2nd edn. CRC Press.

\bibitem[{Zweig \emph{et~al.}(2020)Zweig, Newman \&
  Saunders}]{zweig2020applied}
Zweig, C.~L., Newman, S. \& Saunders, C.~J. (2020).
\newblock Applied use of alternate stable state modeling in restoration
  ecology.
\newblock \emph{Ecological Applications}, doi:10.1002/eap.2195.

\end{thebibliography}


\begin{thebibliography}{57}
\expandafter\ifx\csname natexlab\endcsname\relax\def\natexlab#1{#1}\fi
\expandafter\ifx\csname url\endcsname\relax
  \def\url#1{\texttt{#1}}\fi
\expandafter\ifx\csname urlprefix\endcsname\relax\def\urlprefix{URL }\fi

\bibitem[{Adam \emph{et~al.}(2019)Adam, Griffiths, Leos-Barajas, Meese, Lowe,
  Blackwell, Righton \& Langrock}]{AdamEtAl2019}
Adam, T., Griffiths, C.~A., Leos-Barajas, V., Meese, E.~N., Lowe, C.~G.,
  Blackwell, P.~G., Righton, D. \& Langrock, R. (2019).
\newblock Joint modelling of multi-scale animal movement data using
  hierarchical hidden {M}arkov models.
\newblock \emph{Methods in Ecology and Evolution}, 10, 1536--1550.

\bibitem[{Altman(2007)}]{Altman2007}
Altman, R.~M. (2007).
\newblock Mixed hidden {M}arkov models: an extension of the hidden {M}arkov
  model to the longitudinal data setting.
\newblock \emph{Journal of the American Statistical Association}, 102,
  201--210.

\bibitem[{Barbu \& Limnios(2009)}]{BarbuLimnios2009}
Barbu, V.~S. \& Limnios, N. (2009).
\newblock \emph{Semi-{M}arkov Chains and Hidden Semi-{M}arkov Models Toward
  Applications: Their Use in Reliability and {DNA} Analysis}, vol. 191 of
  \emph{Lecture Notes in Statistics}.
\newblock Springer.

\bibitem[{Bartolucci \emph{et~al.}(2017)Bartolucci, Pandolfi \&
  Pennoni}]{BartolucciEtAl2017}
Bartolucci, F., Pandolfi, S. \& Pennoni, F. (2017).
\newblock {LMest}: An {R} package for latent {M}arkov models for longitudinal
  categorical data.
\newblock \emph{Journal of Statistical Software}, 81, 1--38.

\bibitem[{Benhaiem \emph{et~al.}(2018)Benhaiem, Marescot, Hofer, East,
  Lebreton, Kramer-Schadt \& Gimenez}]{Benhaiem2018}
Benhaiem, S., Marescot, L., Hofer, H., East, M.~L., Lebreton, J.-D.,
  Kramer-Schadt, S. \& Gimenez, O. (2018).
\newblock Robustness of eco-epidemiological capture-recapture parameter
  estimates to variation in infection state uncertainty.
\newblock \emph{Frontiers in Veterinary Science}, 5, 197.

\bibitem[{Braun \emph{et~al.}(2018)Braun, Galuardi \& Thorrold}]{BraunEtAl2018}
Braun, C.~D., Galuardi, B. \& Thorrold, S.~R. (2018).
\newblock Hmmoce: An r package for improved geolocation of archival-tagged
  fishes using a hidden {M}arkov method.
\newblock \emph{Methods in Ecology and Evolution}, 9, 1212--1220.

\bibitem[{Bulla \& Bulla(2013)}]{BullaBulla2013}
Bulla, J. \& Bulla, I. (2013).
\newblock \emph{hsmm: Hidden Semi {M}arkov Models}.
\newblock \urlprefix\url{https://CRAN.R-project.org/package=hsmm}.
\newblock R package version 0.4.

\bibitem[{Choquet \emph{et~al.}(2009)Choquet, Rouan \&
  Pradel}]{ChoquetEtAl2009}
Choquet, R., Rouan, L. \& Pradel, R. (2009).
\newblock Program {E}-{SURGE}: a software application for fitting multievent
  models.
\newblock In: \emph{Modeling demographic processes in marked populations} (eds.
  Thomson, D.~L., Cooch, E.~G. \& Conroy, M.~J.). Springer, pp. 845--865.

\bibitem[{de~Valpine \emph{et~al.}(2017)de~Valpine, Turek, Paciorek,
  Anderson-Bergman, Lang \& Bodik}]{deValpine2017}
de~Valpine, P., Turek, D., Paciorek, C.~J., Anderson-Bergman, C., Lang, D.~T.
  \& Bodik, R. (2017).
\newblock Programming with models: writing statistical algorithms for general
  model structures with {NIMBLE}.
\newblock \emph{Journal of Computational and Graphical Statistics}, 26,
  403--413.

\bibitem[{DeRuiter \emph{et~al.}(2017)DeRuiter, Langrock, Skirbutas, Goldbogen,
  Calambokidis, Friedlaender \& Southall}]{DeRuiterEtAl2017}
DeRuiter, S.~L., Langrock, R., Skirbutas, T., Goldbogen, J.~A., Calambokidis,
  J., Friedlaender, A.~S. \& Southall, B.~L. (2017).
\newblock A multivariate mixed hidden {M}arkov model to analyze blue whale
  diving behaviour during controlled sound exposures.
\newblock \emph{The Annals of Applied Statistics}, 11, 362--392.

\bibitem[{Etienne \& Haegeman(2019)}]{EtienneHaegeman2019}
Etienne, R.~S. \& Haegeman, B. (2019).
\newblock \emph{DDD: Diversity-Dependent Diversification}.
\newblock \urlprefix\url{https://CRAN.R-project.org/package=DDD}.
\newblock R package version 4.1.

\bibitem[{Fiske \& Chandler(2011)}]{FiskeChandler2011}
Fiske, I. \& Chandler, R. (2011).
\newblock {unmarked}: an {R} package for fitting hierarchical models of
  wildlife occurrence and abundance.
\newblock \emph{Journal of Statistical Software}, 43, 1--23.

\bibitem[{Gelman \emph{et~al.}(2004)Gelman, Carlin, Stern \&
  Rubin}]{GelmanEtAl2004}
Gelman, A., Carlin, J.~B., Stern, H.~S. \& Rubin, D.~B. (2004).
\newblock \emph{Bayesian Data Analysis, 2nd Edition}.
\newblock Chapman and Hall, Boca Raton.

\bibitem[{Gelman \emph{et~al.}(2015)Gelman, Lee \& Guo}]{GelmanEtAl2015}
Gelman, A., Lee, D. \& Guo, J. (2015).
\newblock Stan: A probabilistic programming language for bayesian inference and
  optimization.
\newblock \emph{Journal of Educational and Behavioral Statistics}, 40,
  530--543.

\bibitem[{Gimenez \emph{et~al.}(2014)Gimenez, Blanc, Besnard, Pradel,
  Doherty~Jr, Marboutin \& Choquet}]{GimenezEtAl2014}
Gimenez, O., Blanc, L., Besnard, A., Pradel, R., Doherty~Jr, P.~F., Marboutin,
  E. \& Choquet, R. (2014).
\newblock Fitting occupancy models with {E-SURGE}: hidden {M}arkov modelling of
  presence--absence data.
\newblock \emph{Methods in Ecology and Evolution}, 5, 592--597.

\bibitem[{Glennie \emph{et~al.}(2019)Glennie, Borchers, Murchie, Harmsen \&
  Foster}]{glennie2019open}
Glennie, R., Borchers, D.~L., Murchie, M., Harmsen, B.~J. \& Foster, R.~J.
  (2019).
\newblock Open population maximum likelihood spatial capture-recapture.
\newblock \emph{Biometrics}, 75, 1345--1355.

\bibitem[{Goldstein \emph{et~al.}(2019)Goldstein, Turek, Ponisio \& {de
  Valpine}}]{GoldsteinEtAl2019}
Goldstein, B.~R., Turek, D., Ponisio, L. \& {de Valpine}, P. (2019).
\newblock \emph{{nimbleEcology}: Distributions for Ecological Models in
  'nimble'}.
\newblock \urlprefix\url{https://CRAN.R-project.org/package=nimbleEcology}.
\newblock R package version 0.1.0.

\bibitem[{Harte(2017)}]{Harte2017}
Harte, D. (2017).
\newblock \emph{Hidden{M}arkov: Hidden {M}arkov Models}.
\newblock Statistics Research Associates, Wellington.
\newblock \urlprefix\url{http://www.statsresearch.co.nz/dsh/sslib/}.
\newblock R package version 1.8-11.

\bibitem[{Helske \& Helske(2019)}]{HelskeHelske2019}
Helske, S. \& Helske, J. (2019).
\newblock Mixture hidden {M}arkov models for sequence data: The seq{HMM}
  package in {R}.
\newblock \emph{Journal of Statistical Software}, 88, 1--32.

\bibitem[{Himmelmann(2010)}]{Himmelmann2010}
Himmelmann, L. (2010).
\newblock \emph{HMM: Hidden {M}arkov Models}.
\newblock \urlprefix\url{https://CRAN.R-project.org/package=HMM}.
\newblock R package version 1.0.

\bibitem[{Hines(2006)}]{Hines2006}
Hines, J.~E. (2006).
\newblock \emph{{PRESENCE2} - Software to estimate patch occupancy and related
  parameters}.
\newblock \urlprefix\url{http://www.mbr-pwrc.usgs.gov/software/presence.html}.
\newblock USGS--PWRC.

\bibitem[{Jackson(2011)}]{Jackson2011}
Jackson, C.~H. (2011).
\newblock Multi-state models for panel data: the msm package for {R}.
\newblock \emph{Journal of statistical software}, 38, 1--29.

\bibitem[{Jonsen \emph{et~al.}(2005)Jonsen, Flemming \& Myers}]{JonsenEtAl2005}
Jonsen, I.~D., Flemming, J.~M. \& Myers, R.~A. (2005).
\newblock Robust state--space modeling of animal movement data.
\newblock \emph{Ecology}, 86, 2874--2880.

\bibitem[{Kellner \& Swihart(2014)}]{KellnerSwihart2014}
Kellner, K.~F. \& Swihart, R.~K. (2014).
\newblock Accounting for imperfect detection in ecology: a quantitative review.
\newblock \emph{PLoS ONE}, 9, e111436.

\bibitem[{Kendall(2009)}]{Kendall2009}
Kendall, W.~L. (2009).
\newblock One size does not fit all: adapting mark-recapture and occupancy
  models for state uncertainty.
\newblock In: \emph{Modeling demographic processes in marked populations}.
  Springer, pp. 765--780.

\bibitem[{Kendall \emph{et~al.}(2012)Kendall, White, Hines, Langtimm \&
  Yoshizaki}]{KendallEtAl2012}
Kendall, W.~L., White, G.~C., Hines, J.~E., Langtimm, C.~A. \& Yoshizaki, J.
  (2012).
\newblock Estimating parameters of hidden {M}arkov models based on marked
  individuals: use of robust design data.
\newblock \emph{Ecology}, 93, 913--920.

\bibitem[{K{\'e}ry \& Schaub(2011)}]{KerySchaub2011}
K{\'e}ry, M. \& Schaub, M. (2011).
\newblock \emph{Bayesian Population Analysis Using {WinBUGS}: a Hierarchical
  Perspective}.
\newblock Academic Press.

\bibitem[{Kristensen \emph{et~al.}(2016)Kristensen, Nielsen, Berg, Skaug \&
  Bell}]{KristensenEtAl2016}
Kristensen, K., Nielsen, A., Berg, C.~W., Skaug, H.~J. \& Bell, B. (2016).
\newblock Tmb: Automatic differentiation and laplace approximation.
\newblock \emph{Journal of Statistical Software}, 70, 1--21.

\bibitem[{Laake(2013)}]{Laake2013rmark}
Laake, J.~L. (2013).
\newblock \emph{{RM}ark: An {R} Interface for Analysis of Capture-Recapture
  Data with {MARK}}.
\newblock AFSC Processed Rep 2013-01, 25p. Alaska Fish. Sci. Cent., NOAA, Natl.
  Mar. Fish. Serv., 7600 Sand Point Way NE, Seattle WA 98115.

\bibitem[{Laake \emph{et~al.}(2013)Laake, Johnson \& Conn}]{LaakeEtAl2013}
Laake, J.~L., Johnson, D.~S. \& Conn, P.~B. (2013).
\newblock marked: an {R} package for maximum likelihood and {M}arkov chain
  {M}onte {C}arlo analysis of capture--recapture data.
\newblock \emph{Methods in Ecology and Evolution}, 4, 885--890.

\bibitem[{Leos-Barajas \emph{et~al.}(2017)Leos-Barajas, Gangloff, Adam,
  Langrock, Van~Beest, Nabe-Nielsen \& Morales}]{Leos-BarajasEtAl2017}
Leos-Barajas, V., Gangloff, E.~J., Adam, T., Langrock, R., Van~Beest, F.~M.,
  Nabe-Nielsen, J. \& Morales, J.~M. (2017).
\newblock Multi-scale modeling of animal movement and general behavior data
  using hidden {M}arkov models with hierarchical structures.
\newblock \emph{Journal of Agricultural, Biological and Environmental
  Statistics}, 22, 232--248.

\bibitem[{Louvrier \emph{et~al.}(2018)Louvrier, Chambert, Marboutin \&
  Gimenez}]{louvrier_accounting_2018}
Louvrier, J., Chambert, T., Marboutin, E. \& Gimenez, O. (2018).
\newblock Accounting for misidentification and heterogeneity in occupancy
  studies using hidden {Markov} models.
\newblock \emph{Ecological Modelling}, 387, 61--69.

\bibitem[{Lunn \emph{et~al.}(2012)Lunn, Jackson, Best, Spiegelhalter \&
  Thomas}]{LunnEtAl2012}
Lunn, D., Jackson, C., Best, N., Spiegelhalter, D. \& Thomas, A. (2012).
\newblock \emph{The {BUGS} book: A practical introduction to Bayesian
  analysis}.
\newblock Chapman and Hall/CRC.

\bibitem[{Lunn \emph{et~al.}(2009)Lunn, Spiegelhalter, Thomas \&
  Best}]{LunnEtAl2009}
Lunn, D., Spiegelhalter, D., Thomas, A. \& Best, N. (2009).
\newblock The bugs project: Evolution, critique and future directions.
\newblock \emph{Statistics in medicine}, 28, 3049--3067.

\bibitem[{MacKenzie \& Hines(2018)}]{MacKenzieHines2018}
MacKenzie, D. \& Hines, J. (2018).
\newblock \emph{RPresence: R Interface for Program PRESENCE}.
\newblock \urlprefix\url{https://www.mbr-pwrc.usgs.gov/software/presence.html}.
\newblock R package version 2.12.34.

\bibitem[{MacKenzie \emph{et~al.}(2018)MacKenzie, Nichols, Royle, Pollock,
  Bailey \& Hines}]{MackenzieEtAl2018}
MacKenzie, D.~I., Nichols, J.~D., Royle, J.~A., Pollock, K.~H., Bailey, L. \&
  Hines, J.~E. (2018).
\newblock \emph{Occupancy Estimation and Modeling: Inferring Patterns and
  Dynamics of Species Occurrence}.
\newblock 2nd edn. Elsevier.

\bibitem[{Marescot \emph{et~al.}(2018)Marescot, Benhaiem, Gimenez, Hofer,
  Lebreton, Olarte-Castillo, Kramer-Schadt \& East}]{MarescotEtAl2018}
Marescot, L., Benhaiem, S., Gimenez, O., Hofer, H., Lebreton, J.-D.,
  Olarte-Castillo, X.~A., Kramer-Schadt, S. \& East, M.~L. (2018).
\newblock Social status mediates the fitness costs of infection with canine
  distemper virus in {Serengeti} spotted hyenas.
\newblock \emph{Functional Ecology}, 32, 1237--1250.

\bibitem[{Marescot \emph{et~al.}(2019)Marescot, Lyet, Singh, Carter \&
  Gimenez}]{MarescotEtAl2019}
Marescot, L., Lyet, A., Singh, R., Carter, N. \& Gimenez, O. (2019).
\newblock Inferring wildlife poaching in southeast {A}sia with multispecies
  dynamic occupancy models.
\newblock \emph{Ecography}.

\bibitem[{Martin \emph{et~al.}(2005)Martin, Wintle, Rhodes, Kuhnert, Field,
  Low-Choy, Tyre \& Possingham}]{MartinEtAl2005}
Martin, T.~G., Wintle, B.~A., Rhodes, J.~R., Kuhnert, P.~M., Field, S.~A.,
  Low-Choy, S.~J., Tyre, A.~J. \& Possingham, H.~P. (2005).
\newblock Zero tolerance ecology: improving ecological inference by modelling
  the source of zero observations.
\newblock \emph{Ecology Letters}, 8, 1235--1246.

\bibitem[{McClintock \& Michelot(2018)}]{McClintockMichelot2018}
McClintock, B.~T. \& Michelot, T. (2018).
\newblock momentu{HMM}: {R} package for generalized hidden {M}arkov models of
  animal movement.
\newblock \emph{Methods in Ecology and Evolution}, 9, 1518--1530.

\bibitem[{Michelot \emph{et~al.}(2016)Michelot, Langrock \&
  Patterson}]{MichelotEtAl2016}
Michelot, T., Langrock, R. \& Patterson, T.~A. (2016).
\newblock move{HMM}: An {R} package for the statistical modelling of animal
  movement data using hidden {M}arkov models.
\newblock \emph{Methods in Ecology and Evolution}, 7, 1308--1315.

\bibitem[{Miller \emph{et~al.}(2011)Miller, Nichols, McClintock, Grant, Bailey
  \& Weir}]{MillerEtAl2011}
Miller, D.~A., Nichols, J.~D., McClintock, B.~T., Grant, E. H.~C., Bailey,
  L.~L. \& Weir, L.~A. (2011).
\newblock Improving occupancy estimation when two types of observational error
  occur: non-detection and species misidentification.
\newblock \emph{Ecology}, 92, 1422--1428.

\bibitem[{O'Connell \& H\o{}jsgaard(2011)}]{OConnellHojsgaard2011}
O'Connell, J. \& H\o{}jsgaard, S. (2011).
\newblock Hidden semi {M}arkov models for multiple observation sequences: The
  {mhsmm} package for {R}.
\newblock \emph{Journal of Statistical Software}, 39, 1--22.

\bibitem[{Patterson \emph{et~al.}(2017)Patterson, Parton, Langrock, Blackwell,
  Thomas \& King}]{patterson2017statistical}
Patterson, T.~A., Parton, A., Langrock, R., Blackwell, P.~G., Thomas, L. \&
  King, R. (2017).
\newblock Statistical modelling of individual animal movement: an overview of
  key methods and a discussion of practical challenges.
\newblock \emph{AStA Advances in Statistical Analysis}, 101, 399--438.

\bibitem[{Plummer(2003)}]{Plummer2003}
Plummer, M. (2003).
\newblock {JAGS}: A program for analysis of {B}ayesian graphical models using
  {G}ibbs sampling.
\newblock In: \emph{Proceedings of the 3rd international workshop on
  distributed statistical computing}, vol. 124. Vienna, Austria.

\bibitem[{Plummer(2019)}]{Plummer2019}
Plummer, M. (2019).
\newblock \emph{rjags: Bayesian Graphical Models using MCMC}.
\newblock \urlprefix\url{https://CRAN.R-project.org/package=rjags}.
\newblock R package version 4-9.

\bibitem[{Pradel(2005)}]{Pradel2005}
Pradel, R. (2005).
\newblock Multievent: An extension of multistate capture-recapture models to
  uncertain states.
\newblock \emph{Biometrics}, 61, 442--447.

\bibitem[{{R Core Team}(2019)}]{RCoreTeam2019}
{R Core Team} (2019).
\newblock \emph{R: A Language and Environment for Statistical Computing}.
\newblock R Foundation for Statistical Computing, Vienna, Austria.
\newblock \urlprefix\url{https://www.R-project.org/}.

\bibitem[{Santostasi \emph{et~al.}(2019)Santostasi, Ciucci, Caniglia, Fabbri,
  Molinari, Reggioni \& Gimenez}]{SantostasiEtAl2019}
Santostasi, N.~L., Ciucci, P., Caniglia, R., Fabbri, E., Molinari, L.,
  Reggioni, W. \& Gimenez, O. (2019).
\newblock Use of hidden {M}arkov capture--recapture models to estimate
  abundance in the presence of uncertainty: Application to the estimation of
  prevalence of hybrids in animal populations.
\newblock \emph{Ecology and evolution}, 9, 744--755.

\bibitem[{Schmidt \emph{et~al.}(2015)Schmidt, Johnson, Lindberg \&
  Adams}]{SchmidtEtAl2015}
Schmidt, J.~H., Johnson, D.~S., Lindberg, M.~S. \& Adams, L.~G. (2015).
\newblock Estimating demographic parameters using a combination of known-fate
  and open {N}-mixture models.
\newblock \emph{Ecology}, 96, 2583--2589.

\bibitem[{{Stan Development Team}(2019)}]{StanCoreTeam2019}
{Stan Development Team} (2019).
\newblock {RStan}: the {R} interface to {Stan}.
\newblock \urlprefix\url{http://mc-stan.org/}.
\newblock R package version 2.19.2.

\bibitem[{Sturtz \emph{et~al.}(2005)Sturtz, Ligges \& Gelman}]{SturtzEtAl2005}
Sturtz, S., Ligges, U. \& Gelman, A. (2005).
\newblock {R2WinBUGS}: A package for running {WinBUGS} from {R}.
\newblock \emph{Journal of Statistical Software}, 12, 1--16.

\bibitem[{Turek \emph{et~al.}(2016)Turek, de~Valpine \&
  Paciorek}]{TurekEtAl2016}
Turek, D., de~Valpine, P. \& Paciorek, C.~J. (2016).
\newblock Efficient {Markov chain Monte Carlo} sampling for hierarchical hidden
  {M}arkov models.
\newblock \emph{Environmental and Ecological Statistics}, 23, 549--564.

\bibitem[{Visser \& Speekenbrink(2010)}]{VisserSpeenkenbrink2010}
Visser, I. \& Speekenbrink, M. (2010).
\newblock depmix{S}4: an {R} package for hidden {M}arkov models.
\newblock \emph{Journal of Statistical Software}, 36, 1--21.

\bibitem[{White \& Burnham(1999)}]{WhiteBurnham1999}
White, G.~C. \& Burnham, K.~P. (1999).
\newblock Program {MARK}: Survival estimation from populations of marked
  animals.
\newblock \emph{Bird Study}, 46, S120--S138.

\bibitem[{Wilkinson(2019)}]{Wilkinson2019}
Wilkinson, S. (2019).
\newblock aphid: an {R} package for analysis with profile hidden {M}arkov
  models.
\newblock \emph{Bioinformatics}, 35, 3829--3830.

\bibitem[{Zucchini \emph{et~al.}(2016)Zucchini, MacDonald \&
  Langrock}]{ZucchiniEtAl2016}
Zucchini, W., MacDonald, I.~L. \& Langrock, R. (2016).
\newblock \emph{Hidden {M}arkov Models for Time Series: An Introduction Using
  R}.
\newblock 2nd edn. CRC Press.

\end{thebibliography}


\begin{thebibliography}{}

\bibitem[Ellison, 2004]{ellison2004bayesian}
Ellison, A.~M. (2004).
\newblock Bayesian inference in ecology.
\newblock {\em Ecology Letters}, 7(6):509--520.

\bibitem[Fahlbusch and Harrington, 2019]{fahlbusch2019low}
Fahlbusch, J.~A. and Harrington, K.~J. (2019).
\newblock A low-cost, open-source inertial movement {GPS} logger for
  eco-physiology applications.
\newblock {\em Journal of Experimental Biology}, 222(23):jeb211136.

\bibitem[Leos-Barajas et~al., 2017]{Leos-BarajasEtAl2017accelerometer}
Leos-Barajas, V., Photopoulou, T., Langrock, R., Patterson, T.~A., Watanabe,
  Y.~Y., Murgatroyd, M., and Papastamatiou, Y.~P. (2017).
\newblock Analysis of animal accelerometer data using hidden {M}arkov models.
\newblock {\em Methods in Ecology and Evolution}, 8(2):161--173.

\bibitem[Li and Bolker, 2017]{LiBolker2017}
Li, M. and Bolker, B.~M. (2017).
\newblock Incorporating periodic variability in hidden markov models for animal
  movement.
\newblock {\em Movement ecology}, 5(1):1.

\bibitem[MacKenzie et~al., 2002]{MackenzieEtAl2002}
MacKenzie, D.~I., Nichols, J.~D., Lachman, G.~B., Droege, S., Andrew~Royle, J.,
  and Langtimm, C.~A. (2002).
\newblock Estimating site occupancy rates when detection probabilities are less
  than one.
\newblock {\em Ecology}, 83(8):2248--2255.

\bibitem[McClintock and Michelot, 2018]{McClintockMichelot2018}
McClintock, B.~T. and Michelot, T. (2018).
\newblock momentu{HMM}: {R} package for generalized hidden {M}arkov models of
  animal movement.
\newblock {\em Methods in Ecology and Evolution}, 9(6):1518--1530.

\bibitem[Morales et~al., 2004]{MoralesEtAl2004}
Morales, J.~M., Haydon, D.~T., Frair, J., Holsinger, K.~E., and Fryxell, J.~M.
  (2004).
\newblock Extracting more out of relocation data: building movement models as
  mixtures of random walks.
\newblock {\em Ecology}, 85(9):2436--2445.

\bibitem[Myung, 2003]{Myung2003}
Myung, I.~J. (2003).
\newblock Tutorial on maximum likelihood estimation.
\newblock {\em Journal of Mathematical Psychology}, 47(1):90--100.

\bibitem[Newman et~al., 2014]{newman2014modelling}
Newman, K.~B., Buckland, S.~T., Morgan, B. J.~T., King, R., Borchers, D.~L.,
  Cole, D.~J., Besbeas, P., Gimenez, O., and Thomas, L. (2014).
\newblock {\em Modelling population dynamics: model formulation, fitting and
  assessment using state-space methods}.
\newblock Springer.

\bibitem[Patterson et~al., 2009]{PattersonEtAl2009}
Patterson, T.~A., Basson, M., Bravington, M.~V., and Gunn, J.~S. (2009).
\newblock Classifying movement behaviour in relation to environmental
  conditions using hidden {M}arkov models.
\newblock {\em Journal of Animal Ecology}, 78(6):1113--1123.

\bibitem[Patterson et~al., 2017]{patterson2017statistical}
Patterson, T.~A., Parton, A., Langrock, R., Blackwell, P.~G., Thomas, L., and
  King, R. (2017).
\newblock Statistical modelling of individual animal movement: an overview of
  key methods and a discussion of practical challenges.
\newblock {\em AStA Advances in Statistical Analysis}, 101(4):399--438.

\bibitem[Pohle et~al., 2017]{pohle2017selecting}
Pohle, J., Langrock, R., van Beest, F.~M., and Schmidt, N.~M. (2017).
\newblock Selecting the number of states in hidden {M}arkov models: pragmatic
  solutions illustrated using animal movement.
\newblock {\em Journal of Agricultural, Biological and Environmental
  Statistics}, 22(3):270--293.

\bibitem[Qasem et~al., 2012]{qasem2012tri}
Qasem, L., Cardew, A., Wilson, A., Griffiths, I., Halsey, L.~G., Shepard,
  E.~L., Gleiss, A.~C., and Wilson, R. (2012).
\newblock Tri-axial dynamic acceleration as a proxy for animal energy
  expenditure; should we be summing values or calculating the vector?
\newblock {\em PLoS ONE}, 7(2).

\bibitem[{R Core Team}, 2019]{RCoreTeam2019}
{R Core Team} (2019).
\newblock {\em R: A Language and Environment for Statistical Computing}.
\newblock R Foundation for Statistical Computing, Vienna, Austria.

\bibitem[Rabiner, 1989]{Rabiner1989}
Rabiner, L.~R. (1989).
\newblock A tutorial on hidden {M}arkov models and selected applications in
  speech recognition.
\newblock {\em Proceedings of the IEEE}, 77(2):257--286.

\bibitem[Turek et~al., 2016]{TurekEtAl2016}
Turek, D., de~Valpine, P., and Paciorek, C.~J. (2016).
\newblock Efficient {Markov chain Monte Carlo} sampling for hierarchical hidden
  {M}arkov models.
\newblock {\em Environmental and Ecological Statistics}, 23(4):549--564.

\bibitem[White and Burnham, 1999]{WhiteBurnham1999}
White, G.~C. and Burnham, K.~P. (1999).
\newblock Program {MARK}: Survival estimation from populations of marked
  animals.
\newblock {\em Bird Study}, 46:S120--S138.

\bibitem[Yackulic et~al., 2020]{YackulicEtAl2020}
Yackulic, C.~B., Dodrill, M., Dzul, M., Sanderlin, J.~S., and Reid, J.~A.
  (2020).
\newblock A need for speed in {B}ayesian population models: a practical guide
  to marginalizing and recovering discrete latent states.
\newblock {\em Ecological Applications}, 30:e02112.

\bibitem[Zucchini, 2000]{zucchini2000introduction}
Zucchini, W. (2000).
\newblock An introduction to model selection.
\newblock {\em Journal of mathematical psychology}, 44(1):41--61.

\bibitem[Zucchini et~al., 2016]{ZucchiniEtAl2016}
Zucchini, W., MacDonald, I.~L., and Langrock, R. (2016).
\newblock {\em Hidden {M}arkov Models for Time Series: An Introduction Using
  R}.
\newblock CRC Press, second edition.

\end{thebibliography}

\end{document}


\sloppy

\begin{titlepage}
\begin{singlespacing}
\begin{center}
\huge{\textbf{Uncovering ecological state dynamics with hidden Markov models \\ --- \\ Supplementary Material }}\\[0.6cm]
\large{Brett T.\ McClintock}\\
\normalsize{NOAA National Marine Fisheries Service, U.S.A.\\ brett.mcclintock@noaa.gov}\\[0.2cm]
\large{Roland Langrock}\\
\normalsize{Department of Business Administration and Economics, Bielefeld University\\ roland.langrock@uni-bielefeld.de}\\[0.2cm]
\large{Olivier Gimenez}\\
\normalsize{CNRS Centre d'Ecologie Fonctionnelle et Evolutive, France\\ olivier.gimenez@cefe.cnrs.fr}\\[0.2cm]
\large{Emmanuelle Cam}\\
\normalsize{Laboratoire des Sciences de l'Environnement Marin, Institut Universitaire Europ\'een de la Mer, Univ. Brest, CNRS, IRD, Ifremer, France\\ Emmanuelle.Cam@univ-brest.fr}\\[0.2cm]
\large{David L.\ Borchers}\\
\normalsize{School of Mathematics and Statistics, University of St Andrews\\ dlb@st-andrews.ac.uk}\\[0.2cm]
\large{Richard Glennie}\\
\normalsize{School of Mathematics and Statistics, University of St Andrews\\ rg374@st-andrews.ac.uk}\\[0.2cm]
\large{Toby A.\ Patterson}\\
\normalsize{CSIRO Oceans and Atmosphere, Australia\\ toby.patterson@csiro.au}\\[0.45cm]
\end{center}
\end{singlespacing}

\end{titlepage}

\onehalfspacing

\appendix
\renewcommand{\thesection}{Appendix \Alph{section}:}
\setcounter{section}{0}

\section{Dynamic species co-existence HMM}
Here we provide additional details of an HMM formulation for species co-existence dynamics based on presence-absence data \citep[][]{MarescotEtAl2019}. Let the states $S_t=\text{A}$ (respectively $S_t=\text{B}$ and $S_t=\text{AB}$) indicate ``site occupied by species A'' (respectively by species B and by both species) and $S_t=\text{U}$ indicate ``unoccupied site''. Define $X_{t,k} \in \{0,1,2,3\}$, where $0$ indicates neither species was detected, $1$ indicates only species A was detected, $2$ indicates only species B was detected, and $3$ indicates both species were detected on the $k$th visit at time $t$. We could for example have:\bigskip
\begin{center}
\begin{tikzpicture}[node distance = 1.5cm]
\tikzset{state/.style = {circle, draw, minimum size = 35pt, scale = 0.75}}
\node [state,fill=lightgray!25] (3) {$\cdots$};
\node [state,fill=lightgray!25] (4) [right=8mm of 3] {U};
\node [state,fill=lightgray!25] (5) [right=24mm of 4] {B};
\node [state,fill=lightgray!25] (6) [right=24mm of 5] {AB};
\node [state,fill=lightgray!25] (7) [right=8mm of 6] {$\cdots$};
\node [state,fill=lightgray!25] (14) [above left = 6mm of 4] {$0$};
\node [state,fill=lightgray!25] (24) [above = 6mm of 4] {$\cdots$};
\node [state,fill=lightgray!25] (34) [above right = 6mm of 4] {$0$};
\node [state,fill=lightgray!25] (15) [above left = 6mm of 5] {$0$};
\node [state,fill=lightgray!25] (25) [above = 6mm of 5] {$\cdots$};
\node [state,fill=lightgray!25] (35) [above right = 6mm of 5] {$2$};
\node [state,fill=lightgray!25] (17) [above left = 6mm of 6] {$1$};
\node [state,fill=lightgray!25] (27) [above = 6mm of 6] {$\cdots$};
\node [state,fill=lightgray!25] (37) [above right = 6mm of 6] {$3$};
\node [text=black,] (20) [right = 4mm of 7] {hidden $S_t \in \{\text{AB,A,B,U}\}$};
\node [text=black,] (21) [above = 10mm of 20] {\hspace*{2.5mm} observed $X_{t,k} \in \{0,1,2,3\}$ at};
\node [text=black,] (22) [above = 4mm of 20] {\hspace*{2.5mm} multiple visits $k = 1, \ldots, K$};
\draw[->,black, line width=0.25mm,-latex] (3) to (4);
\draw[->,black, line width=0.25mm,-latex] (4) to (5);
\draw[->,black, line width=0.25mm,-latex] (5) to (6);
\draw[->,black, line width=0.25mm,-latex] (6) to (7);
\draw[->,black, line width=0.25mm,-latex] (4) to (14);
\draw[->,black, line width=0.25mm,-latex] (4) to (24);
\draw[->,black, line width=0.25mm,-latex] (4) to (34);
\draw[->,black, line width=0.25mm,-latex] (5) to (15);
\draw[->,black, line width=0.25mm,-latex] (5) to (25);
\draw[->,black, line width=0.25mm,-latex] (5) to (35);
\draw[->,black, line width=0.25mm,-latex] (6) to (17);
\draw[->,black, line width=0.25mm,-latex] (6) to (27);
\draw[->,black, line width=0.25mm,-latex] (6) to (37);
\node [text=black,] (40) [below = 4.2mm of 4] {\it \color{gray} $t-1$};
\node [text=black,] (41) [below = 4.2mm of 5] {\it \color{gray} $t$};
\node [text=black,] (42) [below = 4.2mm of 6] {\it \color{gray} $t+1$};
\node [text=black,] (45) [below = 6mm of 20] {\it \color{gray} time};
\end{tikzpicture}
\end{center}

\begin{equation*}
    \begin{blockarray}{cccccl}
    & S_1=\text{AB} & S_1=\text{A} & S_1=\text{B} & S_1=\text{U} \\
    \begin{block}{c(cccc)l}
    \boldsymbol{\delta}= & \psi^{AB}  & \psi^A  & \psi^B & 1 - \psi^{AB} - \psi^A - \psi^B \\
    \end{block}
    \end{blockarray}
\end{equation*}
\begin{equation*}
    \boldsymbol{\Gamma}=
    \begin{blockarray}{ccccc}
    S_{t+1}=\text{AB} & S_{t+1}=\text{A} & S_{t+1}=\text{B} & S_{t+1}=\text{U} \\
    \begin{block}{[cccc]c}
    1-\epsilon^{AB}-\epsilon^A+\epsilon^B & \epsilon^B & \epsilon^A & \epsilon^{AB} & S_t=\text{AB}\\
    \eta^{B} & 1-\omega^A-\eta^{B}-\nu^{A} & \omega^A & \nu^{A} &  S_t=\text{A}\\
    \eta^{A} & \omega^B & 1-\omega^B-\eta^{A}-\nu^{B} & \nu^{B} &  S_t=\text{B}\\
    \gamma^{AB} & \gamma^{A} & \gamma^{B} & 1 - \gamma^{A} - \gamma^{B} - \gamma^{AB} &  S_t=\text{U}\\
    \end{block}
    \end{blockarray}
\end{equation*}
with diagonal elements of $\mathbf{P}(\mathbf{x}_t)$
\begin{eqnarray*}
    f(\mathbf{x}_t \mid S_t=\text{AB})  & = & \prod_{k=1}^K  r_{Ab}^{I(x_{t,k}=1)}(1-r_{Ab})^{1-I(x_{t,k}=1)} \\
                      & &    + r_{aB}^{I(x_{t,k}=2)}(1-r_{aB})^{1-I(x_{t,k}=2)}\\
                      & &    + r_{AB}^{I(x_{t,k}=3)}(1-r_{AB})^{1-I(x_{t,k}=3)}\\
   f(\mathbf{x}_t \mid S_t=\text{A}) & = & \prod_{k=1}^K p_A^{I(x_{t,k}=1)}(1-p_A)^{1-I(x_{t,k}=1)}\\
    f(\mathbf{x}_t \mid S_t=\text{B}) & = & \prod_{k=1}^K p_B^{I(x_{t,k}=2)}(1-p_B)^{1-I(x_{t,k}=2)}\\
    f(\mathbf{x}_t \mid S_t=\text{U}) & = & \prod_{k=1}^K I(x_{t,k}=0)
\end{eqnarray*}
where $\psi_A$ (respectively $\psi_B$) is the probability of only species A (respectively B) being present, $\psi_{AB}$ is the probability of both species being present, $p_A$ (respectively $p_B$) is probability of detecting species A given only species A is present, $r_{AB}$ is the probability of detecting both species given both species are present, $r_{Ab}$ is the probability of detecting species A, not B, given both species are present, and $r_{aB}$ is the probability of detecting species B, not A, given both species are present. The state transition probability matrix ${\boldsymbol\Gamma}$ is composed of the following parameters:
\begin{itemize}
    \item $\epsilon_{AB}$ is the probability that both species A and B go locally extinct between $t$ and $t+1$;
    \item $\epsilon_{A}$ (respectively $\epsilon_{B}$) is the probability that species A goes locally extinct between $t$ and $t+1$, given both species are present at $t$;
    \item $\nu_A$ (respectively $\nu_B$) is the probability that species A goes locally extinct between $t$ and $t+1$, given species B was absent at $t$ and $t+1$;
    \item $\gamma_{AB}$ is the probability that both species A and B colonise a site between $t$ and $t+1$;
    \item $\gamma_{A}$ (respectively $\gamma_{B}$) is the probability that species A colonises a site between $t$ and $t+1$, given both species are absent at $t$;
    \item $\eta_A$ (respectively $\eta_B$) is the probability that species A colonises a site between $t$ and $t + 1$, given species B was present at $t$ and $t + 1$;
    \item $\omega_A$ (respectively $\omega_B$) is the probability that species A is replaced by B between $t$ and $t + 1$.
\end{itemize}

\section{HMM software}
The computational machinery of 
HMMs, such as the forward and Viterbi algorithms, 
can be coded from scratch by a proficient statistical programmer \citep[e.g.][]{ZucchiniEtAl2016,louvrier_accounting_2018,SantostasiEtAl2019}, but recent advances in computing power and 
user-friendly software have made the implementation of HMMs much more feasible for practitioners. Many different HMM software packages and stand-alone programs are now available, some of which are focused on specific classes of state dynamics within the individual, population, or community level of the ecological hierarchy. However, the features and capabilities of the software are varied, and it can be challenging to determine which software may be most appropriate for a specific objective. Here we will describe some of the most popular HMM software currently available, including potential advantages and disadvantages for ecological applications. We limit our treatment to freely available R \citep{RCoreTeam2019} packages and stand-alone programs that we believe are most accessible to ecologists and non-statisticians.

The Comprehensive R Archive Network (\verb|https://cran.r-project.org|) currently hosts 26 packages that include ``hidden Markov'' in their description. While most HMM packages in R include data simulation, parameter estimation, and state decoding for an arbitrary number of system states
, they differ in many key respects (see Table 
2 in main text). Most of the packages are focused on categorical sequence analysis and are therefore limited in the state-dependent probability distributions that can be implemented \citep[][]{Himmelmann2010,BartolucciEtAl2017,HelskeHelske2019,Wilkinson2019}. However, some of the more general packages provide greater flexibility for specifying state-dependent probability distributions, including commonly used discrete (e.g.\ binomial, Poisson) and continuous (e.g.\ gamma, normal) distributions \citep[][]{VisserSpeenkenbrink2010, Jackson2011, Harte2017, McClintockMichelot2018}. One of the earliest and most flexible HMM packages, \verb|depmixS4| \citep{VisserSpeenkenbrink2010}, includes a broad range of probability distributions and can accommodate multivariate HMMs, multiple observation sequences (e.g.\ from multiple individuals or sites), parameter covariates, parameter constraints, and missing observations. With additional features originally motivated by animal movement HMMs \citep[][]{MichelotEtAl2016}, \verb|momentuHMM| \citep{McClintockMichelot2018} is similar to \verb|depmixS4| in terms of features and flexibility, but can also be used to implement mixed HMMs \citep[][]{DeRuiterEtAl2017}, hierarchical HMMs \citep[][]{Leos-BarajasEtAl2017,AdamEtAl2019}, zero-inflated probability distributions \citep[][]{MartinEtAl2005}, and partially-observed state sequences. However, unlike \verb|depmixS4| and other packages such as \verb|mhsmm| \citep{OConnellHojsgaard2011} and \verb|HiddenMarkov| \citep{Harte2017}, \verb|momentuHMM| does not currently support custom-coded state-dependent probability distributions. To our knowledge, only \verb|hsmm| \citep{BullaBulla2013} and \verb|mhsmm| \citep{OConnellHojsgaard2011} can currently implement hidden semi-Markov models \citep[][]{BarbuLimnios2009}. 

Many R packages are less general and specialise on specific HMM applications within individual- or population-level ecology. The \verb|marked| package \citep{LaakeEtAl2013} implements many of the popular capture-recapture HMMs described in Section 
3.1. Packages that specialise in animal movement behaviour HMMs for telemetry data, such as those described in Section 
3.1.2, include \verb|bsam| \citep{JonsenEtAl2005}, \verb|moveHMM| \citep{MichelotEtAl2016}, and \verb|momentuHMM| \citep{McClintockMichelot2018}. The package \verb|HMMoce| \citep{BraunEtAl2018} is specifically catered for HMMs that infer location from archival tag data (e.g.\ light levels, depth-temperature profiles) such as those described in Section 
3.1.3. Using telemetry and count data, \verb|kfdnm| \citep{SchmidtEtAl2015} can fit HMMs for population abundance and related demographic parameters such as those described in Section 
3.2. The package \verb|DDD| \citep{EtienneHaegeman2019} implements HMMs for macroevolutionary inference about diversification rates from phylogenetic trees such as those described in Section 3.2.2. The package \verb|openpopscr| \citep{glennie2019open} can fit spatial capture-recapture HMMs that account for unobserved animal movements when estimating population-level density and survival, such as those described in Section 
3.2.3. The popular package \verb|unmarked| \citep{FiskeChandler2011} includes many of the HMMs for inferring patterns and dynamics of species occurrence from repeated presence-absence data that were described in Section 
3.2.3. 

There are also several stand-alone, user-friendly software programs that focus on specific HMM applications in ecology. Programs \verb|MARK| \citep{WhiteBurnham1999} and \verb|E-SURGE| \citep{ChoquetEtAl2009} both provide a very general framework for implementing HMMs with individual-level capture-recapture \citep[][]{Pradel2005} or population-level presence-absence \citep[][]{GimenezEtAl2014} data, including observation process error arising from non-detection \citep[][]{KellnerSwihart2014}, state uncertainty \citep[][]{Kendall2009,KendallEtAl2012}, and species misidentfication \citep[][]{MillerEtAl2011}. Program \verb|PRESENCE| \citep{Hines2006} has many of the features of \verb|MARK| and \verb|E-SURGE| but focuses solely on presence-absence data, including models for species co-occurrence dynamics \citep[][]{MackenzieEtAl2018}. The \verb|RMark| \citep{Laake2013rmark} and \verb|RPresence| \citep{MacKenzieHines2018} packages have been developed as R interfaces for Programs \verb|MARK| and \verb|PRESENCE|, respectively.

Although not intended specifically for HMMs, it is worth noting that there are a number of software programs with which these types of models can be relatively easily implemented by users with minimal statistical programming experience. For Bayesian inference using MCMC sampling \citep[][]{GelmanEtAl2004}, these include \verb|WinBUGS|/\verb|OpenBUGS| \citep{LunnEtAl2009,KerySchaub2011,LunnEtAl2012}, \verb|JAGS| \citep{Plummer2003}, and \verb|Stan| \citep{GelmanEtAl2015}. There are R package interfaces for all of these programs, including \verb|R2OpenBUGS| \citep{SturtzEtAl2005}, \verb|rjags| \citep{Plummer2019}, and \verb|rstan| \citep{StanCoreTeam2019}, respectively. The R package \verb|nimble| \citep{deValpine2017} and its \verb|nimbleEcology| extension for common HMMs in ecology \citep{GoldsteinEtAl2019} use a statistical programming language similar to \verb|BUGS| and can be used for Bayesian or maxmimum likelihood inference. The R package \verb|TMB| \citep{KristensenEtAl2016} generally has a steeper learning curve but can be advantageous for maximum likelihood inference \citep[e.g.][]{Benhaiem2018,MarescotEtAl2018}, particularly for mixed HMMs that include continuous-valued random effects \citep[][]{Altman2007}. From a computational point of view, 
neither maximum likelihood estimation nor MCMC sampling is vastly superior, and in practice users will typically adopt the approach they are most comfortable with (cf.\ \citealp{patterson2017statistical}, for a more comprehensive discussion). However, 
MCMC samplers that include both the parameter vector $({\boldsymbol \theta})$ and the latent states $(S_1,\ldots,S_T)$, as commonly implemented in \verb|WinBUGS|/\verb|OpenBUGS| and \verb|JAGS|
, 
are inherently slow; sampling from the parameter vector only 
while applying the forward algorithm to evaluate the likelihood will often be preferable \citep[][]{TurekEtAl2016}.

\bibliographystyle{ecology_letters2}
\bibliography{master}


\sloppy

\begin{titlepage}
\begin{singlespacing}
\begin{center}
\huge{\textbf{Uncovering ecological state dynamics with hidden Markov models \\ --- \\ Supplementary Tutorial }}\\[0.6cm]
\large{Brett T.\ McClintock}\\
\normalsize{NOAA National Marine Fisheries Service, U.S.A.\\ brett.mcclintock@noaa.gov}\\[0.2cm]
\large{Roland Langrock}\\
\normalsize{Department of Business Administration and Economics, Bielefeld University\\ roland.langrock@uni-bielefeld.de}\\[0.2cm]
\large{Olivier Gimenez}\\
\normalsize{CNRS Centre d'Ecologie Fonctionnelle et Evolutive, France\\ olivier.gimenez@cefe.cnrs.fr}\\[0.2cm]
\large{Emmanuelle Cam}\\
\normalsize{Laboratoire des Sciences de l'Environnement Marin, Institut Universitaire Europ\'een de la Mer, Univ. Brest, CNRS, IRD, Ifremer, France\\ Emmanuelle.Cam@univ-brest.fr}\\[0.2cm]
\large{David L.\ Borchers}\\
\normalsize{School of Mathematics and Statistics, University of St Andrews\\ dlb@st-andrews.ac.uk}\\[0.2cm]
\large{Richard Glennie}\\
\normalsize{School of Mathematics and Statistics, University of St Andrews\\ rg374@st-andrews.ac.uk}\\[0.2cm]
\large{Toby A.\ Patterson}\\
\normalsize{CSIRO Oceans and Atmosphere, Australia\\ toby.patterson@csiro.au}\\[0.45cm]
\end{center}
\end{singlespacing}

\end{titlepage}

\onehalfspacing

\tableofcontents

\section{Introduction}  
In the main text, we review how hidden Markov models (HMMs) are used to uncover ecological dynamics that operate at the individual, population, community, and ecosystem levels. The breadth of application of HMMs in ecology shows that their general structure can reflect ecological processes or how we understand and summarise ecological processes. Nevertheless, formulating an HMM that faithfully represents the ecological dynamics under study, fitting this HMM to real data, and checking that the HMM is a good model to use are non-trivial steps in any analysis. In this supplementary tutorial, we introduce the main steps in an HMM analysis, highlighting the decisions and assumptions to be considered along the way.  

\begin{figure}[!h]
\centering 
\includegraphics[width=\textwidth]{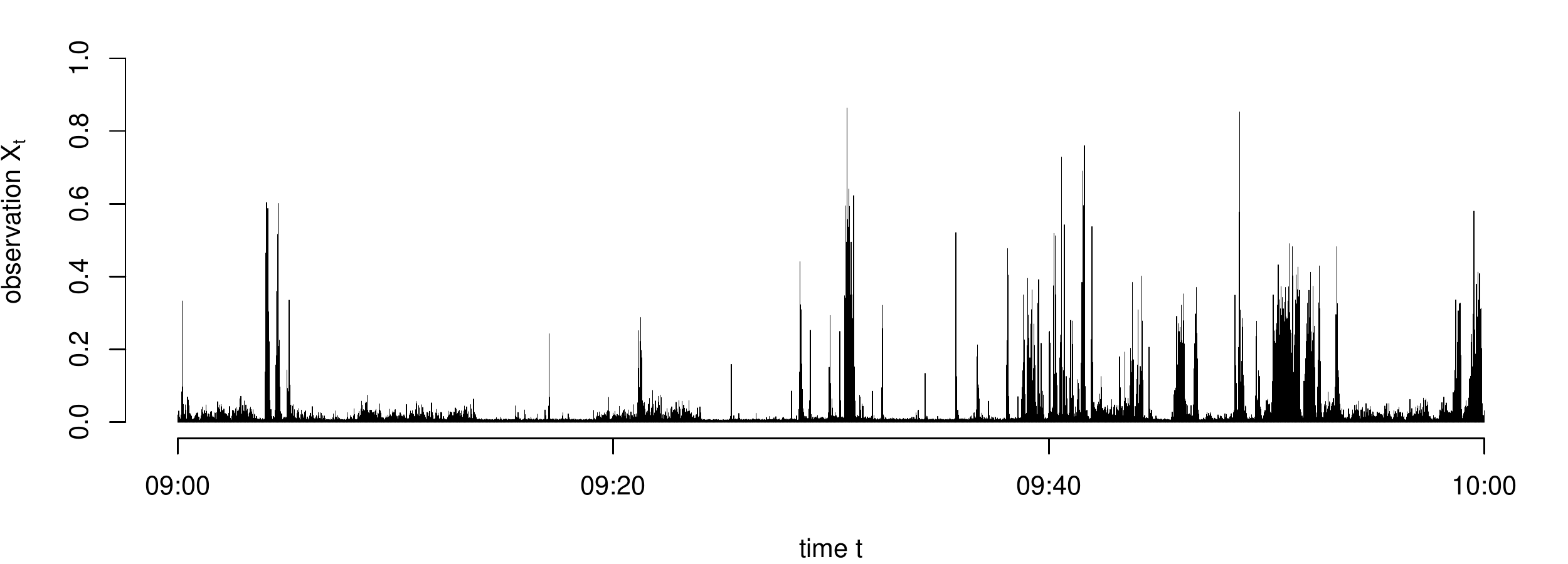}
\caption{Vectorial dynamic body acceleration $X_t$, $t=1,\ldots,3600$, for a striated caracara (\textit{Phalcoboenus australis}) measured over one hour at 1 Hz \citep{fahlbusch2019low}.}
\label{fig:data}
\end{figure}

As demonstrated in the main text, each application of an HMM is particular to the problem under study. Here we illustrate how to tailor an HMM to a given study, thereby showcasing the inferential tools available for HMMs. We consider a single application, introduced in the main text: a time series of the vectorial sum of the dynamic body acceleration (VDBA) of a striated caracara (\textit{Phalcoboenus australis}) measured every second over one hour \citep{fahlbusch2019low}. VDBA is a measure of the overall activity of an animal: it is the length of the three-dimensional vector defined by the dynamic accelerations in all three coordinate directions \citep{qasem2012tri}. Figure \ref{fig:data} shows the VDBA time series considered here.

We consider the full analysis cycle, comprising the formulation of an HMM, the estimation of its parameters, the selection between different candidate models, checking how well the model fits the data, inferring the latent states, and interpreting the final results. The supplemental R \citep{RCoreTeam2019} script \verb|caracaraExample.R| can be used to completely reproduce and further explore this illustrative example using the package \verb|momentuHMM| \citep{McClintockMichelot2018}.

\section{Building the HMM}  

The observations are realisations of the sequential process $X_1, \ldots, X_T$, often a time series, which we call the observation process (or state-dependent process). HMMs comprise two parts: a model for the observation process and a model for the state process underlying the observation process. Each part has associated \emph{assumptions} which induce both the simplicity that HMMs are favoured for and the limitations that may inhibit their ability to faithfully characterise either process. 

\subsection{State process}

The state process $S_1, \ldots, S_T$ is modelled as a Markov chain over a set of $N$ states, i.e.\ $S_t\in \{ 1,\ldots,N\}$ for $t=1,\ldots,T$. The key assumption involved here is that the state process is Markovian: the probability of which state occurs next in time is known conditional on the current state, irrespective of the states that occurred in the past. As a consequence, the way the process evolves over time is completely determined by the $N \times N$ state transition probability matrix, which comprises the probabilities of switching from any state $i$ at time $t$ to any state $j$ at time $t+1$, for $i,j=1,\ldots,N$. The Markov property also implies that the time spent in each state has a geometric distribution. Clearly, this is a strong assumption on the memory of the ecological process and so it is important to remember this specification in light of interpreting results. 

Specifying the state process involves selecting the number of states as well as any potential pre-determined structure in the transition probability matrix. To make either of these decisions, one must first consider what latent ecological process drives the observations and whether such a process, or a simplified version of it, can be described by a Markov chain taking finitely many states. The idea behind specifying the model for the state process is to create a structure that is likely to capture the pattern in the underlying ecological dynamics that the observations evince, with the caveat that beyond any pre-defined structure the state characteristics are driven solely by data and may, in the end, not necessarily conform with our original intentions for them. 

For the VDBA example, we must rely on our understanding of the movement of the study species. It is common to consider the activity level of an animal to vary across different behavioural modes (see Section 3.1.3 in the main text). The model states are thus intended to correspond with these behaviours. Whereas in some applications of HMMs the choice of the number of states can be obvious, such as in the Cormack-Jolly-Seber model with the two states ``alive'' and ``dead'' (see Section 3.1.1 in the main text), there are also such where it is not clear {\it a priori} how many states there are. This is indeed the case in the caracara example considered here. In cases such as these the choice of $N$ should be guided by expert knowledge on the subject matter. In the given VDBA example, the research biologists who collected the data expected $N=4$ states, corresponding to resting behaviour, minimal activity such as preening, moderate activity such as walking or digging, and flight, to most adequately reflect the variation in activity.
In general, selecting the maximum number of states to consider is a trade-off between computational feasibility, ecological knowledge of the system, and fit to the data. \citet{pohle2017selecting} provide a step-by-step guide to selecting the number of states for an HMM and discuss why relying solely on statistical model selection methods, such as the Akaike information criterion (AIC) or the Bayesian information criterion (BIC), may not be desirable. 

Along with the number of states, we specify that it is possible for the animal to switch from any one behaviour to any other and so the transition probability matrix has unknown parameters for all entries: 
\begin{equation*}
\boldsymbol{\Gamma}=   
    \begin{blockarray}{ccccc}
    S_{t+1}=1 & S_{t+1}=2 & S_{t+1}=3 & S_{t+1}=4  & \\
    \begin{block}{[cccc]c}
    \gamma_{1,1} & \gamma_{1,2} & \gamma_{1,3} & \gamma_{1,4} & S_t=1 \\
    \gamma_{2,1} & \gamma_{2,2} & \gamma_{2,3} & \gamma_{2,4} & S_t=2 \\
    \gamma_{3,1} & \gamma_{3,2} & \gamma_{3,3} & \gamma_{3,4} & S_t=3 \\
    \gamma_{4,1} & \gamma_{4,2} & \gamma_{4,3} & \gamma_{4,4} & S_t=4 \\
    \end{block}
    \end{blockarray}.
\end{equation*}
Further to this, we specify the distribution of the states at time $t=1$, hence the probabilities $\delta_i=\Pr(S_1=i)$ of the animal inhabiting any state $i$, $i=1,\ldots,N$, at the start of the sequence, to be the stationary distribution of the Markov chain  (see \citealp{ZucchiniEtAl2016}, p.\ 17). 

This is the most general transition probability matrix to consider. In other contexts, it can be appropriate to limit transitions between states. For example, in Cormack-Jolly-Seber models the states are intended to represent ``alive" and ``dead", so it is necessary to forbid transitions from the ``dead" to ``alive" state by enforcing a zero value in the corresponding entry of $\bm\Gamma$. Additional structure, such as temporal heterogeneity in the transition probability matrix $(\boldsymbol{\Gamma}_t)$, can also be incorporated through the use of explanatory covariates \citep[e.g.][]{MoralesEtAl2004,LiBolker2017}. This is usually accomplished using a multinomal-logit link function:
\begin{equation*}
    \gamma_{t,i,j}=\frac{\exp\left(\beta_{0,i,j}+\sum_{k=1}^K z_{k,t} \beta_{k,i,j}\right)}{\sum_{l=1}^N \exp\left(\beta_{0,i,l}+\sum_{k=1}^K z_{k,t} \beta_{k,i,l}\right)},
\end{equation*}
where $\gamma_{t,i,j}$ is the time-dependent transition probability from state $i$ at time $t$ to state $j$ at time $t+1$, $\beta_{0,i,j}$ is an intercept term, $z_{k,t}$ is the value of the $k$th covariate at time $t$, and $\beta_{k,i,j}$ is a slope term for the $k$th covariate. Because $\sum_{j=1}^N \gamma_{t,i,j}=1$, it is customary to set $\alpha_{i,j}=\beta_{k,i,j}=0$ for $i=j$ and $k=1,\ldots,K$ to avoid overparameterisation when estimating these parameters (see Section \ref{sec:estimation}). This extension is not implemented in the given example, but see \citet{WhiteBurnham1999}, \citet{MackenzieEtAl2002}, and \citet{PattersonEtAl2009} for common examples of link functions in HMMs.

\subsection{Observation process} 

Conditional on the state process, each observation arises as a sample from a distribution --- selected from $N$ possible distributions according to the current underlying state --- that is unrelated to previous states or observations. This assumption of the observations being conditionally independent of each other, given the states, is again a strong assumption and essentially forces all the serial dependence in the observations to be described by the state process alone, leaving the state-dependent distributions to capture the commonality between observations within each state, independent of time. In particular, for any period of time that the state process remains within a single state, the observations within that period are assumed to be independent of each other, which will not always be realistic.

HMMs are extremely flexible with respect to observation type: observations can be discrete or continuous, multivariate, and have distributions with parameters that depend on covariates via suitable link functions \citep[e.g.\ see Table 2 in][]{McClintockMichelot2018}. As with the state process, the observation process can include additional pre-defined structure that reflects the system of interest. For example, in Cormack-Jolly-Seber models any individuals in the ``dead'' state cannot be detected, and the state-dependent distribution for the ``dead'' state is therefore pre-defined as a Bernoulli distribution with a success probability of zero (see Section 3.1.1 in the main text).

In the striated caracara example, VDBA is a positive and continuous quantity. It therefore seems adequate to use for example Weibull or gamma state-dependent distributions in this case. Below we show the results obtained assuming VDBA at time $t$ $(X_t)$ to be gamma-distributed within each state:
\begin{equation*}
X_t \mid S_t = i \sim \text{Gamma}\left(\kappa_i,\rho_i\right),    
\end{equation*}
where $\kappa_i$ and $\rho_i$ are the shape and scale parameters, respectively, of the state-dependent gamma distribution for $i \in \{1,2,3,4\}$, with corresponding mean $\mu_i=\kappa_i \rho_i$ and variance $\sigma_i^2=\kappa_i \rho_i^2$. We have imposed no additional structure for the VDBA observations, and all of the state-dependent parameters are therefore to be freely estimated and entirely data-driven.

\section{Parameter estimation} 
\label{sec:estimation}
\subsection{Overview}

Once a model formulation has been identified, the next step is to estimate the model's parameters, summarised in the parameter vector $\boldsymbol{\theta}$, based on the observation sequence $(x_1,\ldots,x_T)$. There are three main strategies for estimating the parameters of an HMM:
\begin{itemize}
\item maximum likelihood by direct numerical maximisation of the likelihood;
\item maximum likelihood by the expectation-maximisation (EM) algorithm;
\item Bayesian inference using Markov chain Monte Carlo (MCMC).
\end{itemize}
Accessible reviews on the theoretical and practical differences between maximum likelihood and Bayesian analysis methods include \citet{ellison2004bayesian}, \citet[Chapter 4]{newman2014modelling}, and \citet{patterson2017statistical}. Although not intended specifically for ecologists, \citet{Rabiner1989} provides a very accessible introduction to fitting HMMs using the EM algorithm. Within each of these approaches, the existence of efficient recursive schemes for evaluating the likelihood of an HMM is a key asset
. We therefore proceed by providing detailed information on how the likelihood is evaluated. 

\subsection{Likelihood evaluation using the forward algorithm}
\label{subsec:llk}

Using the model assumptions --- i.e.\ the Markov property for the state process $S_1,S_2,\ldots,S_T$, and conditional independence of the observations $X_1,X_2,\ldots,X_T$, given the states ---
the likelihood of an $N$--state HMM can be obtained as
\begin{align*}\label{bruteforceL}
\nonumber 
\mathcal{L}(\boldsymbol{\theta} \mid x_1,\ldots,x_T) & = 
f_{\boldsymbol{\theta}}(x_1,\ldots,x_T) \\
\nonumber & = 
\sum_{s_1=1}^N \ldots \sum_{s_T=1}^N f_{\boldsymbol{\theta}}(x_1,\ldots,x_T | s_1, \ldots, s_T) f_{\boldsymbol{\theta}}(s_1, \ldots, s_T) \\
& = \sum_{s_1=1}^N \ldots \sum_{s_T=1}^N \delta_{s_1} \prod_{t=1}^T f_{\boldsymbol{\theta}} (x_t | s_{t})  \prod_{t=2}^T \gamma_{s_{t-1},s_t}.
\end{align*}
The first step uses the law of total probability, while the second is an immediate consequence of the model's dependence structure. The state-dependent densities (or probabilities, for discrete data) $f_{\boldsymbol{\theta}} (x_t | s_{t})$ depend on the distributional assumption made for the state-dependent (observation) process. In the VDBA example continued from above, with our assumptions made these would be densities of the gamma distribution, with one set of shape and scale parameters for each of the $N$ possible states. 

In this form, the likelihood involves $N^T$ summands, rendering its evaluation infeasible even for a moderate number of observations, $T$. This has led some users to believe that standard (numerical) likelihood maximisation is not feasible for HMMs. This is generally not true, as there is in fact a much more efficient way to calculcate the likelihood, namely the {\em forward algorithm}, which exploits the model's dependence structure to avoid the above brute force summation over all possible state sequences. To explain the inner workings of the forward algorithm, we define the forward variables at time $t$ as 
$$ {\alpha}_{t}(j) = f_{\bm\theta}(x_1,...,x_{t},s_{t}=j),\quad j=1,\ldots,N,$$
which can be summarised in a vector as $ \boldsymbol{\alpha}_{t} = \bigl( {\alpha}_{t}(1), \ldots, {\alpha}_{t}(N) \bigr)$. A close look at $ {\alpha}_{t}(j) $ reveals that this quantity comprises information on both the likelihood of all observations up to time $t$, since
$$ f(x_1,...,x_{t}) = \sum_{j=1}^N f(x_1,...,x_{t},s_{t}=j) = \sum_{j=1}^N {\alpha}_{t}(j), $$
as well as on the conditional probability of state $j$ being active at time $t$, given all observations up to time $t$, since 
$$ \Pr(S_t=j \mid x_1,\ldots,x_t) = \frac{f(x_1,...,x_{t},s_{t}=j)}{f(x_1,...,x_{t})} = \frac{{\alpha}_{t}(j)}{\sum_{k=1}^N {\alpha}_{t}(k)}.$$
The forward algorithm now traverses along the time series, updating $\boldsymbol{\alpha}_{t}$ step-by-step (and hence the likelihood, while retaining information on the probabilities of being in the different states). More specifically, from the dependence assumptions it follows that
$$ 
{\alpha}_{t}(j) = \sum_{i=1}^N  {\alpha}_{t-1}(i) \gamma_{ij}  f_{\bm\theta}(x_t\; |\;s_t=j) .
$$
In matrix notation, this becomes 
$$ \boldsymbol{\alpha}_{t} = \boldsymbol{\alpha}_{t-1} \boldsymbol{\Gamma} \mathbf{P}(x_{t}),$$
where $\mathbf{P}(x_{t})= \text{diag} \bigl( f_{\bm\theta} (x_{t}\;|\;s_t = 1), \ldots, f_{\bm\theta}(x_{t}\;|\;s_t  = N) \big)$. Together with the initial calculation
$ \boldsymbol{\alpha}_{1} = \boldsymbol{\delta} \mathbf{P}(x_{1})$, 
this is the forward algorithm. Note that the exact specification of three model-defining components, $\boldsymbol{\delta}$ (initial state distribution), $\boldsymbol{\Gamma}$ (state transition probability matrix), and $\mathbf{P}({x}_{t})$ (state-dependent observation distributions), depends on the model formulation considered --- as shown in the many detailed examples provided in the main text as well as the specific example of the VDBA series described above. 

The forward algorithm can be applied in order to first calculate $\boldsymbol{\alpha}_{1}$, then $\boldsymbol{\alpha}_{2}$, etc., until one arrives at $\boldsymbol{\alpha}_{T}$, the sum of all elements of which obviously yields the likelihood. Thus,
\begin{equation}\label{eq:lik}
\mathcal{L}(\boldsymbol{\theta} \mid x_1,\ldots,x_T) = \boldsymbol{\delta} \mathbf{P}({x}_1) \boldsymbol{\Gamma} \mathbf{P}({x}_2) \cdots \boldsymbol{\Gamma} \mathbf{P}({x}_{T-1}) \boldsymbol{\Gamma} \mathbf{P}({x}_T) \mathbf{1} \, ,
\end{equation}
where $\mathbf{1}$ is a column vector of ones. The computational cost of evaluating Eq.\ (\ref{eq:lik}) is {\it linear} in the number of observations, $T$, such that likelihood evaluation is typically feasible even for sequences comprising millions of observations. The existence of such recursive techniques for fast evaluation of the likelihood is a key reason for the popularity of HMMs. It is worth pointing out that this step-by-step recursive calculation is possible due to the dependence assumptions made at the process. 


\subsection{Model fitting} 

\begin{figure}[!htb]
\centering
\includegraphics[scale=0.7]{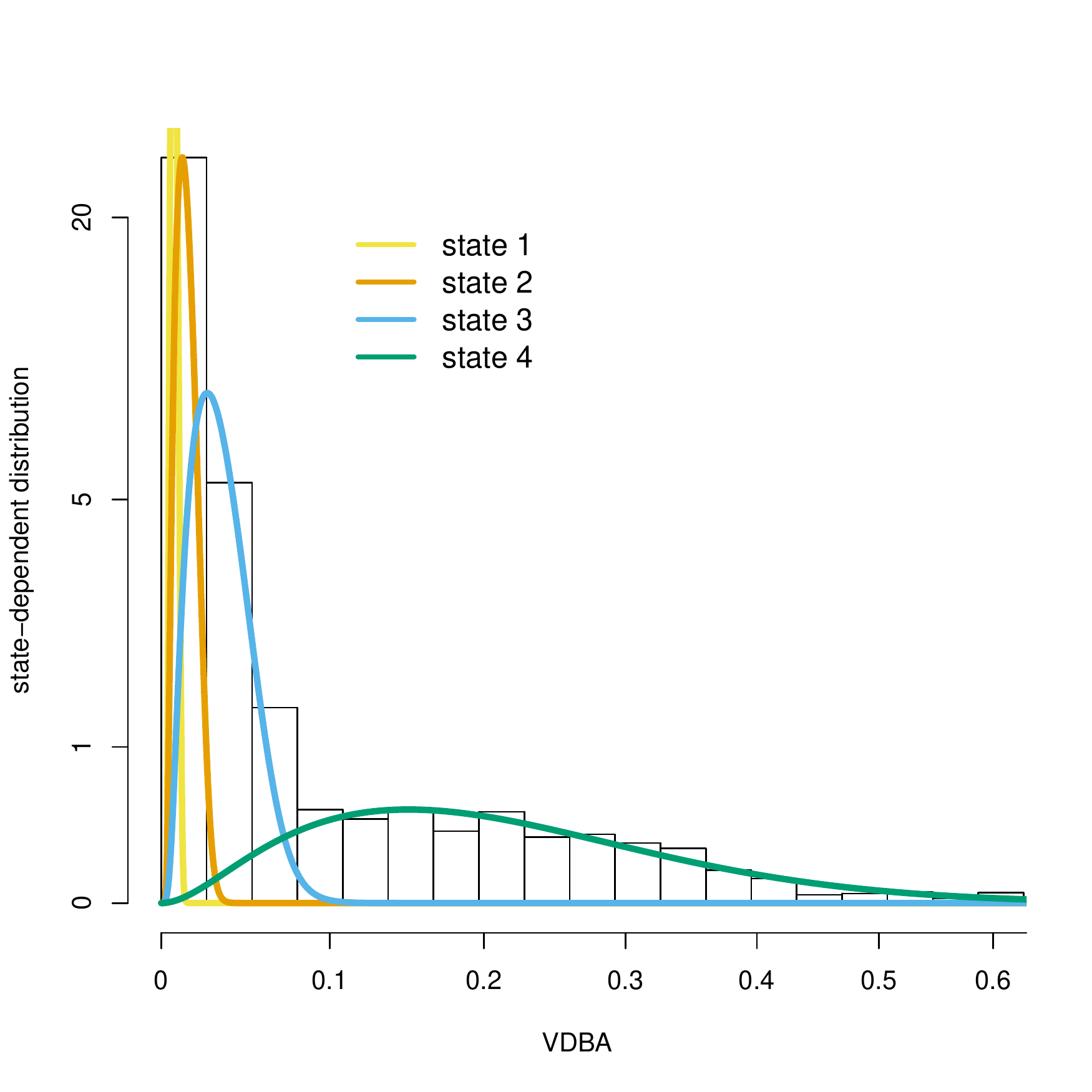}
\caption{Estimated gamma state-dependent distributions within a four-state HMM for a striated caracara.}
\label{fig:statedep}
\end{figure}

Of the three common approaches to fitting an HMM --- numerical likelihood maximisation, EM algorithm, and MCMC --- we used numerical likelihood maximisation for the VDBA data. For example with $N = 4$ states, we obtain estimates for the initial distribution, transition probability matrix, and state-dependent distributions (Figure \ref{fig:statedep}): 

\begin{equation*}
    \begin{blockarray}{ccccc}
    & S_1=1 & S_1=2 & S_1=3 & S_1=4  \\
    \begin{block}{c[cccc]}
    \hat{{\boldsymbol \delta}}=    &  0.30 & 0.28 & 0.28 & 0.14 \\
    \end{block}
    \end{blockarray}
\end{equation*}

\begin{equation*}
\widehat{\boldsymbol{\Gamma}}=   
    \begin{blockarray}{ccccc}
    S_{t+1}=1 & S_{t+1}=2 & S_{t+1}=3 & S_{t+1}=4  & \\
    \begin{block}{[cccc]c}
    0.90 & 0.09 & 0.00 & 0.00 & S_t=1 \\
    0.1 & 0.73 & 0.16 & 0.01 & S_t=2 \\
    0.00 & 0.17 & 0.79 & 0.04 & S_t=3 \\
    0.00 & 0.00 & 0.10 & 0.90 & S_t=4 \\
    \end{block}
    \end{blockarray}
\end{equation*}

\begin{equation*}
    \begin{blockarray}{ccccc}
    & S_{t}=1 & S_{t}=2 & S_{t}=3 & S_{t}=4  \\
    \begin{block}{c[cccc]}
    \hat{{\boldsymbol \mu}}=    &  0.007 & 0.013 & 0.032 & 0.227 \\
    \end{block}
    \end{blockarray}
\end{equation*}

\begin{equation*}
    \begin{blockarray}{ccccc}
    & S_{t}=1 & S_{t}=2 & S_{t}=3 & S_{t}=4  \\
    \begin{block}{c[cccc]}
\hat{{\boldsymbol \sigma}}= &  0.001 & 0.004 & 0.014 & 0.132 \\
    \end{block}
    \end{blockarray}
\end{equation*}

The relatively low computational cost involved in evaluating the likelihood renders numerical maximisation feasible in most cases. For example, fitting the HMM above with $N=4$ states to the VDBA sequence comprising 3600 observations took 45 seconds in R on an octa-core i7 CPU, at 3.4 GHz and with 8 GB RAM --- analysing sequences comprising millions of observations is also generally feasible (depending on model complexity). The computation time can often further substantially be reduced by implementing the forward algorithm in C++, as is done in \verb|momentuHMM| 
, where the same model only requires about 10 seconds to fit. Technical issues arising in the numerical maximisation, such as parameter constraints and numerical underflow, are fairly straightforward to deal with \citep[][pp.\ 50-54]{ZucchiniEtAl2016}. 
For complex models, local maxima of the likelihood function can become problematic \citep[e.g.][]{Myung2003}. In this respect, it is generally advisable to always try a few different random initial parameter vectors in the numerical optimisation.

The EM algorithm and MCMC approach to fitting HMMs are equally feasible approaches that can exploit the recursive schemes available for HMMs, such as the forward algorithm. In some cases, the EM algorithm can lead to a maximisation problem that is easier to solve than numerical maximisation of the likelihood, but EM can be more cumbersome to implement. The MCMC approach easily takes the HMM model into a Bayesian framework, using the forward algorithm to marginalise over states rather than data augmentation \citep{TurekEtAl2016,patterson2017statistical,YackulicEtAl2020}.  

\section{Model selection \& model checking}

The task of selecting a particular HMM formulation from a suite of candidate models --- which may differ, for example, in the number of underlying states, the family of distributions used for the state-dependent process, or the set of covariates that affect the state transition probabilities --- is effectively analogous to any other model selection, say in regression analysis. As a consequence, general recommendations applicable also to such more established modelling classes directly transfer to the class of HMMs. In particular, model selection and checking can be approached in three steps:
\begin{enumerate}
\item consider both exploratory data analysis but also any relevant theory in order to keep the number of candidate models small (not just for computational reasons, but also to minimise the risk of a selection bias; cf.\ \citealp{zucchini2000introduction});  
\item use model selection criteria for guidance, but do not solely rely on such criteria (see \citealp{pohle2017selecting}, for a more detailed discussion of model selection in HMMs);
\item instead, carefully inspect the goodness-of-fit of any promising model to investigate if all patterns in the data pertinent to the study aim are sufficiently well captured by the model (and if not, to identify which components of the model may need to be modified).
\end{enumerate}

\subsection{Model selection} 

For Step 1, in the VDBA example we initially restrict models to four or fewer states based on expert knowledge of the species' behaviour, and further on the observation that it is unlikely that any sensible behaviour classification would lead to five or more genuinely different behavioural modes in a one-hour period. Given this restriction, we can use model selection criteria (Step 2) such as the AIC or the BIC to select between models with up to four states. In this case, both the AIC and the BIC are minimised by the model with $N=4$ states. 

In general, when state characteristics are completely data-driven rather than pre-defined \citep[e.g.][]{Leos-BarajasEtAl2017accelerometer}, there is a tendency for HMMs with more states than seem biologically reasonable to be supported by standard model selection criteria such as AIC or BIC. The reason for this is that any additional state can, to some extent, compensate for any lack of structure or flexibility in the model formulation (cf.\ \citealp{pohle2017selecting}). In general, the judgements taken in Step 1 will often need to be relied upon to justify restricting the complexity of the model in spite of a poorer fit to the data.

\subsection{Model checking} 

Step 3 concerns model checking. Unfortunately, this important step is not as straightforward as for example in a regression analysis. 
The main \textit{formal} approach to model checking in HMMs considers so-called \textit{pseudo-residuals}, defined as 
$$ r_t = \Phi^{-1}(F_{X_t}(X_t)), \quad t=1,\ldots,T,$$
where $\Phi$ is the cumulative distribution function of the standard normal distribution (such that $\Phi^{-1}$ is the corresponding quantile function) and
$$ F_{X_t}(x)=\Pr(X_t\leq x \, | \, X_1=x_1, \ldots X_{t-1}=x_{t-1}, X_{t+1}=x_{t+1}, \ldots X_T=x_T) $$
is the conditional cumulative distribution function of $X_t$, given all other observations, under the fitted model. The conditional distribution $F_{X_t}(x)$ can be obtained using a combination of the forward and the backward variables, respectively (for details, see \citealp{ZucchiniEtAl2016}, pp.\ 82-83). For continuous variables it can be shown that \textit{if the model is correct}, such that $F_{X_t}$ is indeed the conditional cumulative distribution function of $X_t$ given all other observations, then $F_{X_t}(X_t)\sim \text{Uniform}[0,1]$, such that $r_t\sim N(0,1)$. Thus, any deviation of the pseudo-residuals $r_t$ from normality indicates a potential lack of fit, which should be further investigated. While the $r_t$ are not uncorrelated even for the correctly specified model, they should show much reduced correlation compared to the original sequence of observations. The main idea underlying these pseudo-residuals is that the distribution of each $X_t$ within an HMM ought to be seen in the context of neighbouring observations in the sequence, which renders it non-trivial to assess which observations are extreme relative to the model. The transformation above, which is more generally known as the {probability integral transform}, yields a common scale, the standard normal distribution, for all observations, making it easier to identify observations that are extreme relative to the fitted model, and more generally any lack of fit.

\begin{figure}[!htb]
\centering
\includegraphics[scale=0.62]{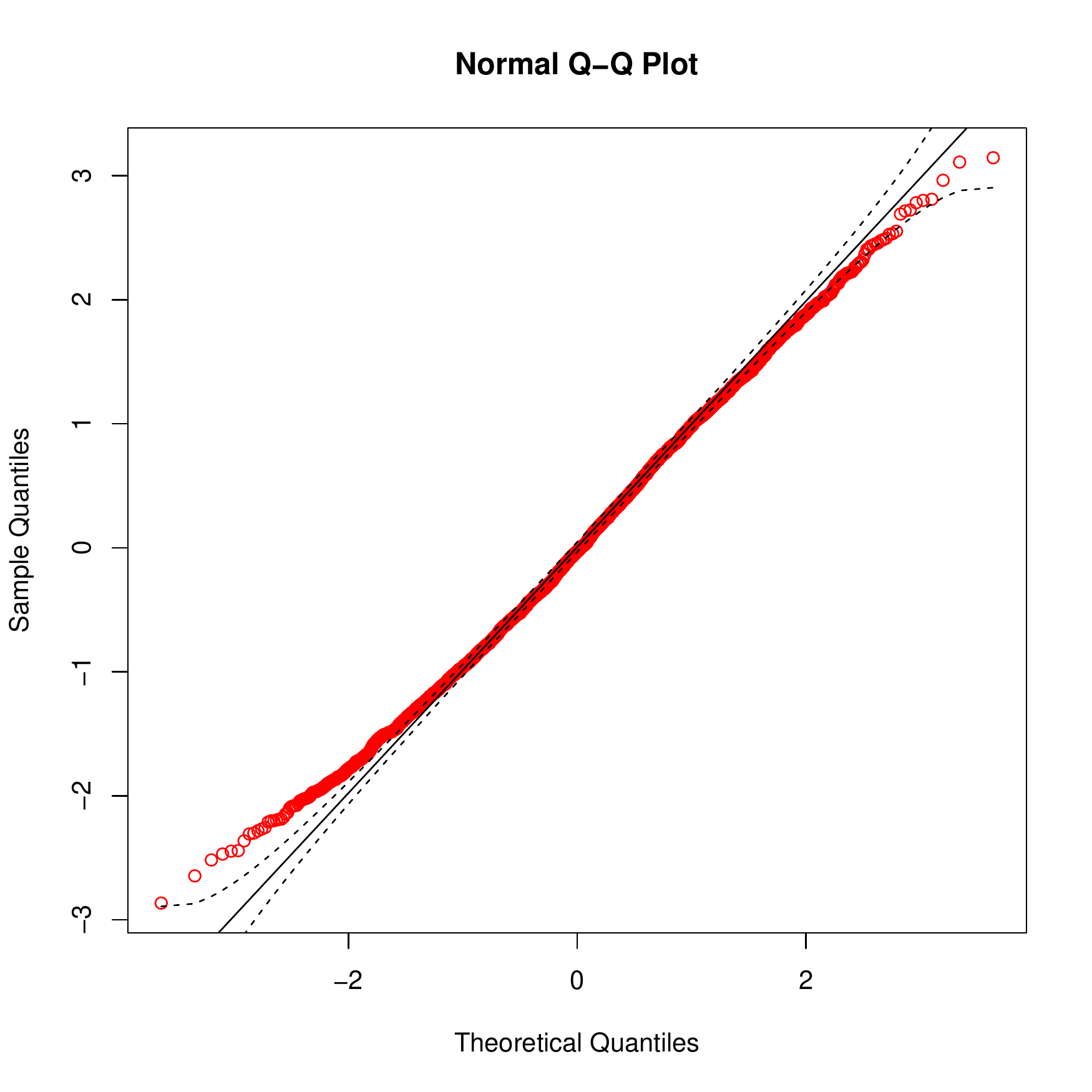}
\caption{QQ-plot for the pseudo-residuals of the VDBA model with four states. Dashed lines indicate point-wise 95\%\ confidence envelopes.}
\label{fig:pseudores}
\end{figure}

For the VDBA example and the model with $N = 4$ states, Figure \ref{fig:pseudores} shows the QQ-plot for the pseudo-residuals. They show deviation from normality in the tails of the distribution, indicating some lack of fit. In particular, the pattern in the residuals indicates that fewer observations of low accelerations were observed than one would expect under the fitted model. Nevertheless, the deviation is relatively minor and not necessarily a cause for concern, unless the focus of the analysis is specifically on these very small VDBA values and the associated behaviours.

\begin{figure}[!htb]
\centering
\includegraphics[scale=0.44]{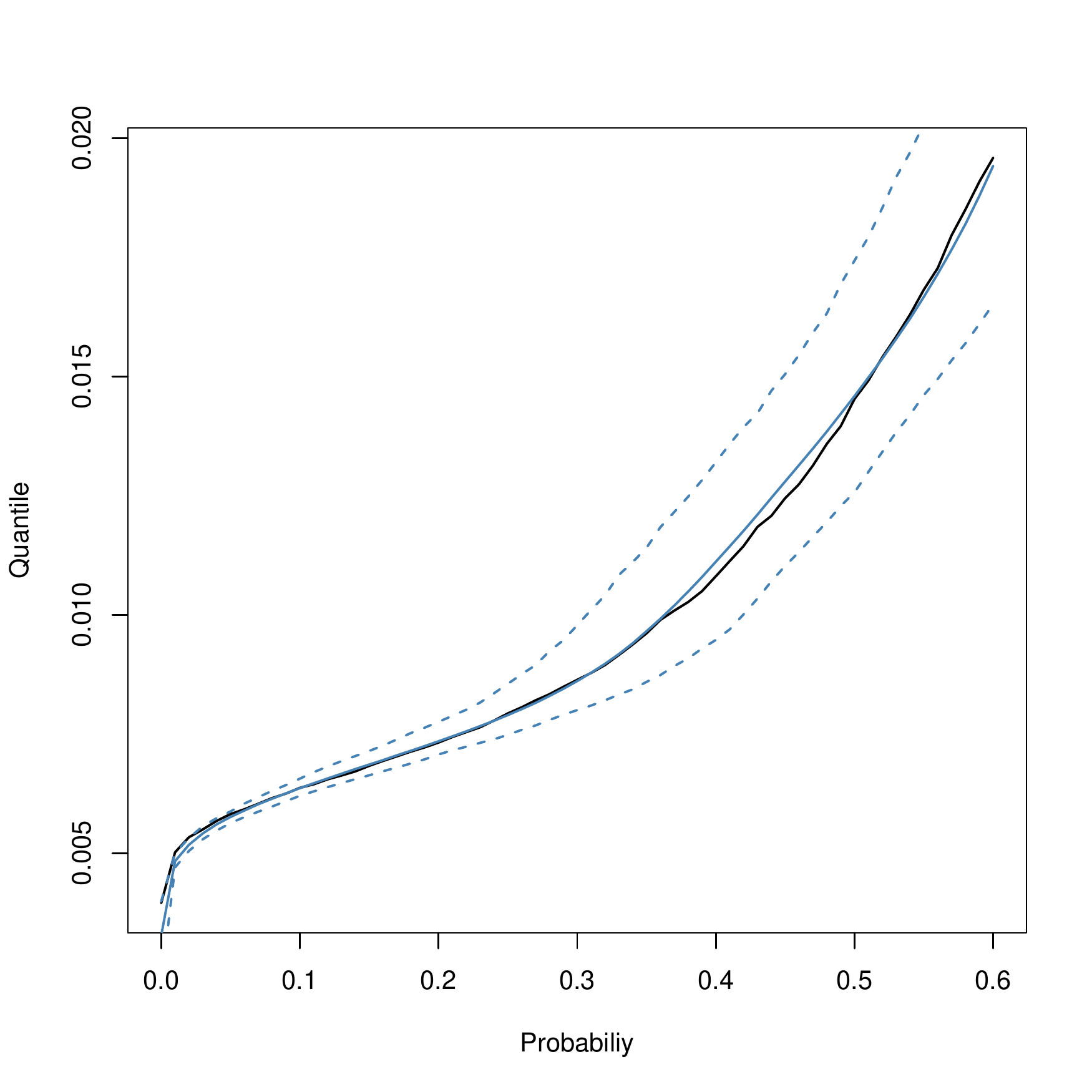}
\includegraphics[scale=0.44]{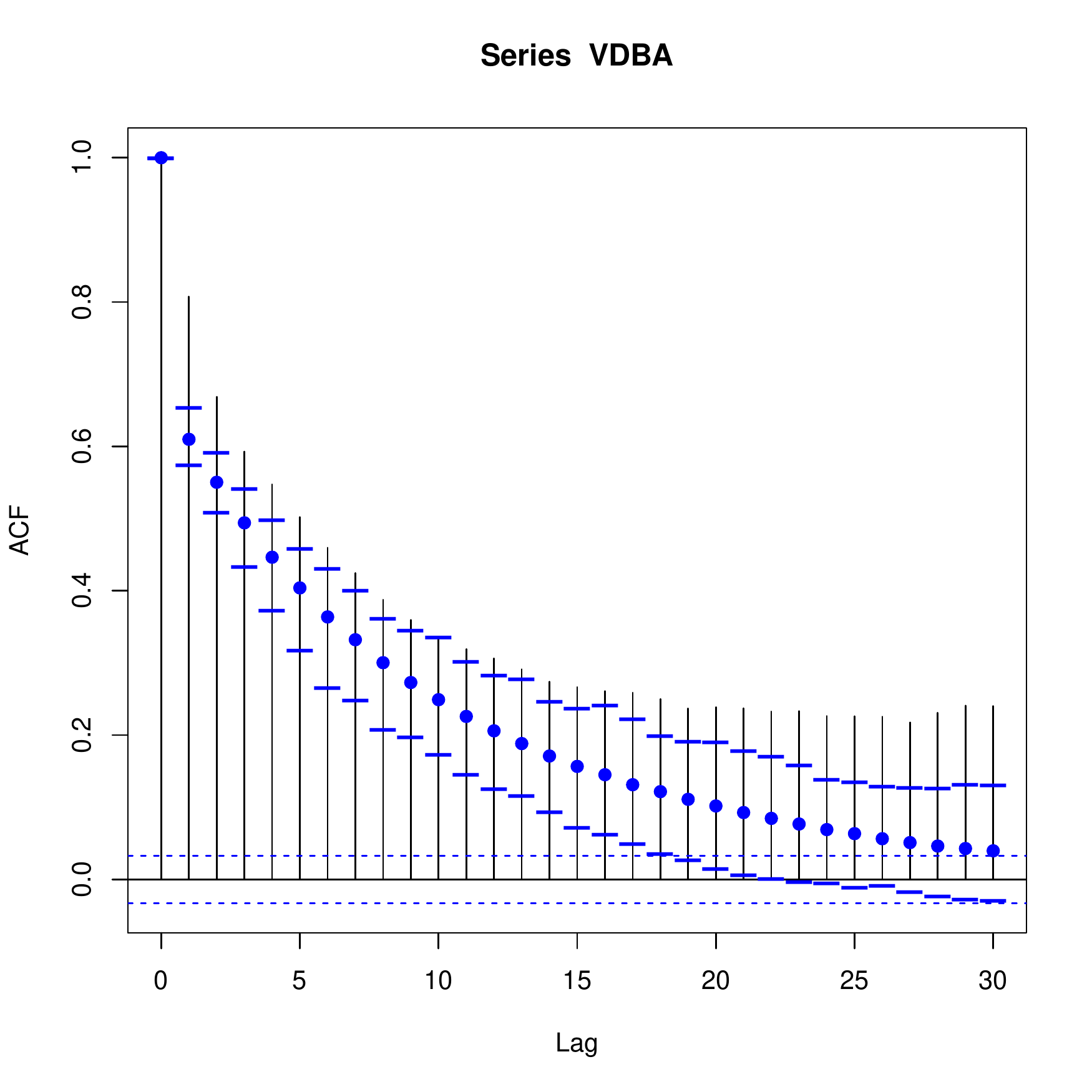}
\caption{Estimated expected (blue) and empirical (black) quantile function (left) and autocorrelation function (right) using $100$ simulated datasets from the VDBA model with four states; \(95\%\) quantile bounds are given for the expectations.}
\label{fig:gof}
\end{figure}

Calculation and interpretation of the pseudo-residuals is somewhat tedious, as both the forward and the backward variables are required, and numerical underflow issues may need to be dealt with. Software packages that specialise in HMMs can be used to perform these calculations (see Section 4.1 in main text), but it can also be appealing to consider less technically involved approaches. An \textit{informal} but often very useful approach to model checking consists of simulating observation sequences from the fitted model, then checking if relevant patterns in the real data are well replicated by the simulated data. The corresponding comparisons between simulated and real data could focus on summary statistics such as specific quantiles of the empirical distribution or the values of the sample autocorrelation function (ACF). 
For example, in the VDBA example, we simulated $100$ datasets under the fitted model with $N = 4$ states. Figure \ref{fig:gof} shows the quantile function and ACF estimated from these simulations. Overall, we find that the HMM fits the data fairly well, capturing the overall patterns in the data. However, we can also see the model's deficiencies: the quantile plot shows a minor lack of fit for very small observations, and the ACF shows that the correlation between neighbouring observations is higher than expected under the model, and that correlation at higher lags is also underestimated. The poorer fit of the ACF may indicate the limitation of the Markov assumption and the assumed geometric distribution in state dwell times, or alternatively a violation of the conditional independence assumption of the observations. Corresponding model extensions could improve the fit --- whether or not the corresponding effort necessary to fit those more complex models is warranted depends on the likely role of the lack of fit given the study aim.

\section{State decoding}  

\subsection{Overview}

Once parameters have been estimated and the model checking is complete, one goal may be to infer
the hidden states $s_1, \ldots, s_T$. This is called ``state decoding". There are two approaches for decoding: local and global. 

\begin{figure}[!h]
\centering 
\includegraphics[width=\textwidth]{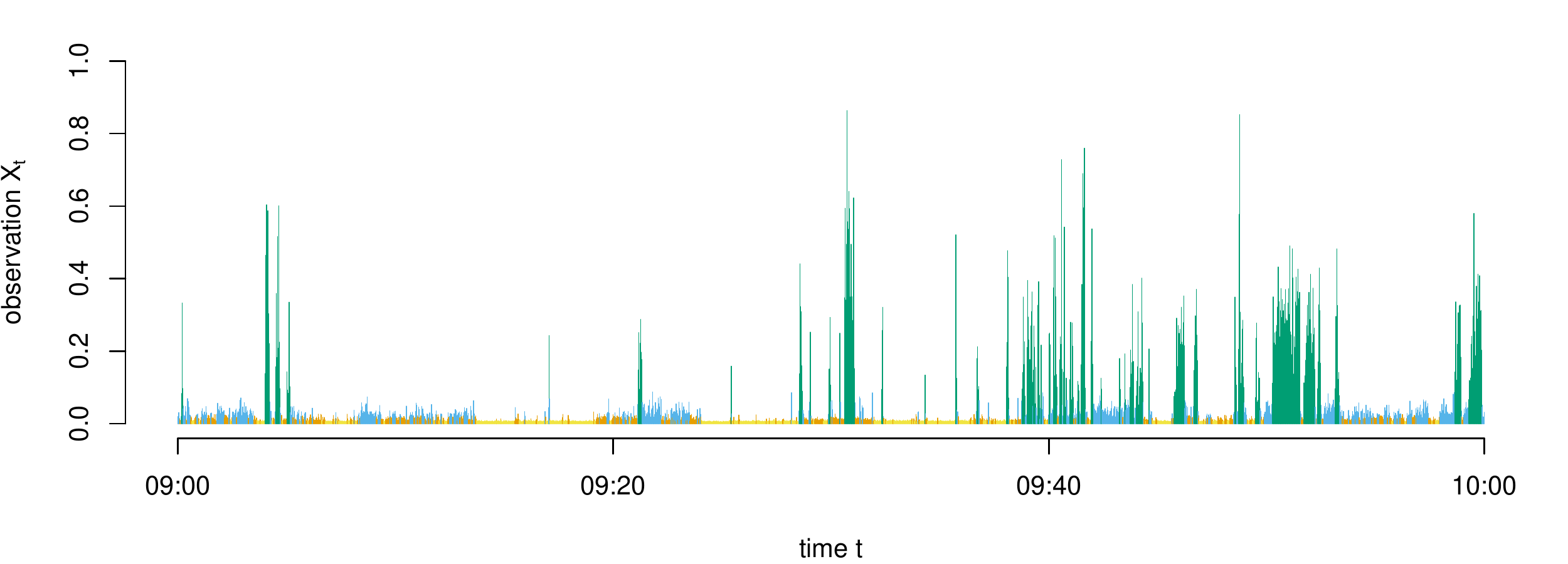}
\caption{Sequence of vectorial dynamic body accelerations $X_t$, $t=1,\ldots,3600$, colour-coded according to the most likely state sequence as inferred using the Viterbi algorithm.}
\label{fig:vit}
\end{figure}

\begin{itemize}
\item \textbf{Local} state decoding is when we infer which value of $s_t$ is \emph{most likely for each time point separately}, that is, we seek the locally decoded state

$$ l_t = \argmax_{s_t} \;\text{Pr}(S_t = s_t \;|\; x_1, \ldots, x_T).$$

Notice that the sequence of locally decoded states $l_1, \ldots, l_T$ ignores the serial dependence in the states: individually most likely does not mean they are jointly most likely. 

\item \textbf{Global} state decoding is when we infer which complete sequence of states is the \emph{jointly most likely}, that is, we seek 

$$(g_1, \ldots, g_T) = \argmax_{(s_1,\ldots,s_T)} \;\text{Pr}(S_1 = s_1, \ldots, S_T = s_T \;|\; x_1, \ldots, x_T).$$
\end{itemize} 
Whether to use locally or globally decoded states depends on the question one is trying to answer: whether one is interested in what state the Markov chain inhabits at a particular time or in the most likely sequence of states the chain followed. In practice the results from both approaches are usually similar. Figure \ref{fig:vit} shows the globally decoded states for the VDBA example. In this case, the locally decoded states agreed with the globally decoded ones over $95\%$ of the time. 

\subsection{Computation} 

Local state decoding involves both the forward probabilities (defined in Section \ref{subsec:llk}) and backward probabilities. Contrasting with the forward probabilities, $\alpha_t(j) = \text{Pr}(x_1, \ldots, x_t, s_t = j)$, the backward probabilities are defined as $\beta_t(j) = \text{Pr}(x_{t+1}, \ldots, x_{T}\;|\; s_t = j)$. The following useful relationship holds for any $t$:  
$$\mathcal{L}(\bm\theta\;|\; x_1, \ldots, x_T) = \bm\alpha_t^{T}\bm\beta_t,$$
since $\alpha_t(j)\beta_t(j) = \text{Pr}(x_1, \ldots, x_T, s_t = j)$. Local decoding involves maximising the quantity $\text{Pr}(S_t = s_t \;|\; x_1, \ldots, x_T)$ with respect to $s_t$, for every $t$. By basic probability laws, it can be shown that
\begin{equation}\label{locdec}
\text{Pr}(S_t = s_t \;|\; x_1, \ldots, x_T) = \dfrac{\alpha_t(s_t)\beta_t(s_t)}{\mathcal{L}}.
\end{equation}
At any time point $t$, this expression can be evaluated for $s_t=1,\ldots,N$, yielding the locally decoded state $l_t$ as the argument maximising the expression.

Global decoding is a more complex optimisation problem, but remarkably can be solved efficiently using the Viterbi algorithm. Let $q_t(j) = \text{Pr}(s_1 = g_1, \ldots, s_{t - 1} = g_{t-1}, s_t = j, x_1, \ldots, x_T)$, that is, suppose we know the optimal sequence up to time $t - 1$ and so $q_t(j)$ is the joint probability (with the data) that the state at time $t$ is $j$. The Viterbi algorithm makes use of the following recurrence: $q_t(j) = \text{max}_i\; q_{t - 1}(i)\gamma_{i,j} f(x_t\;|s_t = j)$. Clearly $q_1(j) = \delta_jf(x_1\;|\;s_1 = j)$  and so $q_t$ can be computed for every $t$. The globally decoded states can then be determined by going backwards in time: choose $g_T= \argmax_j q_T(j)$ and then $g_t = \argmax_j q_t(j)\gamma_{j,g_{t + 1}}$ for $t < T$.

It is worth noting that only local decoding via Eq.\ (\ref{locdec}) provides an uncertainty quantification with respect to the decoded states, i.e.\ probability statements on which state may have been active at any time point. Global decoding via the Viterbi algorithm merely provides the hard decoded most likely state sequence without any probabilistic information on potential state misclassifications.

\section{Interpretation}  

We have formulated the HMM, estimated the parameters, selected between candidate models, checked how well the model fits, inferred the latent states, and now must interpret the final results.  
It is important to remember at this stage two things: 
\begin{itemize}
\item Even though the state process may have been formulated according to certain known (or hypothesised) properties of an underlying ecological process, the parameters of the state-dependent distributions and the transition probability matrix are driven by the data and so may instead lead to states that do not correspond to our original intentions. Our choice of the number of states, the data streams to include, and the structure of the HMM components can encourage the model to reflect the ecological dynamics we are interested in, but this is not formally guaranteed and so results must be interpreted with caution. HMMs are aids to our understanding of these ecological processes, but not substitutes for ecological theory. 

\item The results obtained must be interpreted in light of the assumptions that have been made. The number of assumed states, the assumed form of the state-dependent distributions, the Markov assumption, and the conditional independence assumption. Whether or not these assumptions are well enough respected in any given application must be questioned during model checking, and any deficiencies taken into account when interpreting results. Identifying important issues with an HMM may lead to specifying a more complex model (e.g.\ extensions of HMMs as discussed in Section 2.3 of the main text), or could lead to restricting research conclusions to robust, but weak statements. When the assumptions of HMMs and their extensions fail to adequately describe the underlying ecological dynamics of interest, then other forms of latent variable models may be more appropriate (see Box 1 in main text).
\end{itemize}
For the VDBA example, we settled on an HMM with $N=4$ states. Considering the estimated state-dependent parameters in the context of ecological knowledge available for this species, it is possible for experts to qualitatively associate these states with behavioural modes: ``resting'', ``minimal activity'' (e.g.\ preening), ``moderate activity'' (e.g.\ walking, digging), and ``flying'', which based on global state decoding were respectively assigned to 31\%, 27\%, 27\%, and 14\% of time steps over a period of one hour. In model checking, we found some relatively minor issues with both the marginal distribution and the dependence structure. If our interest is in the broad classification of behaviours and the observations that relate to these behaviours, these issues may not be important. However, if for example our research question concerns how long individuals spend in each state, then the model's failure to capture the full empirical dependence structure may need to be addressed with more sophisticated HMMs or alternative modelling frameworks. 

Overall, this tutorial briefly demonstrates the workflow when trying to uncover underlying ecological dynamics with an HMM. Like all modelling, it comes with assumptions and limitations, but HMMs also bring computational efficiency and a structure that has been found 
to be applicable to an impressively wide range of ecological applications. 

\bibliographystyle{apalike}
\bibliography{master}